\documentclass[12pt]{amsart}

\usepackage{bbold}
\usepackage{amssymb}
\usepackage{graphics}
\usepackage{epsfig}
\usepackage{times}
\usepackage{euscript}
\usepackage{amsfonts}
\def\Comma{\,,\qquad}

\def\12{\frac{1}{2}}
\def\Aut{{\rm Aut}}

\numberwithin{equation}{section}
\numberwithin{figure}{section}

\def\ss{\scriptstyle}

\def\wh{\widehat}

\def\a{\alpha}
\def\b{\beta}

\def\d{\partial}

\def\tr{{\rm tr}\ }

\def\be{\begin{equation}}
\def\ee{\end{equation}}
\def\bea{\begin{eqnarray}}
\def\eea{\end{eqnarray}} 
\def\eqn#1{(\ref{#1})}

\def\nn{\nonumber}

\newtheorem{theorem}{Theorem}[section]
\newtheorem{lemma}[theorem]{Lemma}
\theoremstyle{remark}

\begin{document}

\title[Duality for Matrix Integrals]{
Duality of Orthogonal and Symplectic Matrix Integrals  
and Quaternionic Feynman Graphs}

\author[M.~Mulase]{Motohico Mulase$^{1}$}
\address{
Department of Mathematics\\
One Shields Avenue\\
University of California\\
Davis, CA 95616--8633}
\email{mulase@math.ucdavis.edu}

\author[A.~Waldron]{Andrew Waldron$^{2}$}
\address{Department of Mathematics\\
One Shields Avenue\\
University of California\\
Davis, CA 95616--8633}
\email{wally@math.ucdavis.edu}

\date{\today}
\thanks{$^{1}$Research supported by NSF Grant DMS-9971371 
and the University of California, $\phantom{\, oph}$Davis.}
\thanks{$^{2}$Research supported by
the University of California, Davis.}

\allowdisplaybreaks\setcounter{section}{0}

\begin{abstract}We present an asymptotic expansion for 
quaternionic self-adjoint matrix integrals. The 
Feynman diagrams appearing
in the expansion are  ordinary ribbon graphs and their non-orientable 
counterparts.  The result exhibits 
a striking duality between quaternionic self-adjoint and real 
symmetric matrix
integrals.
The asymptotic expansions of these integrals are given
in terms of summations over topologies of compact surfaces,
both orientable and non-orientable, for all genera and
an arbitrary positive number of marked points on them.
We show that the Gaussian Orthogonal
Ensemble (GOE) and Gaussian Symplectic Ensemble (GSE)
have exactly the same graphical expansion term by term
(when appropriately normalized),
except that the contributions from non-orientable surfaces with
odd Euler characteristic carry the opposite sign. 
As an application, we give a new topological proof of the known
duality 
for correlations 
of characteristic polynomials. Indeed, we show that
this duality is equivalent to Poincar\'e
duality of graphs drawn on a compact surface. 
Another application of our graphical expansion formula
is a simple and simultaneous 
(re)derivation of the \emph{Central Limit Theorem}
for GOE, GUE (Gaussian Unitary Ensemble) and GSE: 
The three cases have exactly the same graphical
limiting formula except for an overall constant that
represents the type of the ensemble.
\end{abstract}

\maketitle

\newpage

\tableofcontents

\section{Introduction}

The purpose of this paper is to establish an asymptotic
expansion for quaternionic matrix integrals
in terms of Feynman diagrams which exhibits a duality between
quaternionic self-adjoint and real symmetric
matrix integrals.

Recent developments in the theory of random matrices 
exhibit particularly rich structures. 
Although originally introduced by Wigner as a model for
heavy nuclei, random matrices appear almost ubiquitously
in mathematics and physics. Applications pertain, for example, to
number theory~\cite{Katz-Sarnak}, combinatorics 
\cite{Baik-Deift-Johansson}, probability theory~\cite{Tracy-Widom},
algebraic geometry~\cite{Kontsevich}  and
quantum gravity~\cite{Witten}. 
't Hooft's discovery that quantum chromodynamics (QCD) simplifies
in the limit where the number of colors $N$ ({\it i.e.} gauge group $SU(N)$)
is large relied on a graphical expansion in terms of ``fat'' or
``ribbon'' graphs~\cite{'tHooft:1974jz}. 
While the three major ensembles ({\it Gaussian
Orthogonal, Gaussian Unitary,} and {\it Gaussian Symplectic})
have important applications, simple graphical expansions
in terms of ribbon graphs and their non-orientable counterparts
are currently available only for the GUE and GOE models.
What type of graphs should be used for the GSE models?
The non-commutativity of matrix entries has defied 
earlier attempts to find an efficient asymptotic expansion formula.
This article provides the missing graphical expansion
for symplectic matrix integrals.

\pagebreak

A $2N\times 2N$ symplectic matrix $M$
\begin{equation}
M C M^\top= C 
\end{equation}
($C=-C^\top=C^{^{-\ss\!\dagger}}$), may be written 
as\footnote{The
Pauli matrices are 
$$
\sigma_1=
\left(
\begin{array}{cc}
0&1\\1&0
\end{array}
\right)\; ,\quad
\sigma_2=
\left(
\begin{array}{cc}
0&-i\\i&0
\end{array}
\right)\; ,\quad
\sigma_3=\left(
\begin{array}{cc}
1&0\\0&1
\end{array}
\right)\; .\quad
$$  
}
\begin{equation}
M=I_{2\times 2}\otimes S+i\sigma_1\otimes A_1
+i\sigma_2\otimes A_2
+i\sigma_3\otimes A_3\, ,
\end{equation} 
where $S$ is a symmetric and $A_1$, $A_2$, $A_3$ are antisymmetric
$N\times N$ matrices. In turn, representing $(i\sigma_1,i\sigma_2,
i\sigma_3)$ as quaternionic units $(i,j,k)$ so that
\begin{equation}
M=S+iA_1+jA_2+kA_3\, ,\label{pauli}
\end{equation}
the problem is to 
write the Feynman rules for a quaternionic matrix integral.

{}'t Hooft's insightful fat graph notation arises from identifying 
each index $a$ and $b$ of an $N\times N$ matrix $(X_{ab})$
as an edge of a ribbon. 
The most naive approach therefore, is to append additional indices 
$\alpha,\beta=1,2$ corresponding to the $2\times2$ Pauli matrices
in~\eqn{pauli}. Unfortunately this method leads only to 
partial results~\cite{Itoi}. The next most obvious idea is to 
employ the quaternionic realization, in much the same way that 
an integration over hermitian matrices $X=X^\dagger$
is immensely simplified when one views the entries as complex numbers
rather than $X=S+iA$ (where the real matrices $S$ and $A$ are symmetric
and antisymmetric, respectively). The drawback is that the quaternions
do not commute, so a non-commutative Feynman calculus is required.
Indeed it is possible to generalize the usual Schwinger trick
to quaternionic source terms, and represent a Gaussian symplectic
integral as non-commutative quaternionic differentiations.
The result is a sum over both orientable and non-orientable
ribbon graphs, and is easily verified to agree with ours  for 
simple graphs. This method is described in Appendix~\ref{Qcalculus}.

Our result for the quaternionic matrix integral is
\begin{multline}
\log\left(
\frac{\textstyle\int [dX]\,\exp\Big(\!-N\ \tr X^2
+\sum_{j}  \frac{2N t_j}{j}\ \tr X^j \Big)}{
\textstyle\int [dX]\,\exp\Big(\!-N\ \tr 
X^2\Big)}\right)
\\
=
\sum_{\Gamma\in\mathfrak G}
\frac{(-2N)^{\chi(S_\Gamma)}
}{|{\rm Aut}(\Gamma)|} \prod_{j}
t_j^{v^{(j)}_\Gamma}\! .
\label{eq: intro}
\end{multline}
Exact conventions are given later,  
at present it suffices to indicate that the sum is over all graphs
$\Gamma$ drawn on compact orientable and non-orientable 
surfaces $S_\Gamma$, and $\chi(S_\Gamma)$
is the Euler characteristic of the surface $S_\Gamma$ 
uniquely defined by the graph
$\Gamma$. Our proof of this result is
based on viewing the GOE integral over real symmetric matrices
as fundamental. The crucial observation is that 
the contribution of any given graph $\Gamma$ in 
the GOE, GUE and GSE
is a topological invariant of the surface 
$S_\Gamma$ with $f_\Gamma$ \emph{marked points} on it,
where $f_\Gamma$ denotes the number of faces of the cell-decomposition
of $S_\Gamma$ defined by the graph $\Gamma$. 
The connectivity of the space of 
triangulations 
of two 
dimensional surfaces  then allows any graphical contribution to 
be calculated from a simple representative graph for any given topology.

Writing the results for all three ensembles in a uniform 
notation (see~\eqn{eq: invariant formula}) makes a new duality
\begin{equation}
\begin{array}{ccc}
{\rm GOE}&\longleftrightarrow&\widetilde{\rm GSE}\\ \\
{\rm GUE}&\longleftrightarrow&{\rm GUE}\\ \\
{\rm GSE}&\longleftrightarrow&\widetilde {\rm GOE}\ ,
\end{array}
\end{equation} 
manifest. The middle line for the GUE is a (trivial) self-duality. The
tilde on the right hand side indicates that equality holds 
upon doubling/halving the matrix size and
an overall sign change for contributions of graphs where the 
Euler characteristic
$\chi(S_\Gamma)$ is odd.

Asides from the computational utility of this duality, 
as an immediate corollary, it provides a simple
proof of the known duality 
for $k$-fold correlations of characteristic polynomials
of $N\times N$ matrices~\cite{baker,bh,bh1,mehta}.
For the GUE model expressed in terms of ribbon graphs, 
this $N$-$k$ duality~\cite{bh,bh1} is precisely Poincar\'e
duality of graphs drawn on a compact oriented surface. Similarly, the 
relation between GOE and GSE correlations
stems from the combination of 
Poincar\'e duality, this time including 
non-orientable surfaces, and our new graphical expansion formula.
It is interesting to note that the machinery of fermionic
integrations employed in~\cite{bh,bh1} is equivalent to a very 
simple switch from a graph on a surface to its dual graph.

If we reduce our integral~\eqn{eq: intro} to a symplectic
Penner model by setting $t_1=t_2=0$ and 
$$
t_j = -z^{\frac{j}{2}-1}, \quad j\ge 3,
$$
then we can explicitly compute the asymptotic expansion in $z$ of 
\begin{equation}
\label{eq: intro Penner}
\lim_{m\rightarrow\infty}
\log\left(\frac{
\int_{{\mathbb R}^N}|\Delta(k)|^{2\alpha}\prod_{i=1}^N
\exp\Big(-\sum_{j=2}^{2m} 
\frac{k_i^j}{j}\left(\frac{z}{\alpha N}\right)^{j/2-1}\Big) dk_i}
{\int_{{\mathbb R}^N}|\Delta(k)|^{2\alpha}\prod_{i=1}^N
\exp\Big(- \frac{k_i ^2}{2}\Big) dk_i}
\right)
\end{equation}
utilizing the Selberg integration formula and the asymptotic
analysis technique of~\cite{Mulase95}, where $\alpha$ is
either a positive integer or its reciprocal. 
We demonstrate that
the duality we found for GOE and GSE of 
\eqn{eq: intro} extends to the same type of duality
between an arbitrary positive integer
$\alpha$ and $1/\alpha$ for~\eqn{eq: intro Penner}
with the sign change for all terms with odd powers of $N$
in the asymptotic expansion.

The orthogonal Penner model
gives the orbifold Euler characteristic of the moduli spaces
of smooth real algebraic curves with an arbitrary number of
marked points~\cite{Goulden-Harer-Jackson}. Also, the
original Penner model~\cite{Penner} provides the orbifold
Euler characteristic of the moduli spaces of pointed algebraic
curves over $\mathbb{C}$~\cite{Harer-Zagier}. Therefore 
it is natural to ask what the
symplectic Penner model gives. 
Our duality shows that actually the symplectic Penner
model is identical to the orthogonal Penner model,
except for a change of the matrix size and an overall sign
for contributions from surfaces of odd Euler characteristic.
This equivalence has been observed independently in~\cite{Ooguri-Vafa}.

Our final application, is a simple and unified (re)derivation of the 
{\it Central Limit Theorem} for Gaussian random matrix ensembles.
This result follows as 
a direct consequence of 't Hooft's original large $N$ limit in which planar
ribbon graphs dominate: we derive a precise limiting formula for
GOE, GUE and GSE matrix ensembles
in terms of planar two-vertex ribbon graphs. The formula is the
same for all three ensembles, except for an overall constant.

The material is organized as follows: In Section~\ref{matrix}
we introduce the matrix integrals studied in this paper.
Our conventions for the topological data of surfaces are given
in Section~\ref{graphs} as well as our theorem 
and its proof for the graphical expansion of matrix integrals.
Examples, including a comparison with the first few terms 
of the Penner model, are given
in Section~\ref{examples}. 
The new duality appears in Section~\ref{Duality}. The central formula of
this paper is equation~\eqn{duality} which gives the graphical 
expansion for the GOE, GUE and GSE simultaneously in a manifestly duality
invariant form.
Its application to characteristic polynomial duality is in Section~\ref{BZ}.
The extended version of our duality for Penner type models is found in
Section~\ref{Penner_Model} while detailed derivations of the
formul\ae~there are presented in Appendix~\ref{don't-think--type}.
Section~\ref{central} concerns the
the {\it Central Limit Theorem} for Gaussian random matrix ensembles.
In the Conclusions (Section~\ref{conclusions}) we discuss possible
further
generalizations. In particular, the construction of a graphical topological
invariant of surfaces with marked points necessary for the proof of
our main result, is rather general and may be applied to higher
algebraic structures.

\section{Matrix Integrals}

\label{matrix}

The object of our study is the integral over self-adjoint 
matrices\footnote{We employ various normalizations 
throughout the paper, so it is  convenient to divide through
by the free matrix integral. }
\begin{equation}
Z^{(\b)}(t,N)=\frac{\textstyle\int [dX]_{_{\!\ss(\b)}}\!\,\exp\Big(\!-\frac{1}{4}\ \tr X^2
+\sum_{j=1} ^\infty \frac{ t_j}{2j}\ \tr X^j \Big)}{
\textstyle\int [dX]_{_{\!\ss(\b)}}\!\,\exp
\Big(\!-\frac{1}{4}\ \tr 
X^2\Big)}
\label{Z}
\end{equation}
as a function of the ``coupling constants'' 
$t = (t_1, t_2, t_3,  \ldots)$ and the size $N$ of the matrix
variable $X$. Here
\begin{equation}
X=S+\sum_{i=1}^{\b-1} e_i A_i\label{X}
\end{equation}
is built from real, $N\times N$, symmetric and antisymmetric matrices
$S$ and $A_i$, respectively. The parameter $\beta$ takes
values $1$, $2$ or $4$ depending whether we study 
real, complex or quaternionic self-adjoint
matrices and in turn Gaussian orthogonal, unitary or symplectic ensembles 
(GOE, GUE, GSE).
The imaginary units $e_i$
are then drawn from one of three sets
\begin{equation}
e_i\in
\left\{
\begin{array}{cl}
\emptyset\Comma&\beta=1\, ,\\ \\
\{i:\,i^2=-1\}\Comma&\beta=2\, ,\\ \\
\{i,j,k:i^2=j^2=k^2=ijk=-1\}\Comma&\beta=4\, .
\end{array}
\right.
\end{equation}
The self-adjoint condition 
\begin{equation}
X^\dagger\equiv \overline X^{\top}=X\Comma
\overline{e_i}=-e_i\, ,
\end{equation} 
is implied by antisymmetry of the matrices $A_i$.
Finally, the measure  $[dX]_{_{\!\ss(\b)}}\!$ is the translation invariant
Lebesgue measure of the vector space of real dimension
$\12N(\b(N-1)+2)$ 
spanned by independent matrix elements of~$S$ and~$A_i$.
This measure is invariant, respectively,  under 
orthogonal, unitary and symplectic
transformations
\begin{equation}
X\longmapsto U^\dagger XU,
\end{equation}
where $U^\dagger U=1$ and
$U=U_0+\sum_{i=1}^{\beta-1} e_i
U_i$ for real $N\times N$ matrices~$U_0$ and~$U_i$. 

The matrix integral (\ref{Z}) is a holomorphic function
in $(t_1, t_2, \ldots, t_{2m})$ if we fix $m>0$ and
restrict the coupling constants to satisfy
$$
{\rm Re}(t_{2m})<0\qquad
{\text{and}}
\qquad
t_{2m+1} = t_{2m+2} = t_{2m+3} = \cdots = 0.
$$ 
Under this restriction, $Z^{\beta}(t,N)$ has a unique
Taylor expansion in $(t_1$, $t_2$, $\ldots$, $t_{2m-1})$ and
an \emph{asymptotic} expansion in $t_{2m}$ as 
$t_{2m}\rightarrow 0$ while keeping ${\rm Re}(t_{2m})<0$.
Let us introduce a weighted degree of the coupling constants
by $\deg(t_n)=n$. Then the asymptotic expansion of 
the truncated integral $Z^{\beta}(t,N)$ has a well-defined limit as
$m\rightarrow \infty$ in the ring
$$
\left(\mathbb{Q}[N]\right)[[t_1,t_2,t_3,\ldots]]
$$
of formal power series in infinitely many variables with coefficients
in the polynomial ring of  $N$ with rational coefficients~\cite{Mulase95}. 
The subject of our study in what follows is
this asymptotic expansion of $Z^{\beta}(t,N)$ as a function in
infinitely many variables.

\section{Graphical Expansion}

\label{graphs}

The graphs appearing in our asymptotic expansion of the
matrix integrals~(\ref{Z}) are those drawn on
orientable as well as \emph{non-orientable} surfaces. To avoid
confusion with an already well-established convention that 
ribbon graphs are drawn on orientable surfaces, we
propose the terminology  {\it M\"obius  graphs}.
Let us recall that a ribbon graph  $\Gamma$ is a
graph with a cyclic order chosen at each vertex
for half-edges adjacent to it. Equivalently,  
it  is a graph drawn on a compact
oriented surface $S$ giving a  cell-decomposition of it.
The complement $S\setminus\Gamma$ of the graph $\Gamma$
on $S$ is the disjoint union of $f_\Gamma$ open disks (or faces)
of the surface. Since a ribbon graph $\Gamma$ defines a unique
oriented surface on which it is drawn as the 1-skeleton of a 
cell-decomposition,
we denote the surface by $S_\Gamma$.

Similarly, a M\"obius graph is drawn on a compact surface, orientable
or non-orientable, giving a cell-decomposition of the surface.
It can be viewed as a ribbon graph with twisted edges. A M\"obius 
graph $\Gamma$ also uniquely 
defines the surface $S_\Gamma$ in which it is embedded.

Let $\mathfrak{G}$
be the set of connected M\"obius graphs.
A graph $\Gamma\in\mathfrak{G}$ consists of a finite
number of vertices and edges. Let $v_\Gamma^{(j)}$ 
denote the number of $j$-valent vertices of  $\Gamma$. 
Then the number of vertices and edges are given by
\begin{equation}
v_\Gamma=\sum_{j} v_\Gamma^{(j)}
\qquad {\text{and}}\qquad
e_\Gamma=\frac12\ \sum_{j} jv_\Gamma^{(j)}.
\label{vedge}
\end{equation}
The unique compact surface $S_\Gamma$ has $f_\Gamma$
faces and its Euler characteristic is
\begin{equation}
\chi(S_\Gamma)=v_\Gamma-e_\Gamma + f_\Gamma\, .
\end{equation}
We will also need the number of faces with a given number of edges, so denote
the number of 
$j$-gons in the cell-decomposition of $S_\Gamma$ by $f_\Gamma^{(j)}$ whereby
the Poincar\'e dual formul\ae~to the equations~\eqn{vedge} are
\begin{equation}
f_\Gamma=\sum_{j} f_\Gamma^{(j)}
\qquad {\text{and}}\qquad
e_\Gamma=\frac12\ \sum_{j} jf_\Gamma^{(j)}.
\end{equation}

A M\"obius graph $\Gamma$ also determines the orientability of $S_\Gamma$
and we define
\begin{equation}
\natural_\Gamma=\left\{
\begin{array}{rl}
1&S_\Gamma \mbox{ orientable}\, ,\\ \\
-1&S_\Gamma \mbox{ non-orientable}\, .\\
\end{array}
\right.
\end{equation}
By {\it genus} of $S_\Gamma$ we mean
\begin{equation}
g(S_\Gamma)= 1- 2^{-\frac{1+\natural_\Gamma}{2}}\; \chi(S_\Gamma)\, .
\end{equation}
Thus $\chi(S_\Gamma) = 2 - 2 g(S_\Gamma)$ for an orientable surface
and $\chi(S_\Gamma) = 1 -  g(S_\Gamma)$ if it is non-orientable.
We also define the parity of $\chi(S_\Gamma)$ by
\begin{equation}
\sharp_\Gamma=(-1)^{\chi(S_\Gamma)}\, ,
\end{equation}
while many results can be written compactly in terms of
\begin{equation}
\Sigma_\Gamma
=\frac12\ 
\Big(1+\sharp_\Gamma\Big)-\natural_\Gamma
=
\left\{
\begin{array}{cl}
0&\Gamma \mbox{ orientable}\, ,\\ \\ 
1&\Gamma\mbox{ non-orientable, } \chi(S_\Gamma) \mbox{ odd}\, ,\\ \\
2&\Gamma\mbox{ non-orientable, } \chi(S_\Gamma) \mbox{ even}\, .
\end{array}
\right.
\end{equation}
Our main result is:

\begin{theorem} \label{theorem}
The logarithm of the asymptotic expansion of the
matrix integral $Z^{(\beta)}(t,N)$
is expressed as a sum over connected M\"obius graphs:
\begin{multline}
\!\!\!
\log\Big(Z^{(\beta)}(t,N)\Big)\\=\sum_{\Gamma\in\mathfrak G}
\frac{\textstyle (-4+6\beta-\beta^2)^{^{\ss 
1-\12 \Sigma_\Gamma-\12 \chi(S_\Gamma)}} 
(2-\beta)^{^{\ss \Sigma_\Gamma}} 
\beta^{^{\ss f_\Gamma-1}}
N^{^{\ss f_\Gamma}}}{|{\rm Aut}(\Gamma)|} \prod_{j}
t_j^{v^{(j)}_\Gamma}\!\!\! \\
\in \left(\mathbb{Q}[N]\right)[[t_1,t_2,t_3,\ldots]].
\label{eq: main formula}
\end{multline}
\end{theorem}

\noindent{\it Remark:}
\begin{enumerate}
\item We define $(2-\beta)^{\Sigma_\Gamma}= 1$ when $\beta=2$ and
 $\Sigma_\Gamma=0$.
\item For every connected M\"obius graph $\Gamma$, the monomial
$\prod_{j}
t_j^{v^{(j)}_\Gamma}$ is a finite product of total degree
$2e_\Gamma$. 
\item The only reason to consider $\log Z^{\beta}(t,N)$ 
is because it yields a compact formula (\ref{eq: main formula}): 
$Z^{\beta}(t,N)$ itself has an expansion in terms of
graphs although a given summand may have a mixture of 
orientable and non-orientable connected components. 
\end{enumerate}

The automorphism group $\Aut(\Gamma)$ 
 of a M\"obius graph $\Gamma$ is a group of automorphisms
of the cellular complex $S_\Gamma$ consisting of $v_\Gamma$
vertices, $e_\Gamma$ edges and $f_\Gamma$ faces. 
When $S_\Gamma$ is orientable, the group $\Aut(\Gamma)$ may
contain orientation-reversing automorphisms as well.
We note that a cyclic rotation of half-edges
around a vertex corresponds to the 
invariance of the trace under a cyclic permutation
\begin{equation}
\tr(M_1M_2M_3\cdots M_n)=\tr(M_nM_1M_2\cdots M_{n-1}),
\label{Michael}
\end{equation}
and an orientation-reversing flip of vertex with adjacent
half-edges corresponds to 
the invariance the trace of symmetric matrices 
\begin{equation}
\tr(S_1S_2S_3\cdots S_n)=\tr(S_nS_{n-1}S_{n-2}\cdots S_1)\
\label{Jackson}
\end{equation}
under reversing the order of multiplication.

\begin{figure}[htb]
\centerline{\epsfig{file=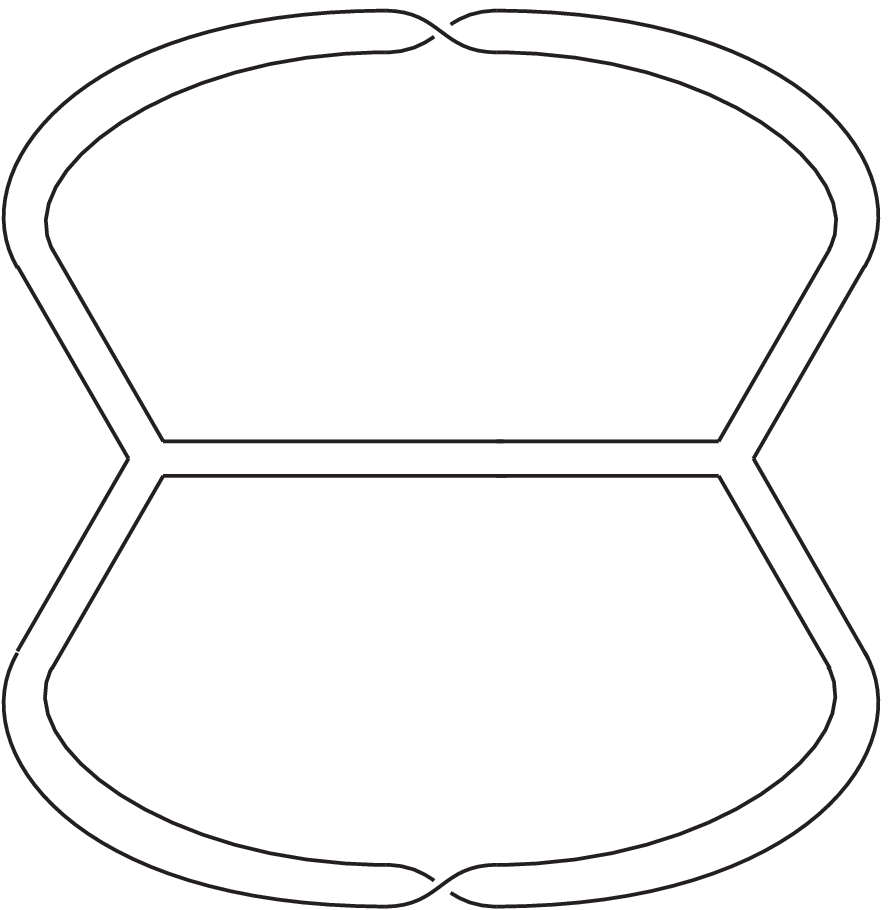, width=1.8in}
\hskip0.8in\epsfig{file=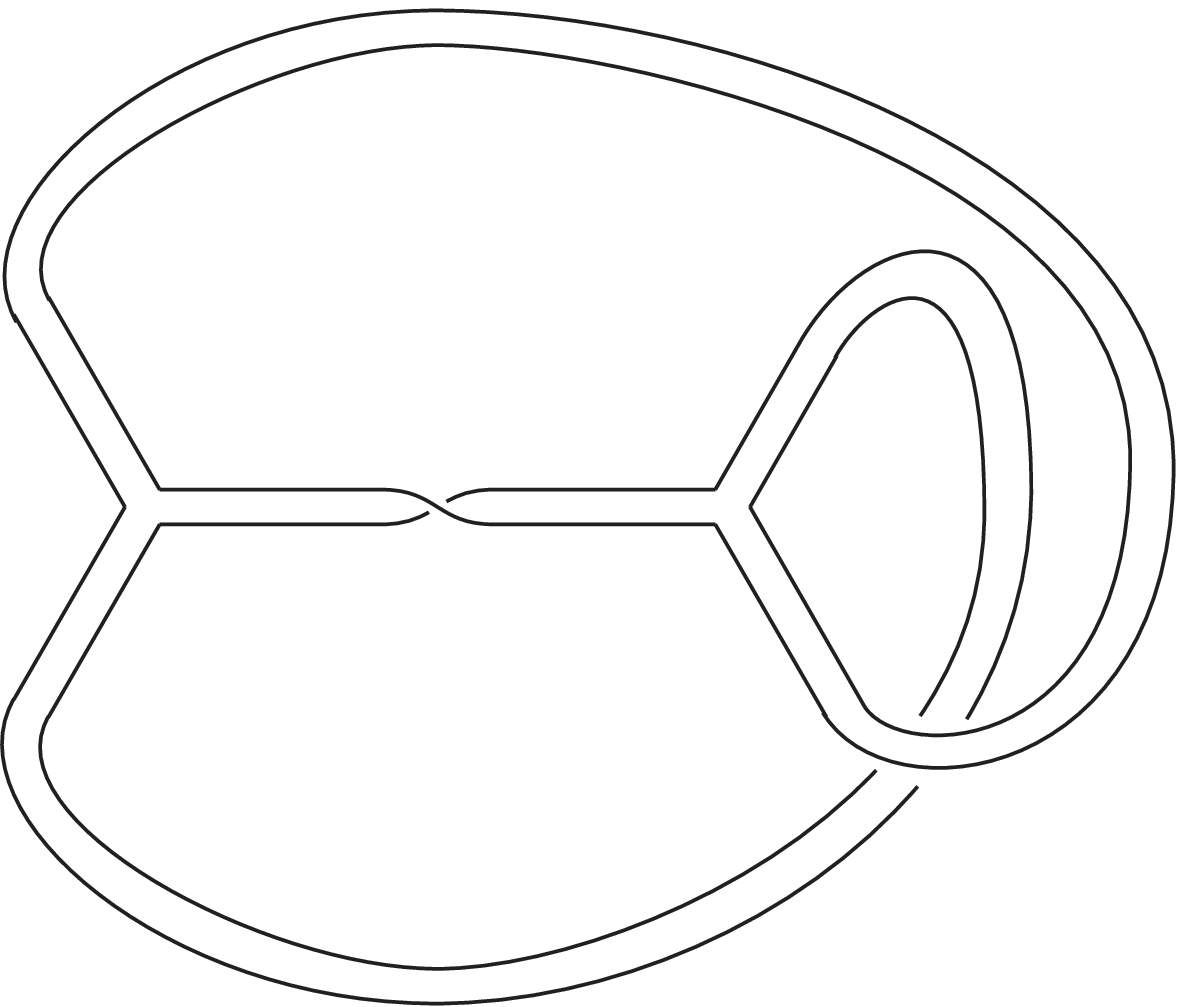, width=2.0in}}
\caption{Two equivalent M\"obius graphs 
consisting of two vertices, three
edges, and one face. The graphs are interchanged
by a \emph{vertex flip}.}
\label{fig:Moebius}
\end{figure}

\begin{figure}[htb] 
\centerline{\epsfig{file=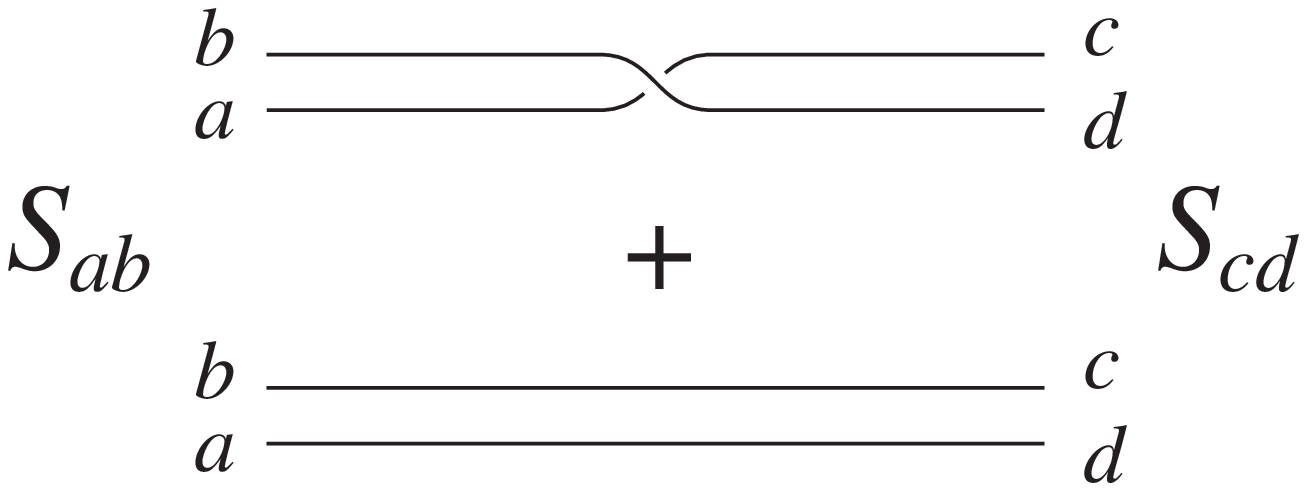, width=2.5in}}
\caption{Propagator for symmetric matrices.}
\label{fig:symedge} 
\end{figure}

\begin{figure}[htb] 
\centerline{\epsfig{file=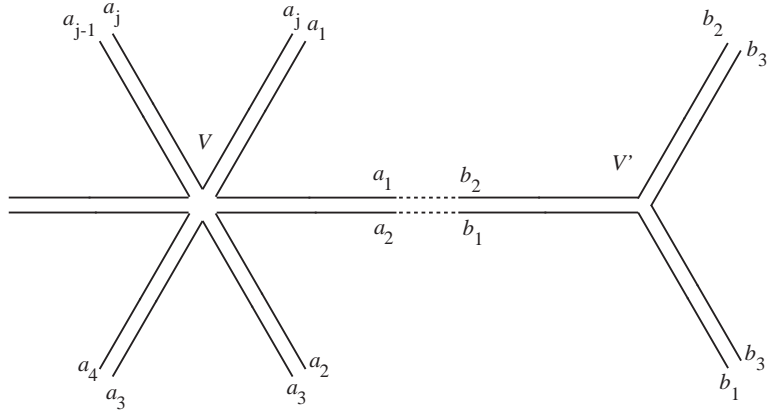, width=4in}} 
\caption{Two vertices connected  
at the half edge labeled by $a_1a_2$ 
to the half edge  labeled  by $b_2b_1$. For the GOE a second
graph with a twisted edge connecting $a_1a_2$ to $b_1b_2$ is also present
since the ribbon edges are not directed.} 
\label{fig:vertices} 
\end{figure}

The proof of the theorem involves two main ingredients. The first is
to view GUE and GSE matrix integrals as the coupling of a singlet or triplet
of skew-symmetric matrix integrals to the fundamental GOE integral.
When $\beta=1$, equation~\eqn{eq: main formula} is the
M\"obius graphical expansion of a symmetric matrix 
integral~\cite{Goulden-Harer-Jackson,Verbaarschot,Silvestrov,Janik}:
\begin{equation}
\log\Big(
Z^{(\beta=1)}(t,N)\Big)=\sum_{\Gamma\in\mathfrak G}
\frac{\textstyle N^{f_\Gamma}}{|{\rm Aut}(\Gamma)|}\; \prod_{j}
t_j^{v^{(j)}_\Gamma}\, .
\label{GOE}
\end{equation}
This formula follows immediately from the fact that the Wick contraction
of any pair of symmetric matrices $S=(S_{ab})$ obeys
\begin{equation}
\langle S_{ab}S_{cd}\rangle=\delta_{ac}\delta_{bd}+\delta_{ad}\delta_{bc}\, , 
\label{S}
\end{equation}
which is denoted graphically as an edge of a M\"obius graph (see
Figure~\ref{fig:symedge}).
These edges connect vertices of the type
\begin{equation}
\frac{t_j}{2j}\,\tr S^j=\frac{t_j}{2j}\,
\sum_{a_1,\ldots,a_j=1}^N\,S_{a_1a_2}S_{a_2a_3}\cdots S_{a_j a_1}.
\label{green}
\end{equation}
as depicted in Figure~\ref{fig:vertices}.
The factor $(2j)^{-1}$ in~\eqn{green} is precisely the one required to 
cancel the over-counting implied by the identities~\eqn{Michael} 
and~\eqn{Jackson}. Therefore the overall weight of any given
graph is as quoted in~\eqn{GOE}.

\begin{figure}[htb] 
\centerline{\epsfig{file=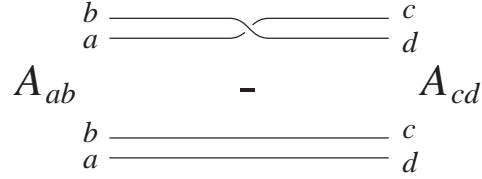, width=2.5in}}
\caption{Propagator for antisymmetric matrices} 
\label{fig:antiedge} 
\end{figure} 
In contrast, for a pair of antisymmetric
matrices $A=(A_{ab})$ the Wick contraction yields
\begin{equation}
\langle A_{ab} A_{cd}\rangle=\delta_{ac}\delta_{bd}-\delta_{ad}\delta_{bc}.
\label{A}
\end{equation}
This is depicted in Figure~\ref{fig:antiedge}.

The minus sign for untwisted edges will play a crucial r\^ole in what follows.
Notice that since the exponent of the Gaussian part of~\eqn{Z} is
\begin{equation}
-\frac{1}{4}\sum_{a=1}^N S_{aa}^2
-\frac{1}{2}\sum_{1\leq a<b\leq N}\Big(S_{ab}^2+\sum_{i=1}^{\beta-1}
A_{iab}^2\Big)\, , 
\end{equation}
there are no non-vanishing Wick contractions between symmetric and
antisymmetric matrices. 
Therefore the model~\eqn{Z} is a sum over
graphs with edges of either of the two types~\eqn{S} or~\eqn{A}.
Furthermore $\langle A_i A_j\rangle$
is only non-vanishing when $i=j$ so edges ``emitted'' at vertices are only
connected when they carry the same imaginary units $e_i$.

\begin{figure}
\begin{equation*}
\begin{split}
\raisebox{-.6in}{
\epsfig{file=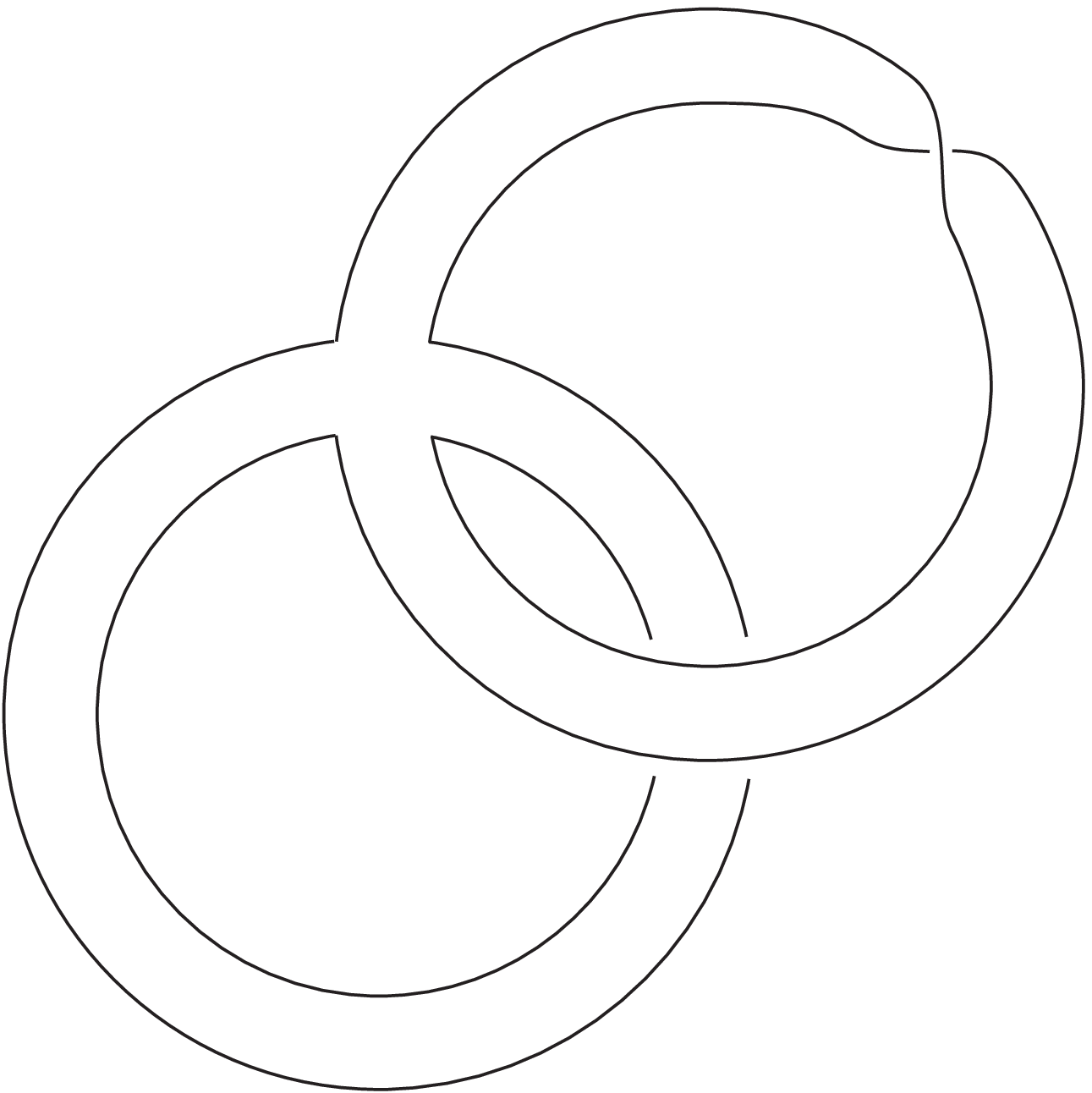,width=1.3in}}
&=
\raisebox{-.6in}{
\epsfig{file=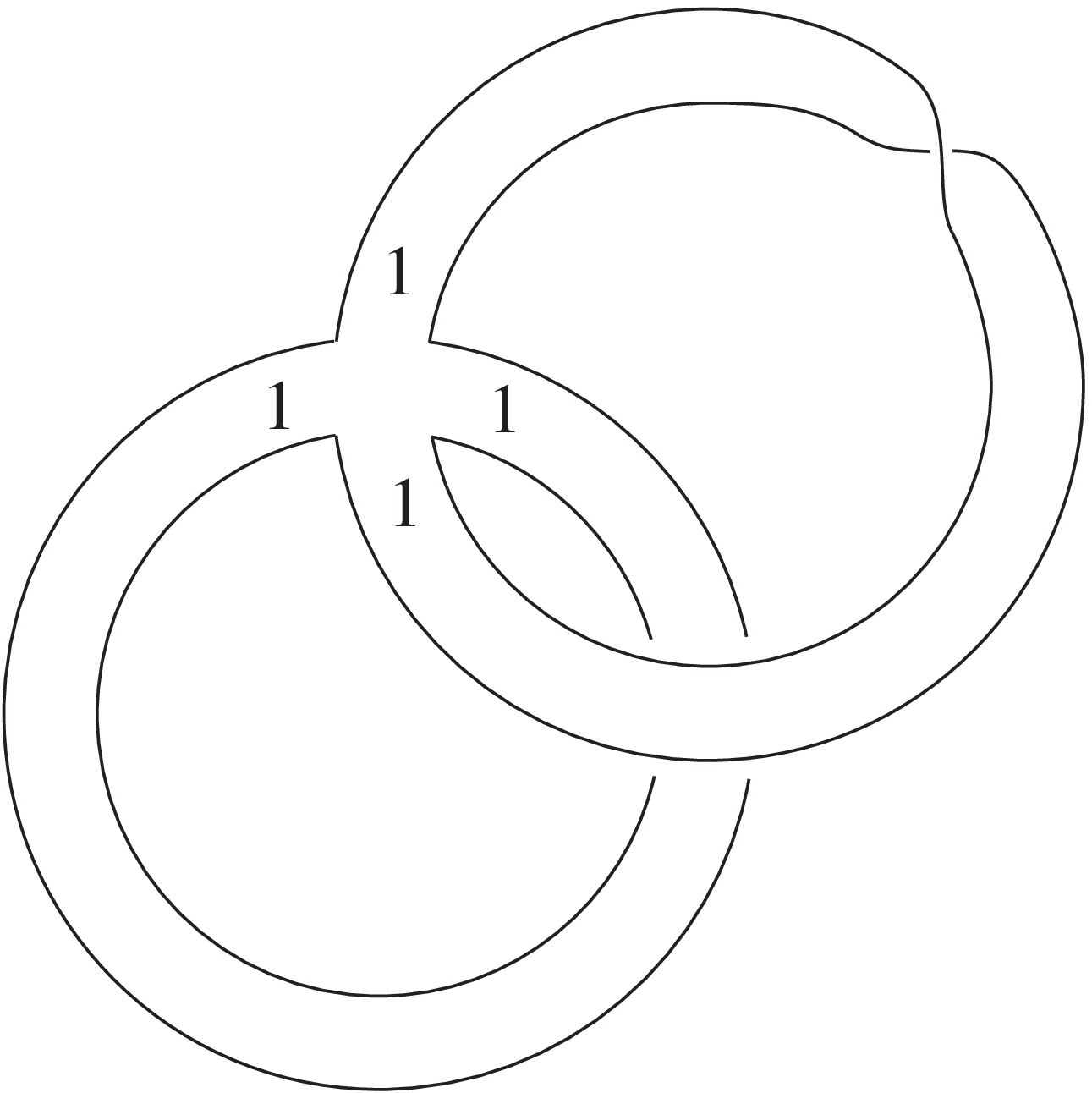,width=1.3in}}
\hspace{-.9cm}
\raisebox{-.9cm}{\large +1}
\hspace{.4cm}\\
&\hspace{-3cm}+\raisebox{-.6in}{
\epsfig{file=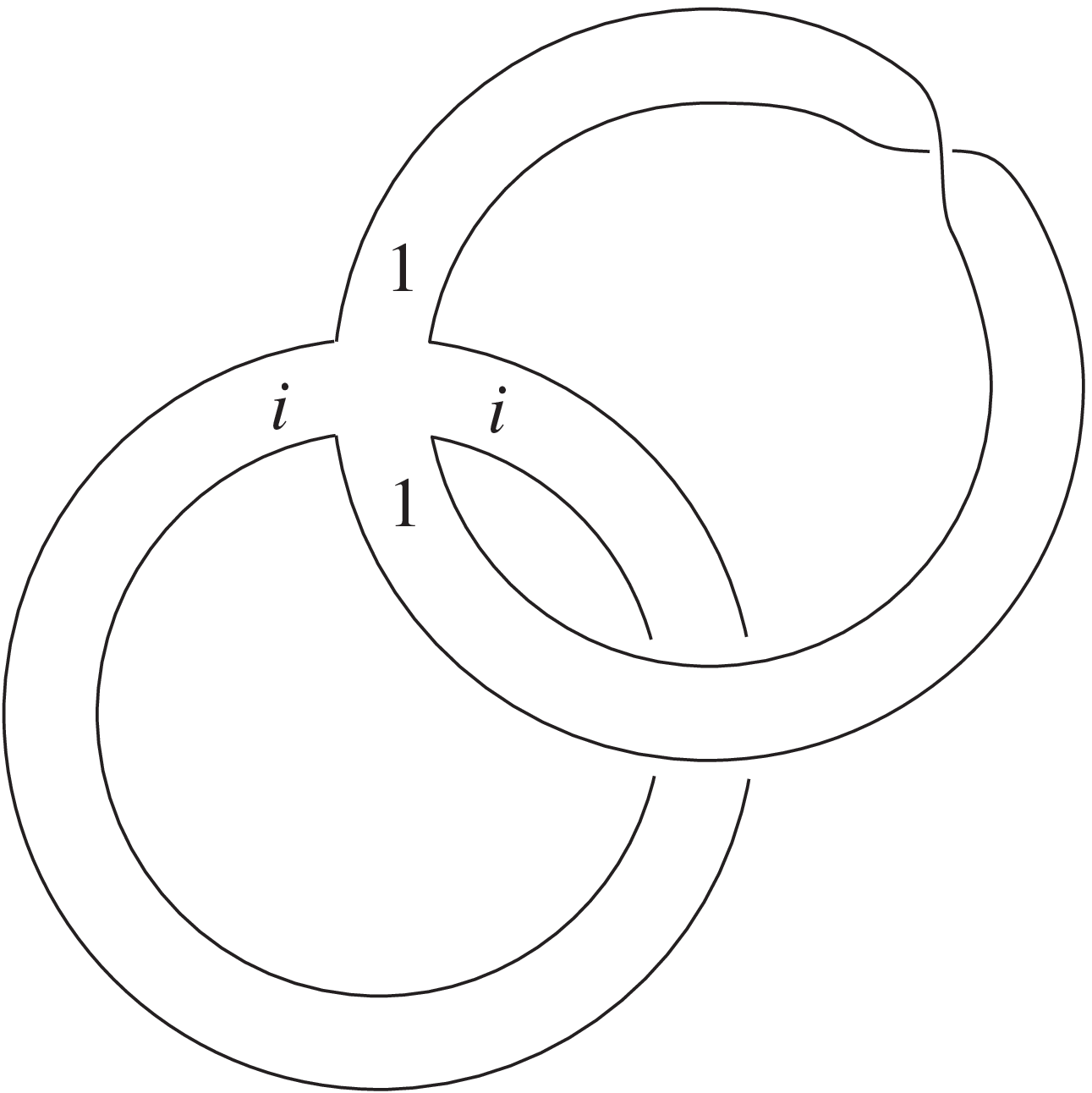,width=1.3in}}\hspace{-.9cm}
\raisebox{-.9cm}{\large +1}
\hspace{.4cm}
+\raisebox{-.6in}{
\epsfig{file=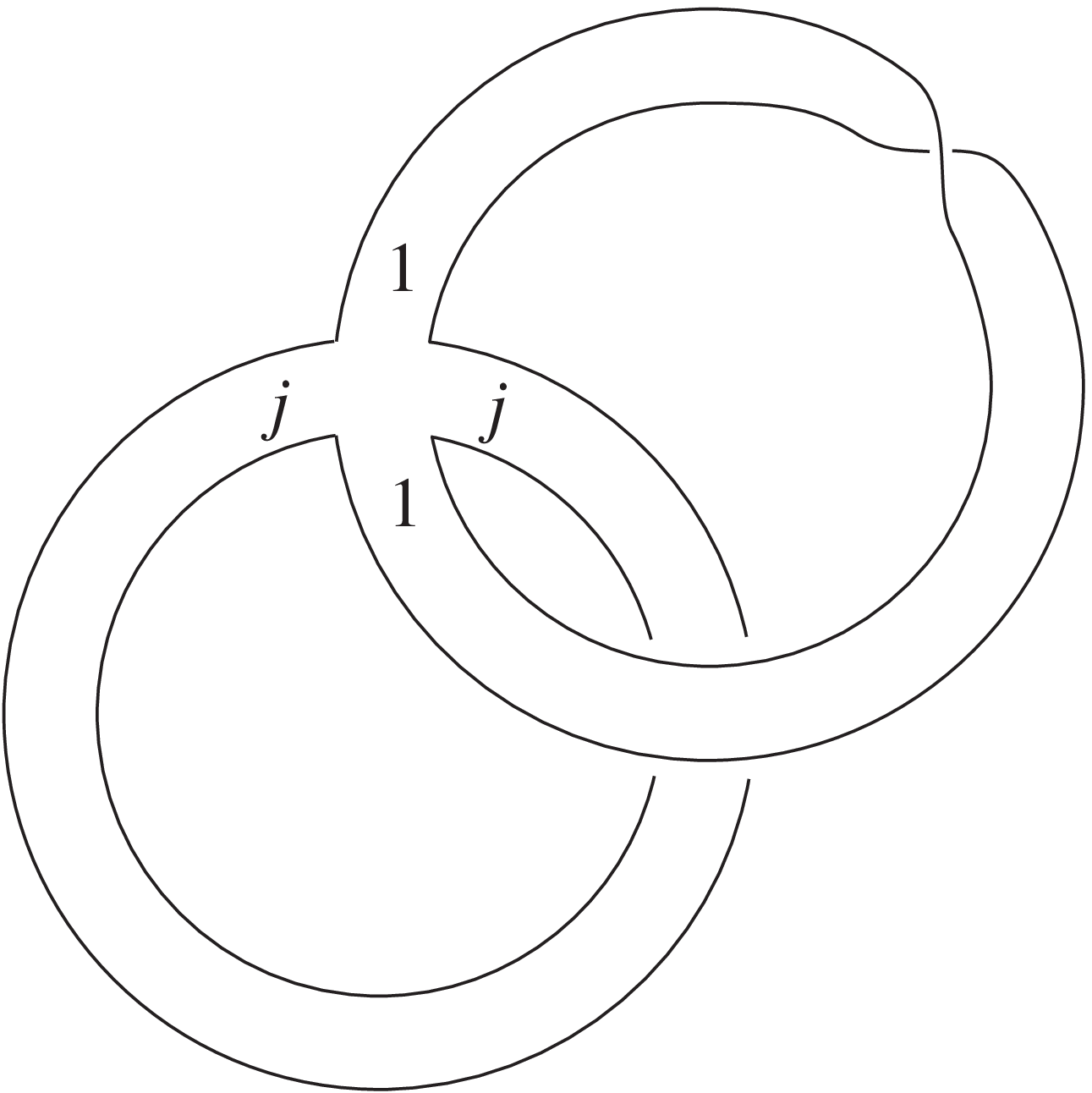,width=1.3in}}\hspace{-.9cm}
\raisebox{-.9cm}{\large +1}
\hspace{.4cm}
+\raisebox{-.6in}{
\epsfig{file=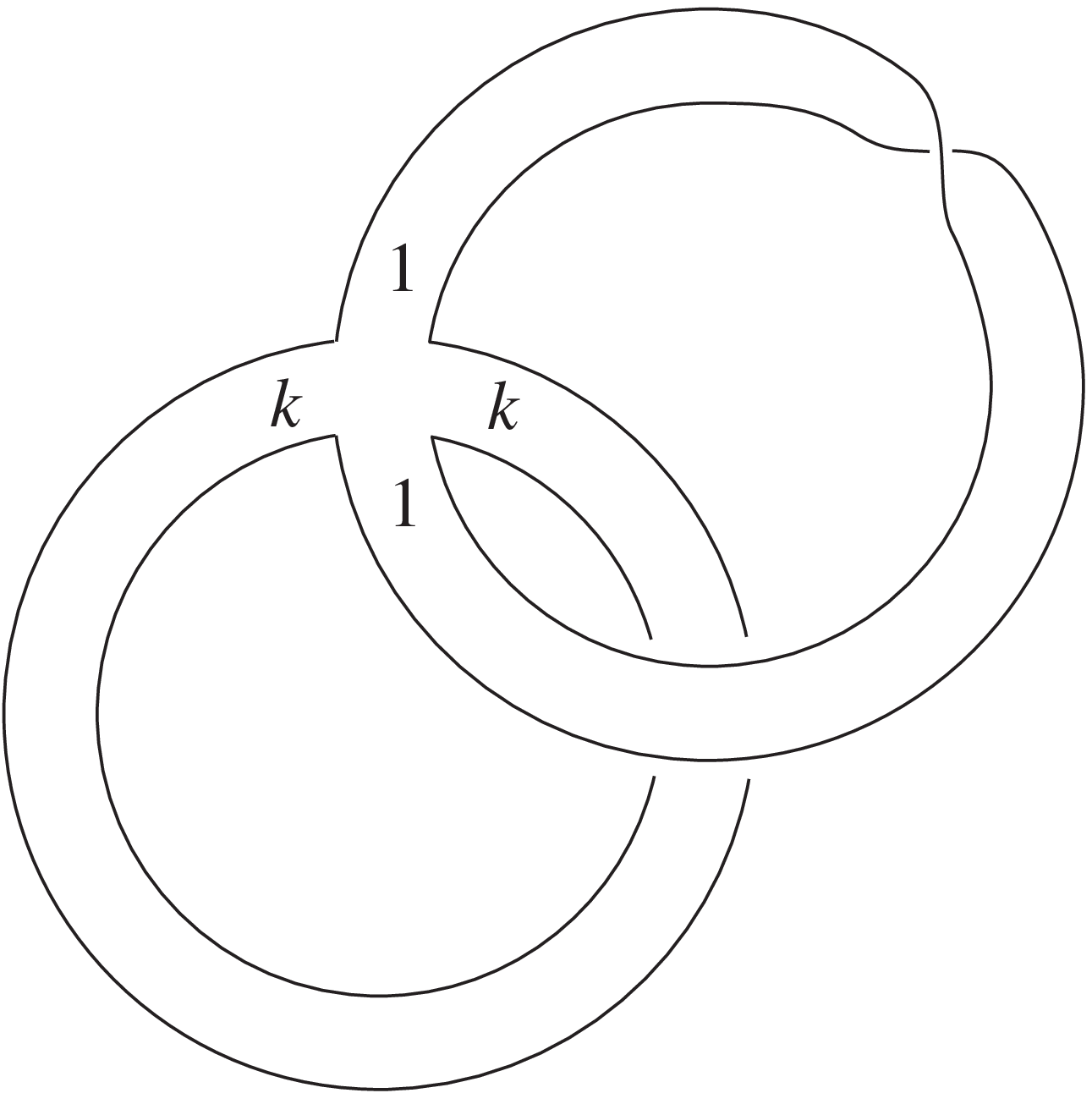,width=1.3in}}\hspace{-.9cm}
\raisebox{-.9cm}{\large +1}
\hspace{.4cm}\\
&\hspace{-3cm}+\raisebox{-.6in}{
\epsfig{file=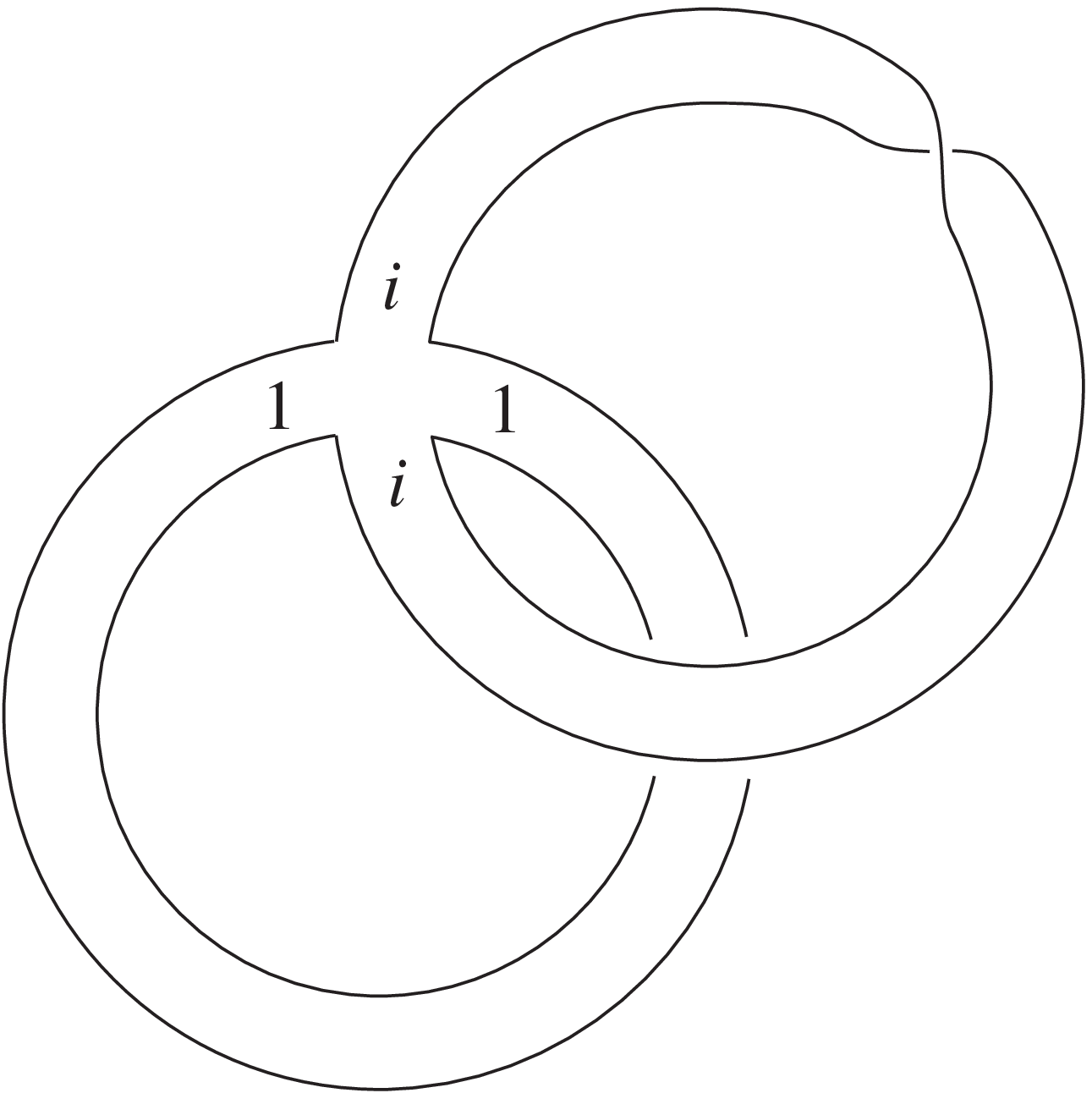,width=1.3in}}\hspace{-.9cm}
\raisebox{-.9cm}{\large --1}
\hspace{.4cm}
+\raisebox{-.6in}{
\epsfig{file=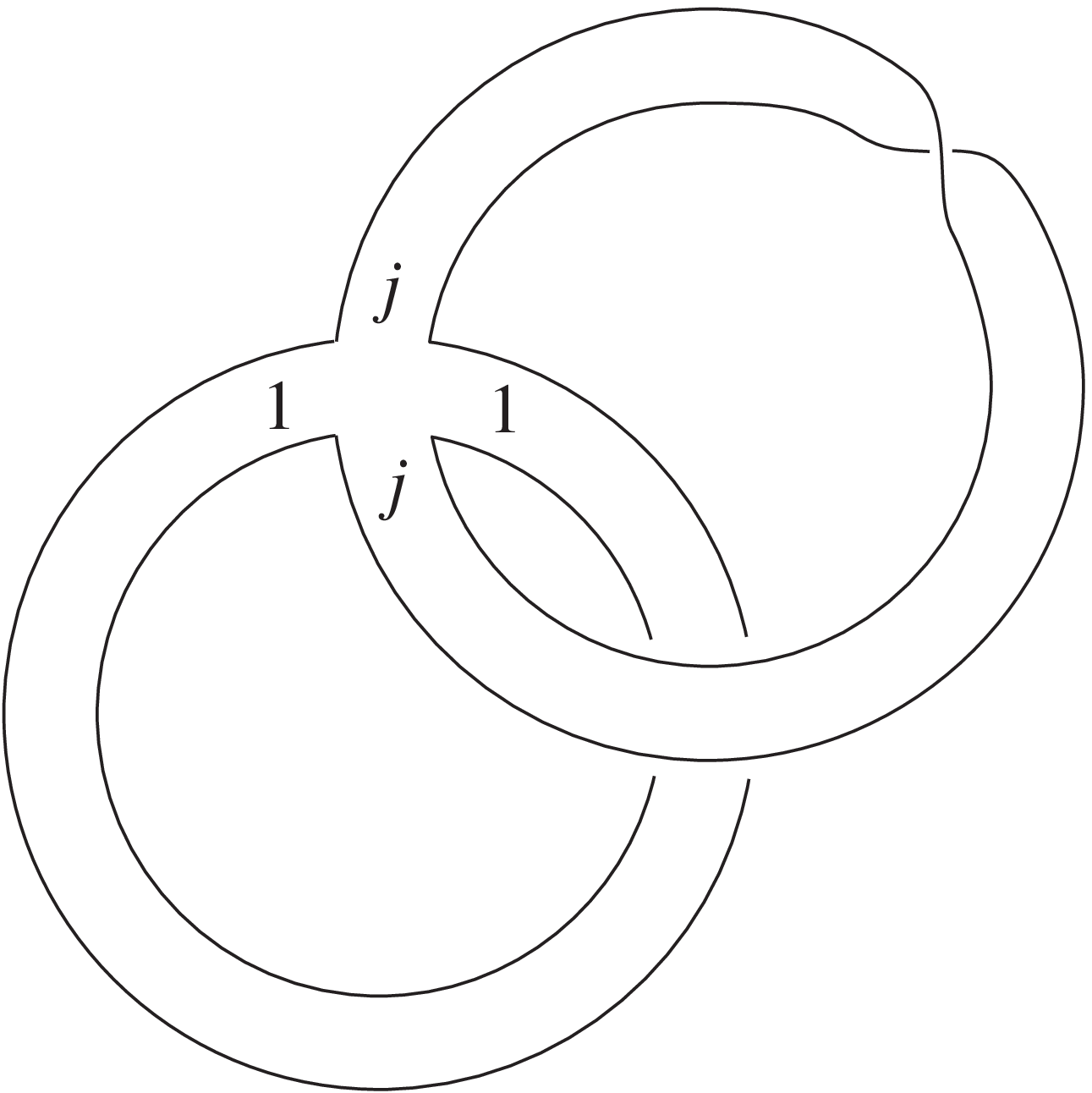,width=1.3in}}\hspace{-.9cm}
\raisebox{-.9cm}{\large --1}
\hspace{.4cm}
+\raisebox{-.6in}{
\epsfig{file=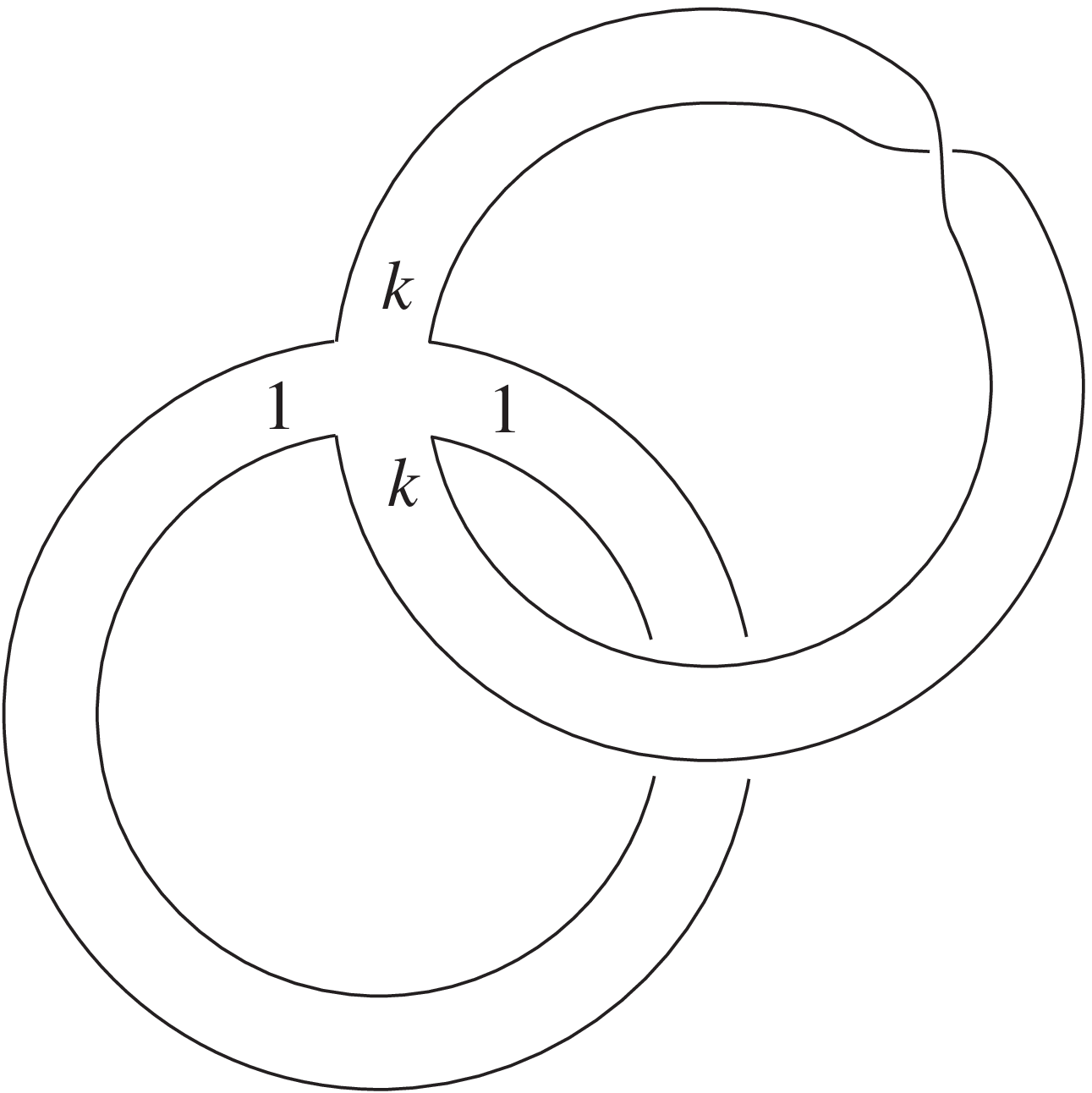,width=1.3in}}\hspace{-.9cm}
\raisebox{-.9cm}{\large --1}
\hspace{.4cm}\\
&\hspace{-3cm}+\raisebox{-.6in}{
\epsfig{file=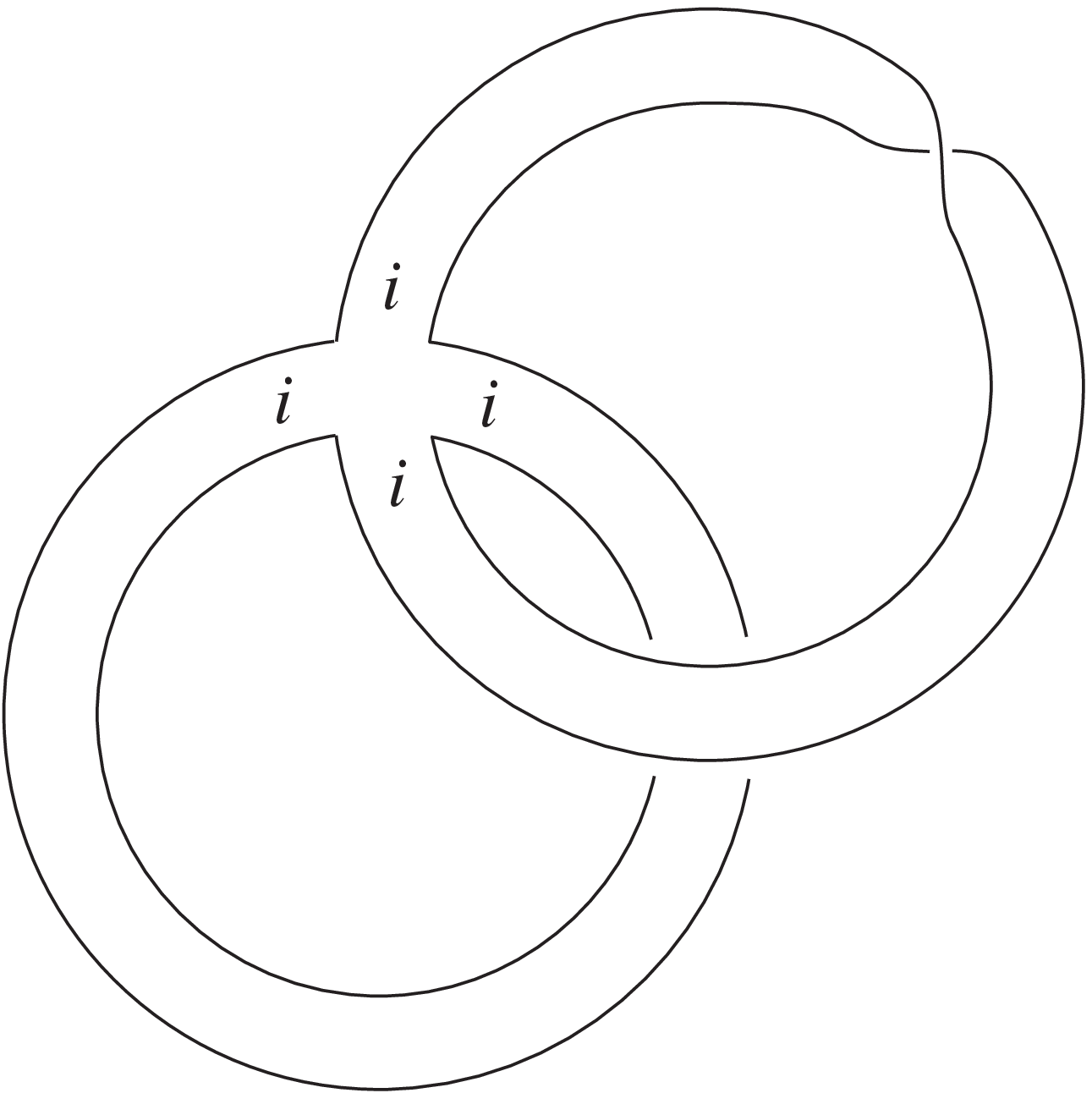,width=1.3in}}\hspace{-.9cm}
\raisebox{-.9cm}{\large --1}
\hspace{.4cm}
+\raisebox{-.6in}{
\epsfig{file=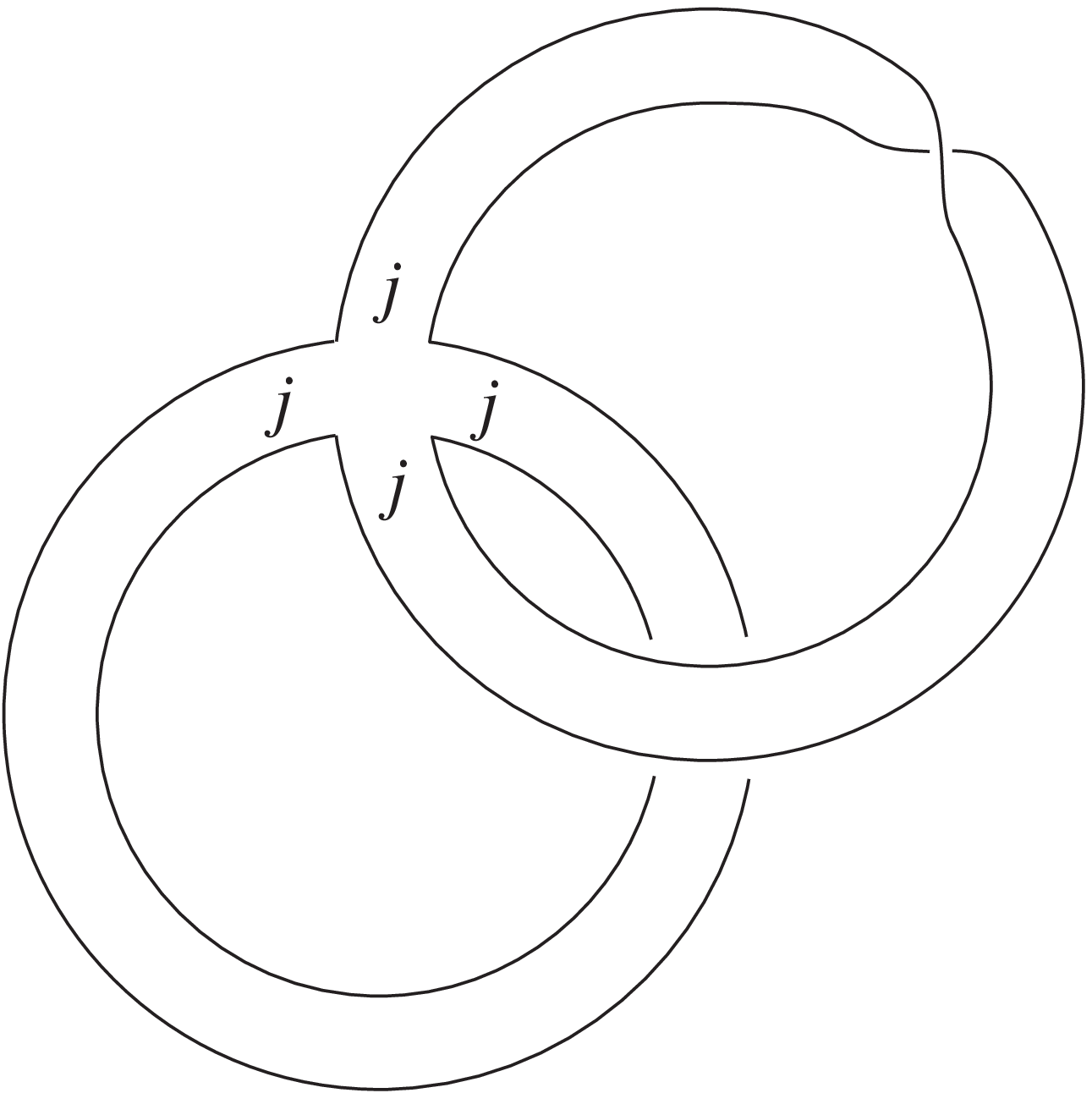,width=1.3in}}\hspace{-.9cm}
\raisebox{-.9cm}{\large --1}
\hspace{.4cm}
+\raisebox{-.6in}{
\epsfig{file=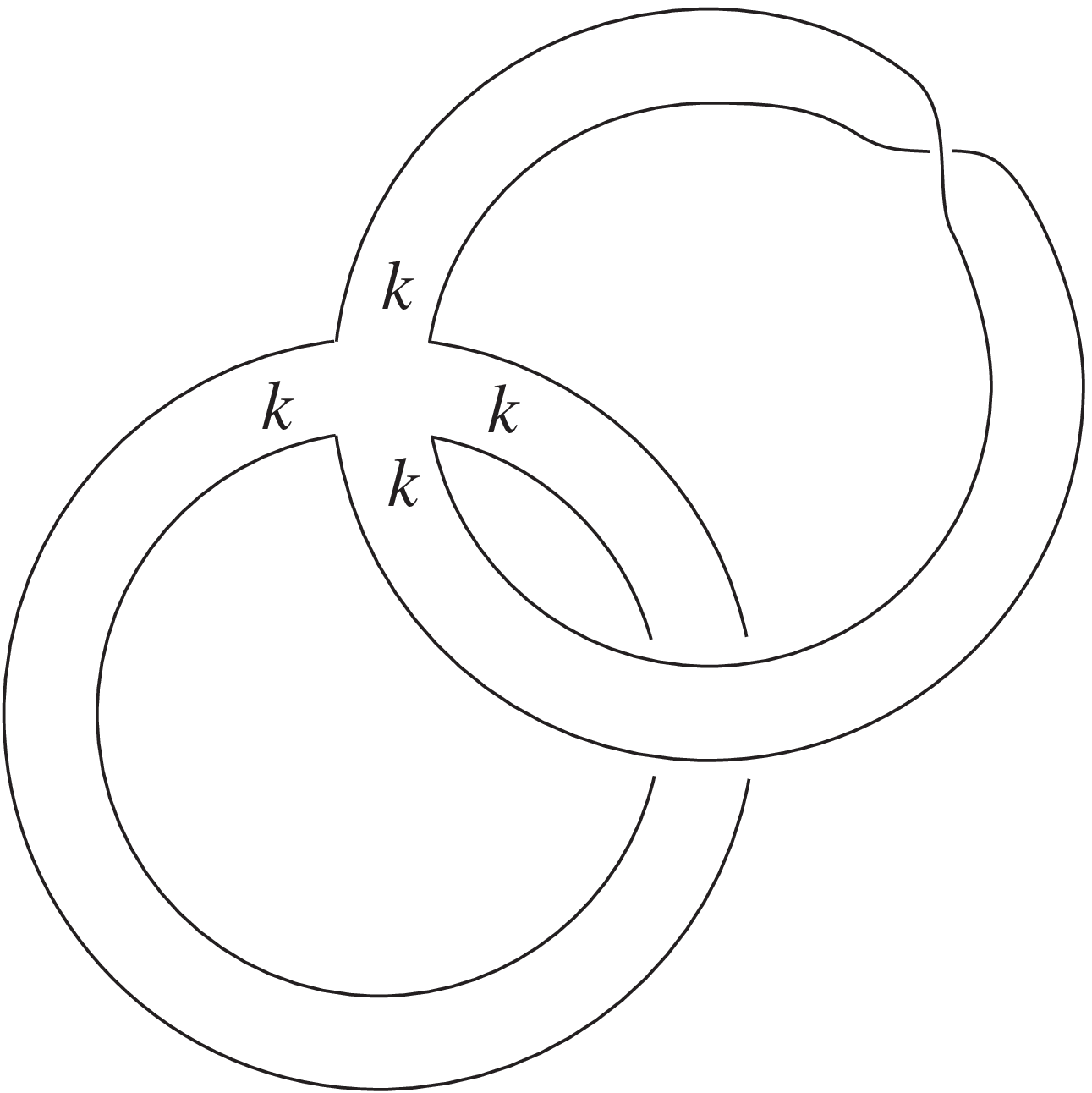,width=1.3in}}\hspace{-.9cm}
\raisebox{-.9cm}{\large --1}
\hspace{.4cm}\\
&\hspace{-3cm}+\raisebox{-.6in}{
\epsfig{file=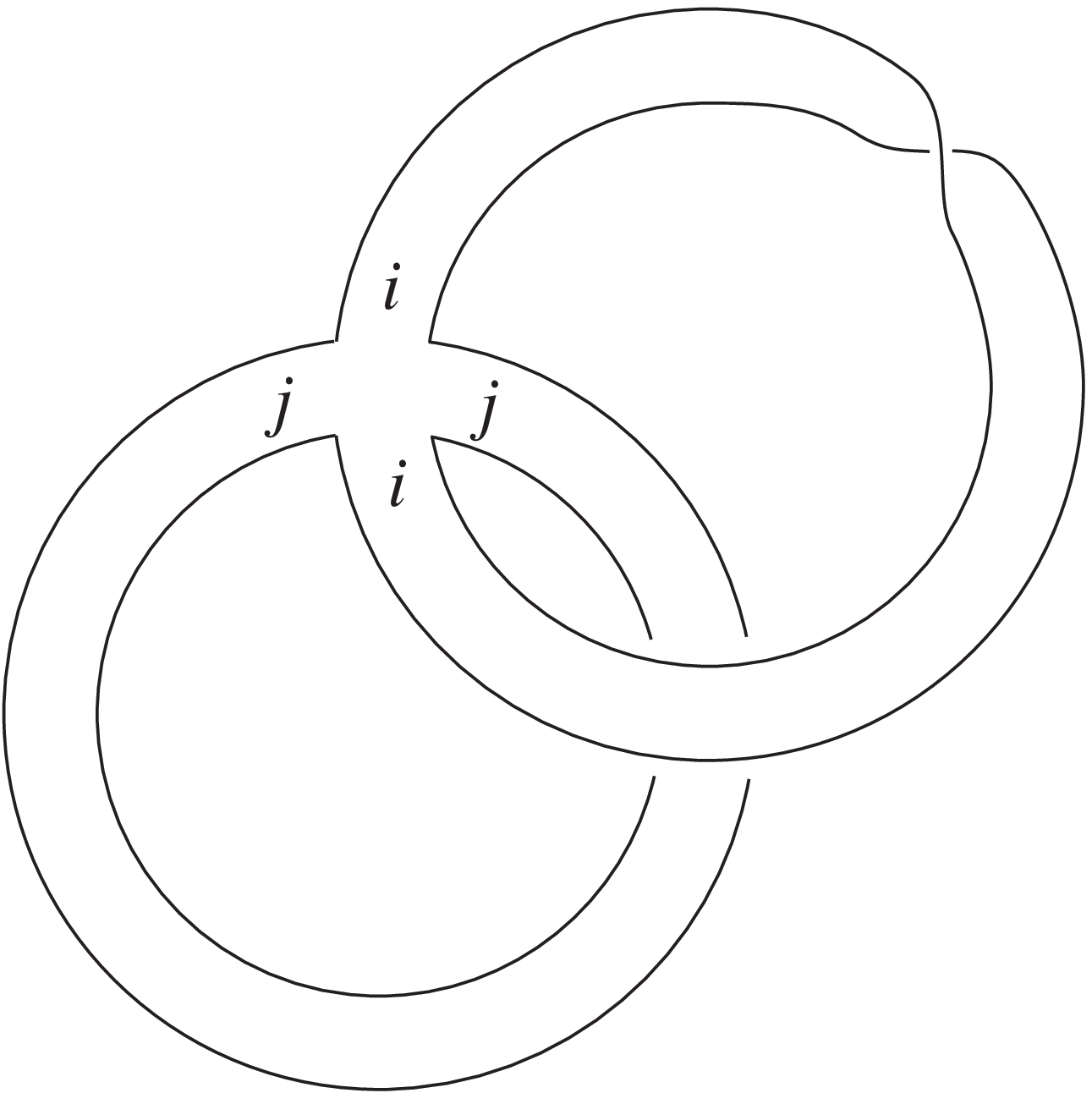,width=1.3in}}\hspace{-.9cm}
\raisebox{-.9cm}{\large +1}
\hspace{.4cm}
+\raisebox{-.6in}{
\epsfig{file=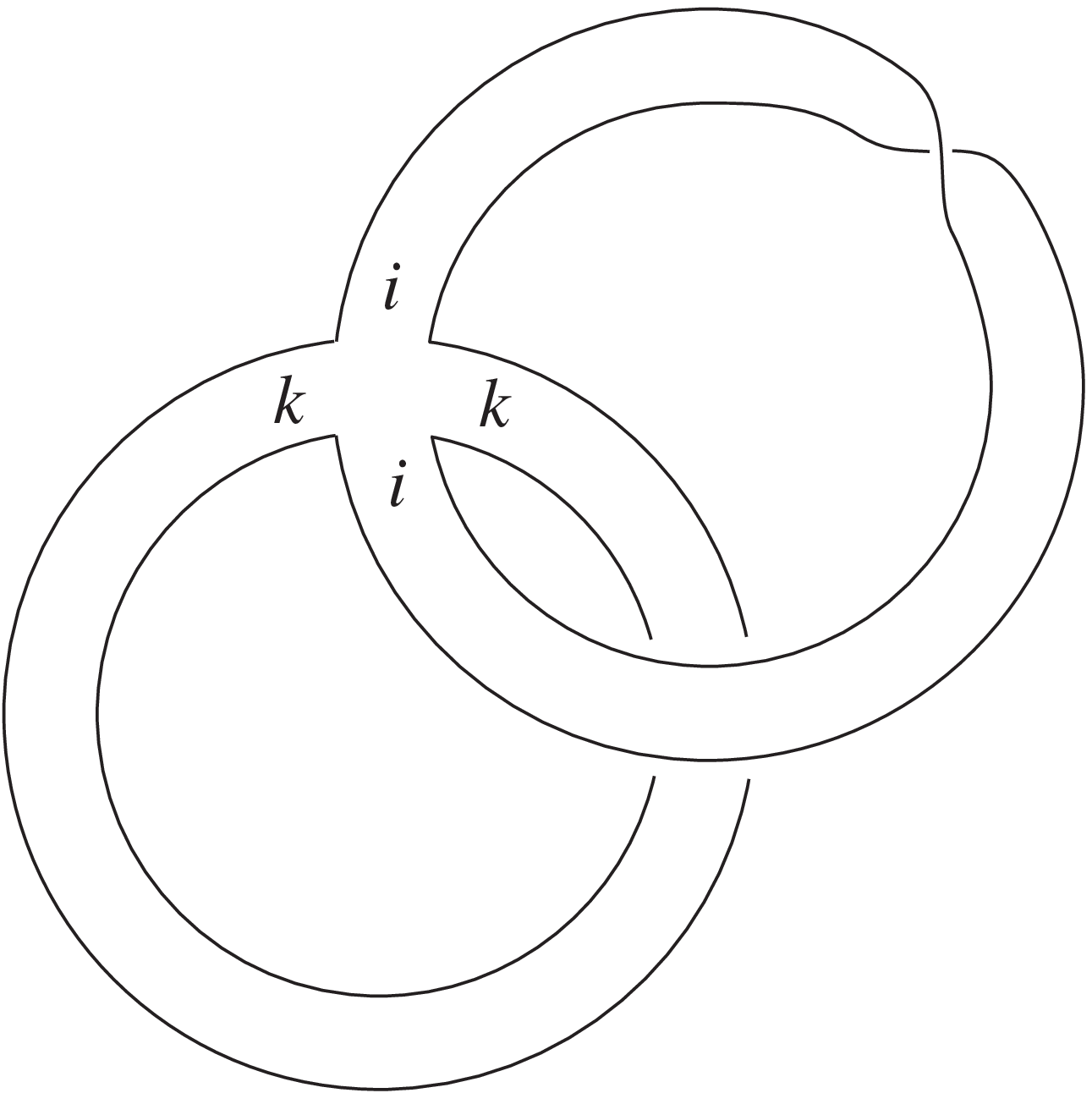,width=1.3in}}\hspace{-.9cm}
\raisebox{-.9cm}{\large +1}
\hspace{.4cm}
+\raisebox{-.6in}{
\epsfig{file=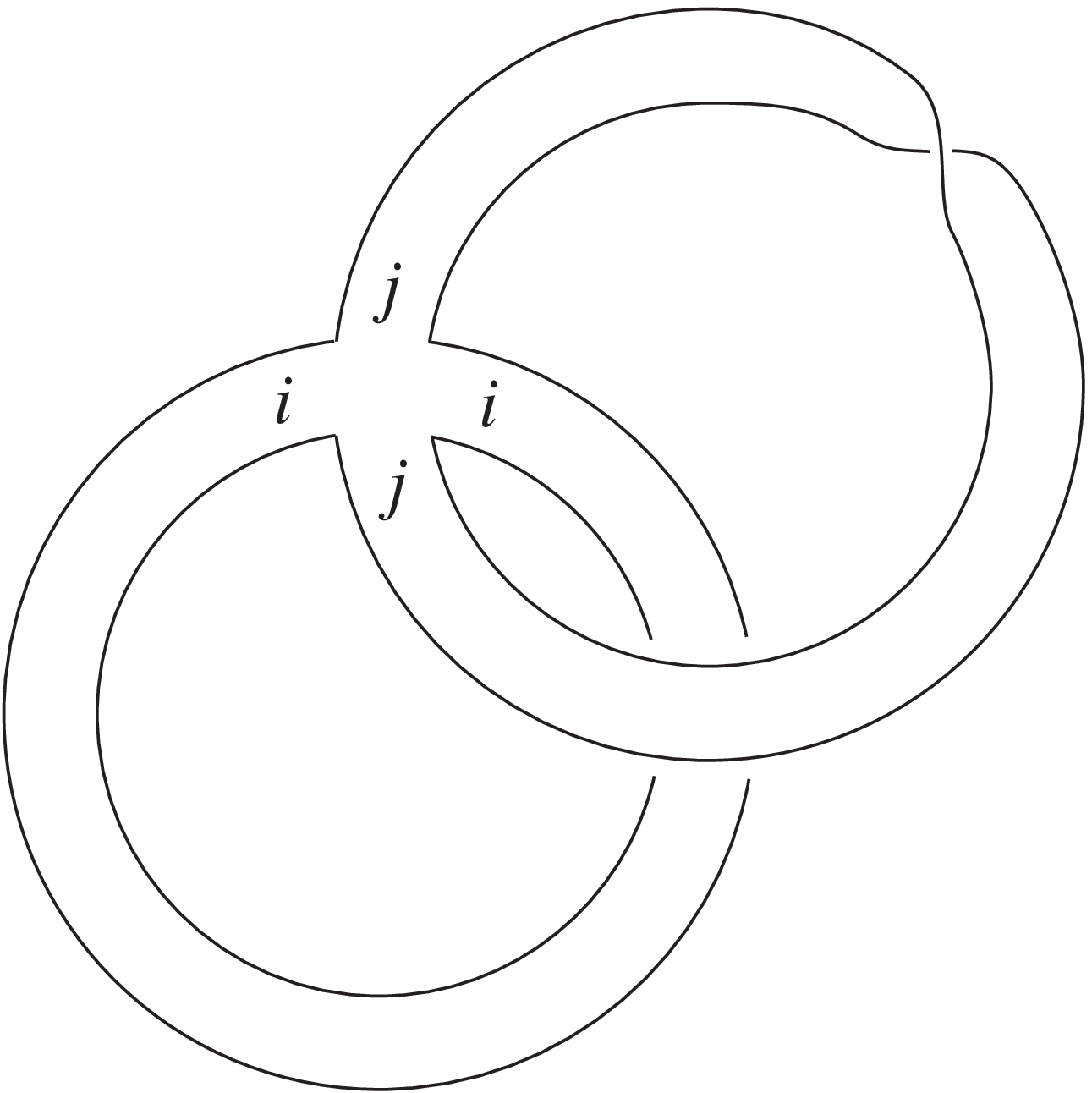,width=1.3in}}\hspace{-.9cm}
\raisebox{-.9cm}{\large +1}
\hspace{.4cm}\\
&\hspace{-3cm}+\raisebox{-.6in}{
\epsfig{file=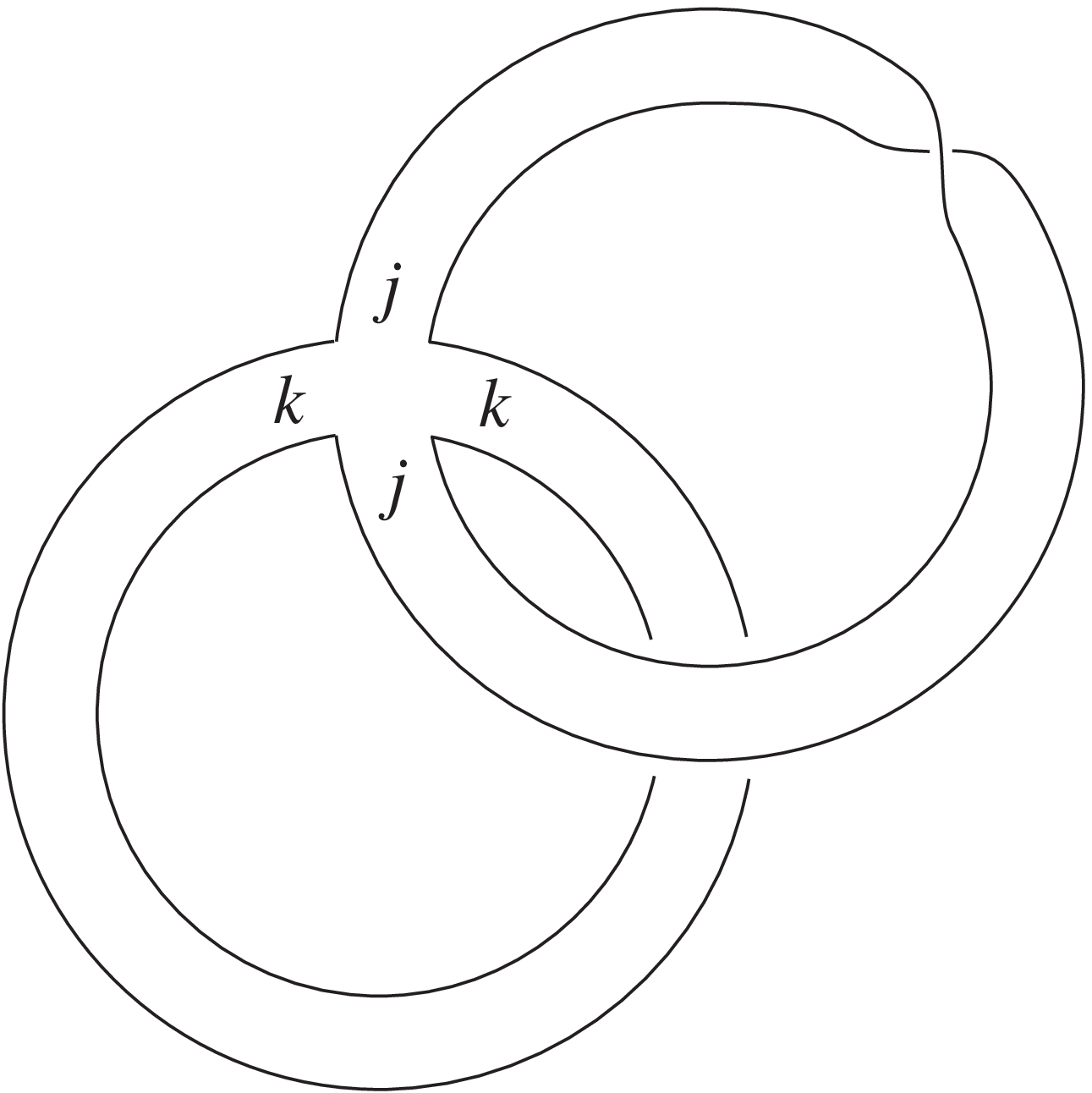,width=1.3in}}\hspace{-.9cm}
\raisebox{-.9cm}{\large +1}
\hspace{.4cm}
+\raisebox{-.6in}{
\epsfig{file=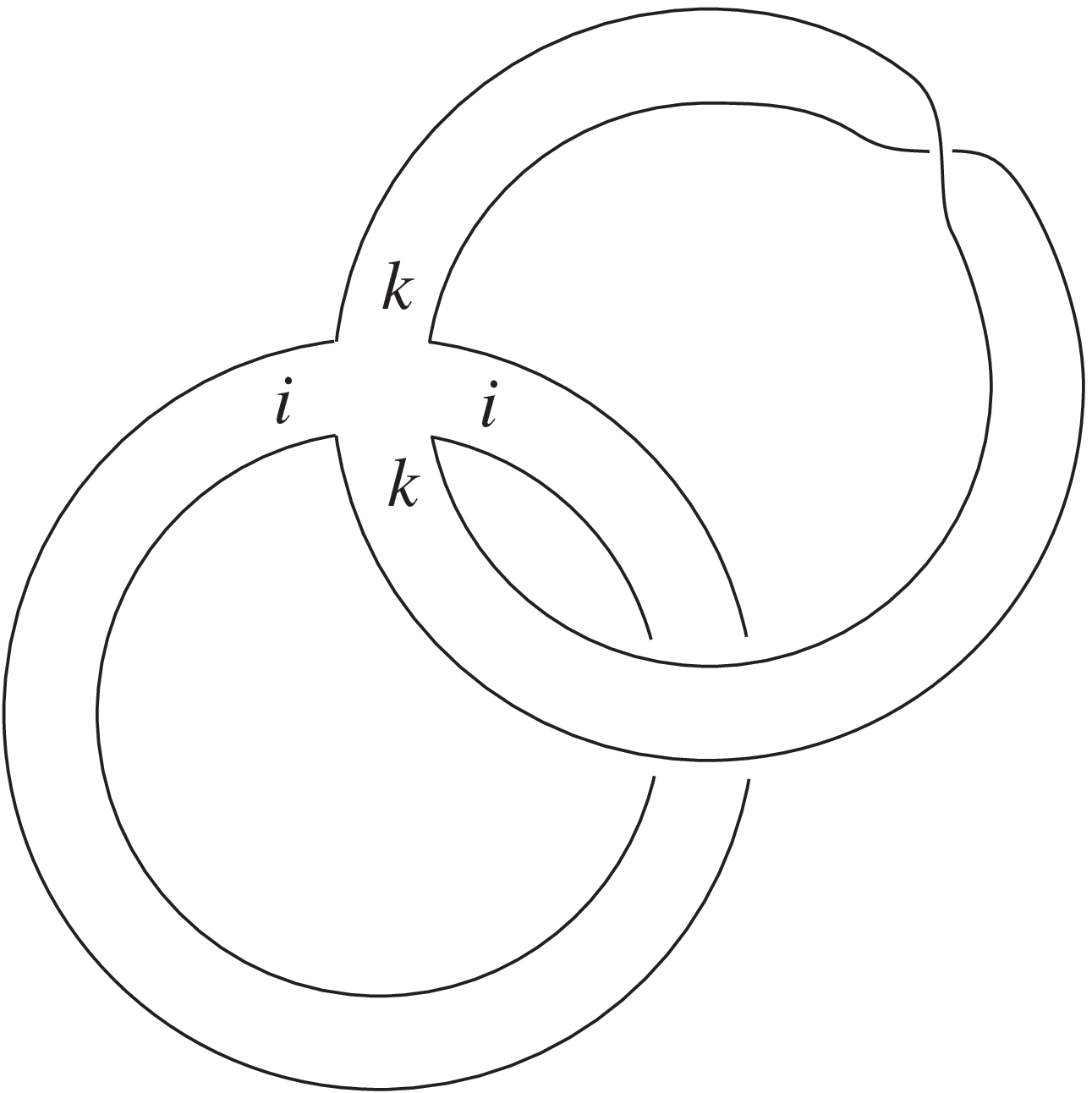,width=1.3in}}\hspace{-.9cm}
\raisebox{-.9cm}{\large +1}
\hspace{.4cm}
+\raisebox{-.6in}{
\epsfig{file=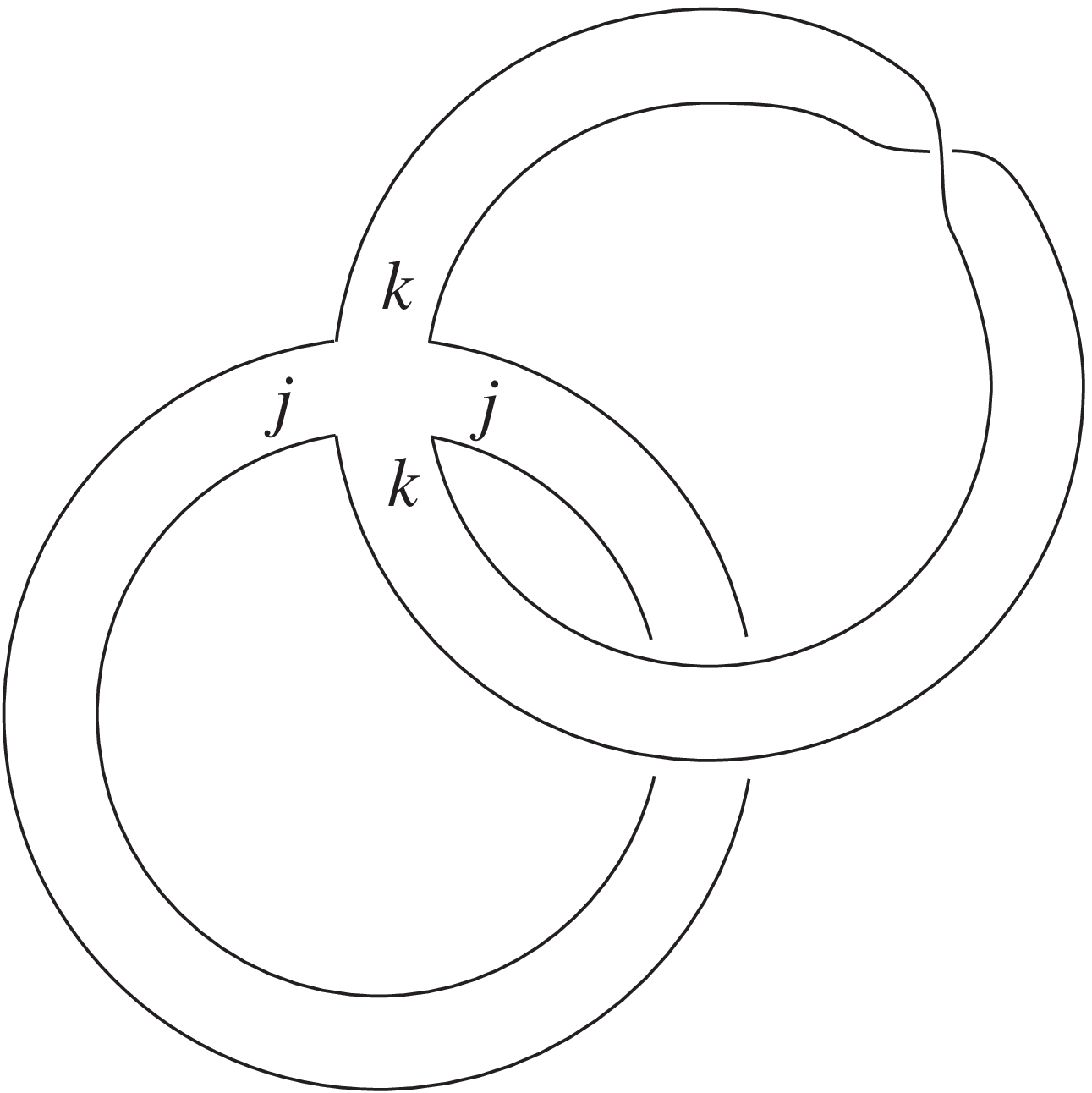,width=1.3in}}\hspace{-.9cm}
\raisebox{-.9cm}{\large +1}
\hspace{.4cm}\\
\end{split}
\end{equation*}
\caption{Computation of $\mu_\Gamma$ for a non-orientable
M\"obius graph $\Gamma$. Here $\chi(S_\Gamma)=0$.
The surface $S_\Gamma$ is a Klein bottle.
The contributions~$\pm 1$ add to yield
$\mu_\Gamma
=4=(2-\beta)^2
$ 
(for $\beta=4$).
}
\label{klein}
\end{figure}

If we call $e_0=1$ and $A_0=S$, a $j$-valent 
vertex now looks like
\begin{equation}
\frac{1}{2j}\,\tr X^j=\frac{1}{2j}\,\tr \sum_{\a=0}^{\beta-1} e_\a
A_\a
=\sum_{\{(\a_1,\ldots,\a_j)\}/\sim}\frac{\rho(\a_1,\ldots,\a_j)}{2j}\, 
\tr \prod_{k=1}^j A_{\a_k}e_{\a_k}  \, .
\label{vertex}
\end{equation}
The last  sum runs over all $j$-tuples of integers 
$0,1,\ldots,\beta-1$ modulo the equivalence relation
\begin{equation}
(\a_1,\ldots,\a_j)\sim(\b_1,\ldots,\b_j)
\quad\mbox{iff}\quad \tr( A_{\a_1}\cdots  A_{\a_j})=
\tr (A_{\b_1}\cdots A_{\b_j})\, .
\end{equation}
The multiplicities $\rho(\a_1,\ldots,\a_j)$ 
are non-zero {\it only}
when the product 
\begin{equation}
e_{\a_1}\cdots e_{\a_j}=\pm 1\, .
\end{equation}
In other words, the vertex~\eqn{vertex} is real. 
Observe that the numbers 
\begin{equation}
\frac{2j}{\rho(\a_1,\ldots,\a_j)}
\end{equation}
count the number of automorphisms of a vertex with a given 
configuration of units~$e_\a$ sprinkled at every edge, with respect
to rotations and vertex flips.
For example, for $\beta=4$ and $j=3$
\bea
\frac{1}{6}\;\tr X^3&=&\frac{1}{6}\;\tr (1.S)^3\nn\\ \nn\\
&+&\frac{1}{2}\;\tr\Big(1.S\ i.A_1\ i.A_1
                       +1.S\ j.A_2\ j.A_2
                       +1.S\ k.A_3\ k.A_3\Big)\nn\\ \nn \\
&+&\frac{1}{1}\;\tr\Big(i.A_1\ j.A_2\ k.A_3\Big)\, .
\eea
On the first line, the trace of three identical symmetric matrices can be 
cycled and reversed, so $2j=6$ is the correct automorphism factor.
On the second line the terms with a single symmetric and a pair of 
antisymmetric matrices can only be flipped yielding two automorphisms.
The term on the third line has no automorphisms. 
We will keep track of the matrix type by ``sprinkling'' units
$\{1,e_i\}$ over the set of M\"obius graphs (see Figure~\ref{klein}).

\vspace{.4cm}
Orchestrating the above observations yields the following: 
\begin{lemma} The matrix integral $\log\left(Z^{(\beta)}(t,N)\right)$ may be
computed as a sum over 
M\"obius graphs $\Gamma$ with weight 
\begin{equation}
\frac{\mu_\Gamma N^{f_{\Gamma}}}{|{\rm Aut}(\Gamma)|}\; 
\end{equation}
multiplied by a single power of $t_j$ for each $j$-valent vertex in $\Gamma$.
The factor $\mu_\Gamma$ is calculated by
\begin{enumerate}
\item Writing down all possible configurations of units
$e_\a\in\{1,e_1,\ldots e_{\beta-1}\}$ 
at each vertex such that their product is $\pm 1$.
\item Counting these signed configurations with an additional
minus sign for every twisted
edge with imaginary units $e_i$ at each end. 
\end{enumerate}
\end{lemma}
The proof of this lemma follows from our previous remarks and by
noting that including $e_\a$ from the vertices, the Wick contractions 
become (no sum on $i$)
\bea
\langle 1.S_{ab}\ 1.S_{cd}\rangle&=&
\phantom{-}\delta_{ac}\delta_{bd}+\delta_{ad}\delta_{bc}\, ,\\
\langle e_i.A_{iab}\ e_i.A_{icd}\rangle&=&
-\delta_{ac}\delta_{bd}+\delta_{ad}\delta_{bc}\, .
\eea
A sample computation of $\mu_\Gamma$ is given in Figure~\ref{klein}.

\begin{figure}[htb]
\centerline{\epsfig{file=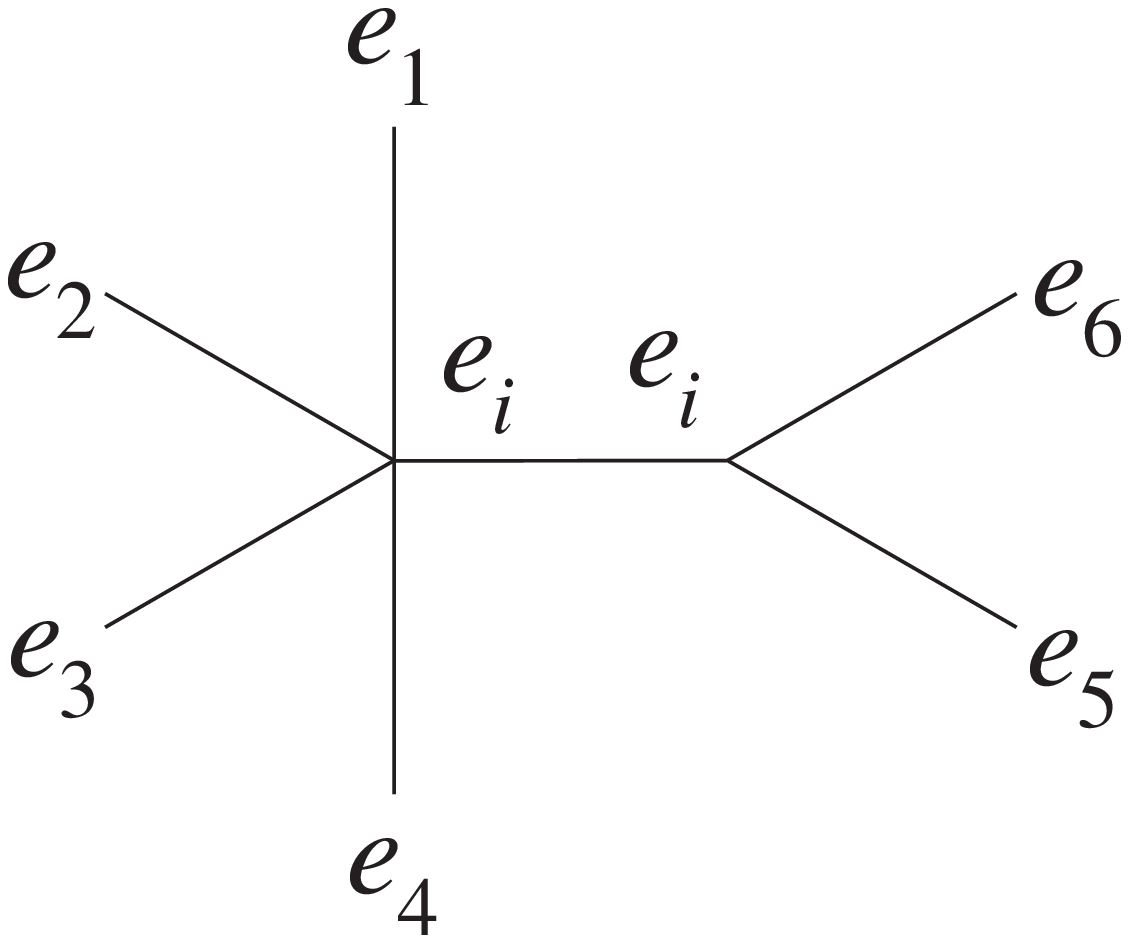, width=1.3in}
\hskip0.3cm \raisebox{1.5cm}{$=$} \hskip0.3cm
\epsfig{file=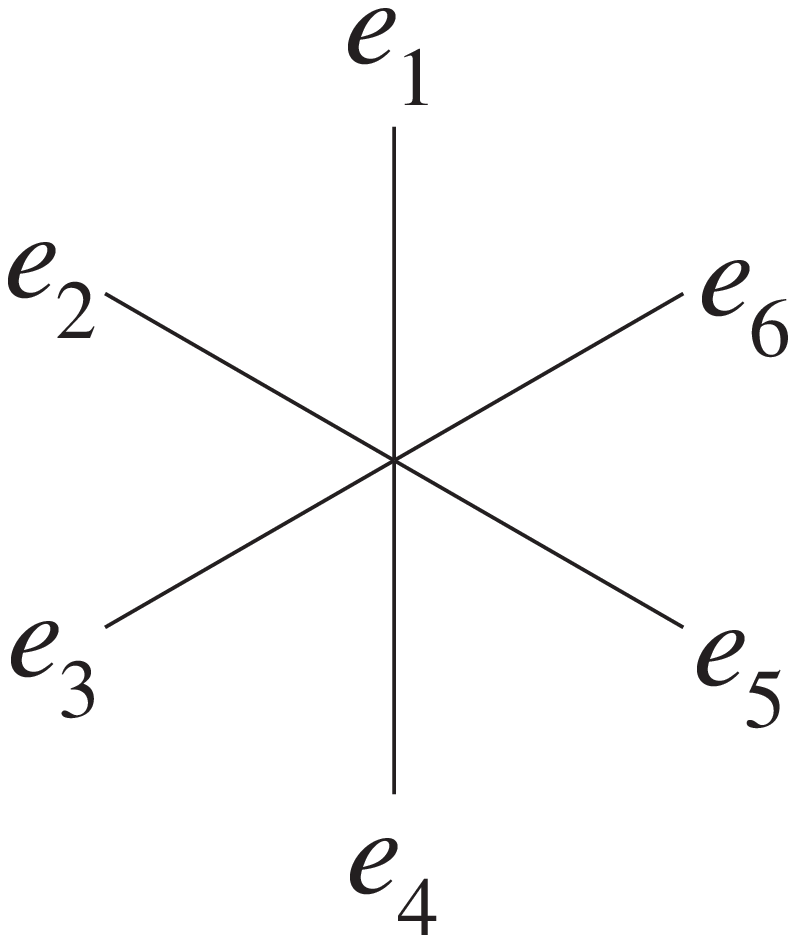, width=1.1in}} \caption{Edge contraction} 
\label{fig:contraction} \end{figure}

Our proof is completed by computing the numbers $\mu_\Gamma$,
which requires a second major ingredient:
\begin{lemma}\label{puncture}
The quantity $\mu_\Gamma$ is a topological invariant of the 
$n$-punctured surface
$S_\Gamma \setminus \{p_1,p_2,\ldots,p_n\}$, where 
the removed points are all distinct and $n=f_\Gamma$ is the 
number of~{$\phantom{.\!\!}$}~faces of the cell decomposition defined 
by $\Gamma$.
\end{lemma}
The proof of this crucial Lemma is almost a triviality.
Graphs are identified if they are equivalent under rotations
and flips of their vertices (Figure~\ref{fig:Moebius}),
so it suffices to show that $\mu_\Gamma$ is invariant
under contraction of an untwisted edge. This operation is depicted in 
Figure~\ref{fig:contraction}

This relation is obvious since the products of units
at either vertex must be real and the Wick contraction corresponding
to an untwisted edge with units $e_{\a}$ at either end comes with
an overall factor $+1$.
We remark that an analogous relation involving a vertex flip 
for the contraction of a twisted edge follows.

\begin{figure}
\centerline{\epsfig{file=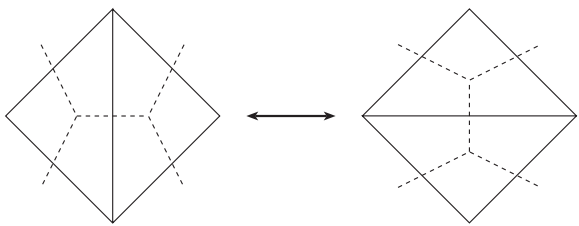, width=3in}}
\caption{The diagonal flip operation for a triangulation.}
\label{diagonal_flip}
\end{figure}

We now compute the invariant $\mu_\Gamma$ for a graph of arbitrary topology.
Since $\mu_\Gamma$ is topological, we may employ a standard graph
for each distinct topology labeled by the orientability $\natural_\Gamma$, 
number of faces $f_\Gamma$ and genus $g(S_\Gamma)$.
The key theorem from topology we need here is the connectivity
of the space of all triangulations of a compact surface
with a fixed number of vertices
under the action of a \emph{diagonal flip}~\cite{Hatcher} (see
Figure~\ref{diagonal_flip}).  A diagonal
flip of a triangulation is exactly the \emph{fusion move} of
the dual M\"obius graph.
For orientable
surfaces, the connectivity in concern implies the path connectivity
of the moduli space $\mathfrak{M}_{g,n}$ of smooth complete algebraic
curves defined over $\mathbb{C}$ with $n$ marked points. 
The connectivity of triangulations of non-orientable surfaces has
been also established.

Now note that because of the invariance of $\mu_\Gamma$ under
edge contraction, we can expand all vertices of valence greater than
3 and contract all vertices of valence 1 and 2 to create a trivalent
graph. Since the dual of a trivalent graph is a triangulation of the
surface and the number of vertices of the triangulation is 
the number $n=f_\Gamma$ of faces of the M\"obius graph $\Gamma$,
the connectivity of the space of triangulations implies that 
our invariant $\mu_\Gamma$ is a constant for each 
topology of a given $n$-punctured surface. In the actual computation,
it is convenient to use the following representatives for each of the
three topological classes:

\begin{enumerate}
\item Orientable ($\natural=1$); a standard graph is 
given in Figure~\ref{fig:oriented}.

\begin{figure}[htb] 
\centerline{\epsfig{file=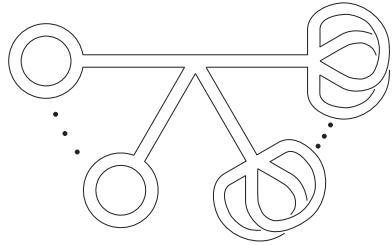, width=2in}} 
\caption{A standard  M\"obius graph
that represents an orientable surface of genus $g$ with
$n$ marked points.
It consists of $n-1$ tadpoles on the left and $g$ bi-petal flowers
on the right.} \label{fig:oriented} \end{figure}

\item Non-orientable ($\natural=-1$), 
odd genus; with standard graph given in Figure~\ref{fig:nonorodd}.

\begin{figure}[htb] 
\centerline{\epsfig{file=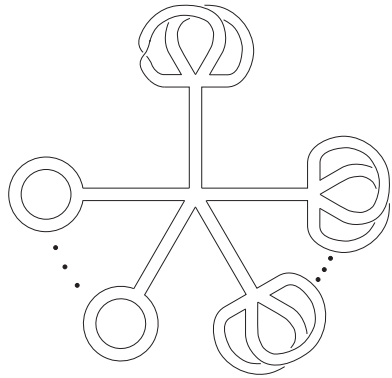, width=2in}} 
\caption{A standard M\"obius 
graph drawn on a non-orientable surface of genus $g=2k+1$
with $n$ marked points. In addition to $n-1$ tadpoles on the left it 
has $k$ orientable bi-petal
flowers on the right  and  a non-orientable one
 added at the top.} \label{fig:nonorodd} \end{figure}

\item Non-orientable ($\natural=-1$), even genus; a standard graph is in 
Figure~\ref{fig:nonoreven}.

\begin{figure}[htb] 
\centerline{\epsfig{file=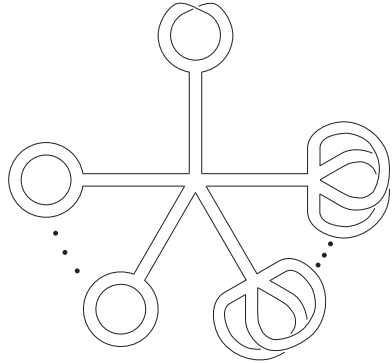, width=2in}} \caption{A
standard  M\"obius graph representing  
a non-orientable surface of genus $g=2k$ with $n$ marked points.
It has $n-1$ orientable tadpoles on the left, $k$ bi-petal orientable flowers
on the right and a non-orientable
tadpole added at the top.} \label{fig:nonoreven} \end{figure}

\end{enumerate}
Finally, it is easy to calculate $\mu_\Gamma$ for each 
one particle irreducible components (subgraphs which remain connected after
cutting a single edge):

\begin{equation}
\centerline{\epsfig{file=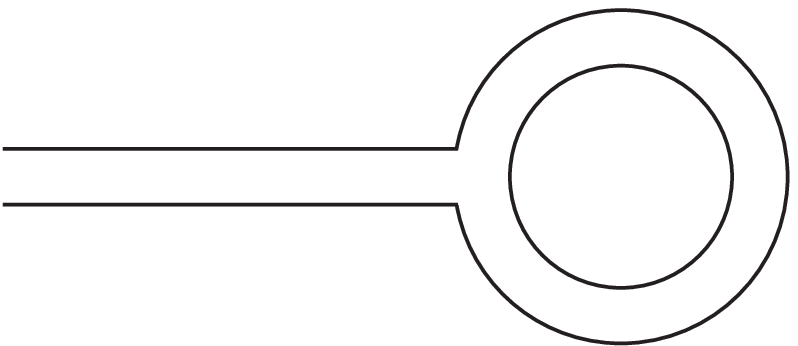, 
width=1in}\hskip0.3cm \raisebox{0.4cm}{$=\beta\, ,$}}
\label{fig:tadpole} 
\end{equation}

\begin{equation}
\centerline{\epsfig{file=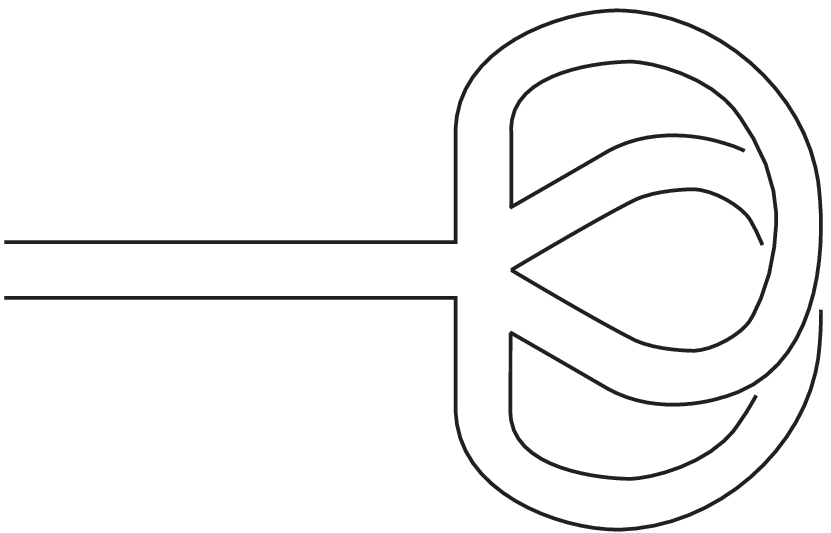, 
width=1in}\hskip0.3cm 
\raisebox{0.7cm}{$=1+3(\beta-1)-(\beta-1)(\beta-2)\, ,$}}
\label{fig:flower} 
\end{equation}

\begin{equation}
\centerline{\epsfig{file=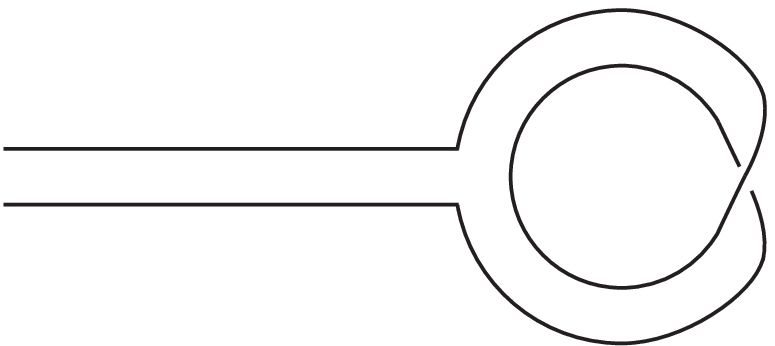, 
width=1in}\hskip0.3cm 
\raisebox{0.4cm}{$=2-\beta\, ,$}}  \label{fig:nonortad} 
\end{equation}

\begin{equation}
\centerline{\epsfig{file=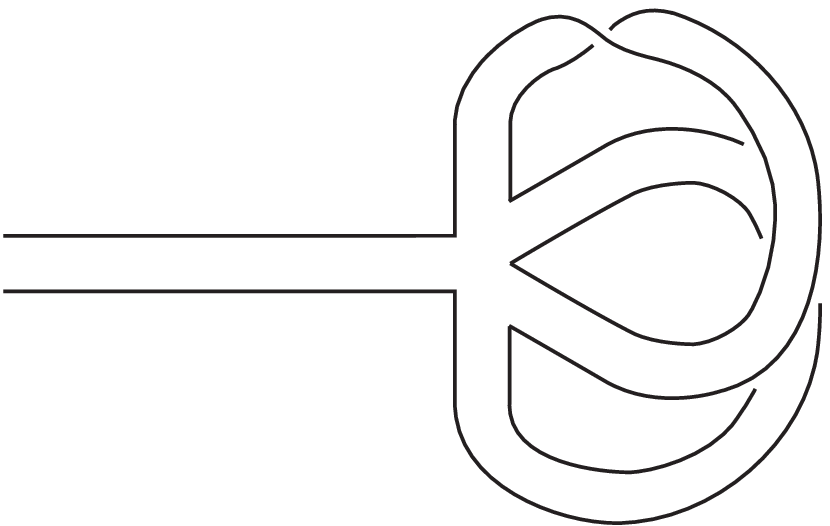, 
width=1in}\hskip0.3cm 
\raisebox{0.7cm}{$=(2-\beta)^2\, .$}}  \label{fig:nonorfl} 
\end{equation}

\noindent
Each of the above calculations is rather similar: A single
line emitted from a tadpole subgraph can only ever carry the unit 
$e_0=1$. As an example, \eqn{fig:flower}
 arises from (i)~placing all 1's
on remaining four lines (ii)~1's on one loop and imaginary units
$e_i$ on the other which can be done in $2(\beta-1)$ ways, 
(iii)~the same imaginary unit on both loops giving $\beta-1$ possibilities,
(iv) different imaginary units which can only be achieved
for the quaternions and in $6=(\beta-1)(\beta-2)$ different ways 
incurring a minus sign since $-1=ijij$ (say).

Therefore we find 
\begin{equation}
\mu_\Gamma=(-4+6\beta-\beta^2)^{^{\ss 1 
- \12\Sigma_\Gamma-\12\chi(S_\Gamma)}} 
(2-\beta)^{^{\ss \Sigma_\Gamma}} 
\beta^{^{\ss f_\Gamma-1}}\, .
\end{equation}
This result combined with the first Lemma above completes the proof
of our main theorem.

\section{Examples}

\label{examples}

A simple consistency check is the case $\beta=2$. 
Clearly, non-orientable graphs give a vanishing
contribution as required for hermitian matrix integrals. 
To obtain a more conventional normalization, we
rescale $X\longrightarrow \beta^{1/2}\ X$ in~\eqn{Z} and absorb all
but one power of $\beta$ in the couplings $t$, so that
\begin{multline}
\log\left(
\frac{\textstyle\int [dX]_{_{\!\ss(\b)}}\!\,\exp\Big(\!-\frac{\beta}{4}\ \tr X^2
+\sum_{j=1}^\infty \frac{\beta t_j}{2j}\ \tr X^j \Big)}{
\textstyle\int [dX]_{_{\!\ss(\b)}}\!\,\exp\Big(\!-\frac{\beta}{4}\ \tr 
X^2\Big)}\right)
\\
=
\sum_{\Gamma\in\mathfrak G}
\frac{\textstyle 
(-4+6\beta-\beta^2)^{^{\ss 1 - \12\Sigma_\Gamma-
\12\chi(S_\Gamma)}} 
(2-\beta)^{^{\ss \Sigma_\Gamma}}
\beta^{^{\ss \chi(S_\Gamma)-1}}
N^{^{\ss f_\Gamma}}}{|{\rm Aut}(\Gamma)|} \prod_{j}
t_j^{v^{(j)}_\Gamma}\! .
\label{nearly}
\end{multline}
Hence when $\beta=2$
\begin{equation}
\log\left(
\frac{\textstyle\int [dX]_{_{\!\ss(2)}}\!\,\exp\Big(\!-\frac{1}{2}\ \tr X^2
+\sum_{j=1}^\infty \frac{t_j}{j}\ \tr X^j \Big)}{
\textstyle\int [dX]_{_{\!\ss(2)}}\!\,\exp\Big(\!-\frac{1}{2}\ \tr 
X^2\Big)}\right)
=
\sum_{\Gamma\in\mathfrak G}
\frac{
2 N^{^{\ss f_\Gamma}}}{|\Aut(\Gamma)|} \prod_{j}
t_j^{v^{(j)}_\Gamma}\! .
\label{herm}
\end{equation}
We note that
\begin{equation}
\label{eq: ribbon}
\sum_{\Gamma\in\mathfrak G}
\frac{
2 N^{^{\ss f_\Gamma}}}{|\Aut(\Gamma)|} \prod_{j}
t_j^{v^{(j)}_\Gamma}
= \sum_{\Gamma\in\mathfrak{R}}
\frac{
 N^{^{\ss f_\Gamma}}}{|\Aut_\mathfrak{R}(\Gamma)|} \prod_{j}
t_j^{v^{(j)}_\Gamma},
\end{equation}
where $\mathfrak{R}$ denotes the set of all connected ribbon
graphs and $\Aut_\mathfrak{R}(\Gamma)$ the automorphism
group of a ribbon graph disallowing orientation-reversing
automorphisms. To see~\eqn{eq: ribbon}, let $\Gamma$
be an oriented ribbon graph. Then either (a) $\Gamma$ and 
its flip $\check{\Gamma}$ are isomorphic as a ribbon graph; or
(b) they are different ribbon graphs. In case (a), we have
$|\Aut(\Gamma)| = 2|\Aut_\mathfrak{R}(\Gamma)|$. If (b)
is the case, then $\Aut(\Gamma) \cong \Aut_\mathfrak{R}(\Gamma)$,
but $\Gamma$ and $\check{\Gamma}$ appear as different graphs
on the right hand side while they are the same as a M\"obius graph
in $\mathfrak{G}$.

Together, equations~\eqn{eq: ribbon}
and~\eqn{herm} constitute the well-known result for the hermitian
matrix integral~\cite{BIZ}.

A less trivial test of our new result for the
$\beta=4$, GSE model is a comparison with the Penner model.
To begin with we re-express~\eqn{eq: main formula} in yet another 
normalization, together with a specialization $t_1=t_2=0$:

\begin{multline}
\log\left(
\frac{\textstyle\int [dX]_{_{\!\ss(4)}}\!\,\exp\Big(\!-\frac{1}{2}\ \tr X^2
+\sum_{j=3}^\infty \frac{t_j}{j}\ \tr X^j \Big)}{
\textstyle\int [dX]_{_{\!\ss(4)}}\!\,\exp\Big(\!-\frac{1}{2}\ \tr
X^2\Big)}\right)
\\
=
\sum_{\Gamma\in\mathfrak G}
\frac{
(-1)^{\chi(S_\Gamma)}
}{|{\rm Aut}(\Gamma)|}(2N)^{^{\ss f_\Gamma}} \prod_{j\ge 3}
t_j^{v^{(j)}_\Gamma}\! .
\label{old}
\end{multline}
Here we have employed the useful identity
\begin{equation}
\label{eq: useful}
(-1)^{\Sigma_\Gamma}=(-1)^{\chi(S_\Gamma)}\, .
\end{equation}

The Penner substitution
\begin{equation}
\label{eq: Penner substitution}
t_j\longrightarrow -z^{j/2-1}, \quad j\ge 3
\end{equation}
yields the graphical expansion
\begin{multline}
\lim_{m\rightarrow\infty}
\log\left(
\frac{\textstyle\int [dX]_{_{\!\ss(4)}}\!\,\exp\Big(\!-\sum_{j=2}^{2m}
\frac{z^{j/2-1}}{j}\ \tr X^j \Big)}{
\textstyle\int [dX]_{_{\!\ss(4)}}\!\,\exp\Big(\!-\frac{1}{2}\ \tr
X^2\Big)}\right)
\\
=
\sum_{\Gamma\in\mathfrak G}
\frac{
(-1)^{e_\Gamma}
}{|{\rm Aut}(\Gamma)|}(2N)^{^{\ss f_\Gamma}}
(-1)^{\chi(S_\Gamma)}(-z)^{e_\Gamma-v_\Gamma}\, .
\label{penner_graphs}
\end{multline}
Equation~\eqn{penner_graphs} is valid as an asymptotic 
expansion of the integral 
for $z\rightarrow 0$ while keeping $z>0$, as an element of a formal
power series ring
$(\mathbb{Q}[N])[[z]]$.

The integral on the left hand side can be evaluated explicitly
(see~\eqn{ernie} and~\eqn{eq:K} of Section~\ref{Penner_Model}).
The leading terms as an expansion in $z$ are
\begin{multline}
\lim_{m\rightarrow\infty}
\log\left(
\frac{\textstyle\int [dX]_{_{\!\ss(4)}}\!\,\exp\Big(\!-\sum_{j=2}^{2m}
\frac{z^{j/2-1}}{j}\ \tr X^j \Big)}{
\textstyle\int [dX]_{_{\!\ss(4)}}\!\,\exp\Big(\!-\frac{1}{2}\ \tr
X^2\Big)}\right)
\\
=\Big(-\frac{1}{12}N-\frac12N^2+\frac23N^3\Big)\ z
+{\mathcal O}(z^2)\, .
\end{multline}

\pagebreak

On the other hand, our graphical expansion yields
\begin{equation}
\sum_{\Gamma\in\mathfrak G}
\frac{
(-)^{e_\Gamma}
(-2N)^{^{\ss b_\Gamma}}z^{e_\Gamma-v_\Gamma}}{|{\rm
Aut}(\Gamma)|}\hspace{7cm}
\end{equation}
\begin{equation*}
\begin{split}
\qquad\;\;\;
&=(-2N)^3z\;
\left\{
\frac{(-1)^2}{4}\;\raisebox{-.2cm}{\epsfig{file=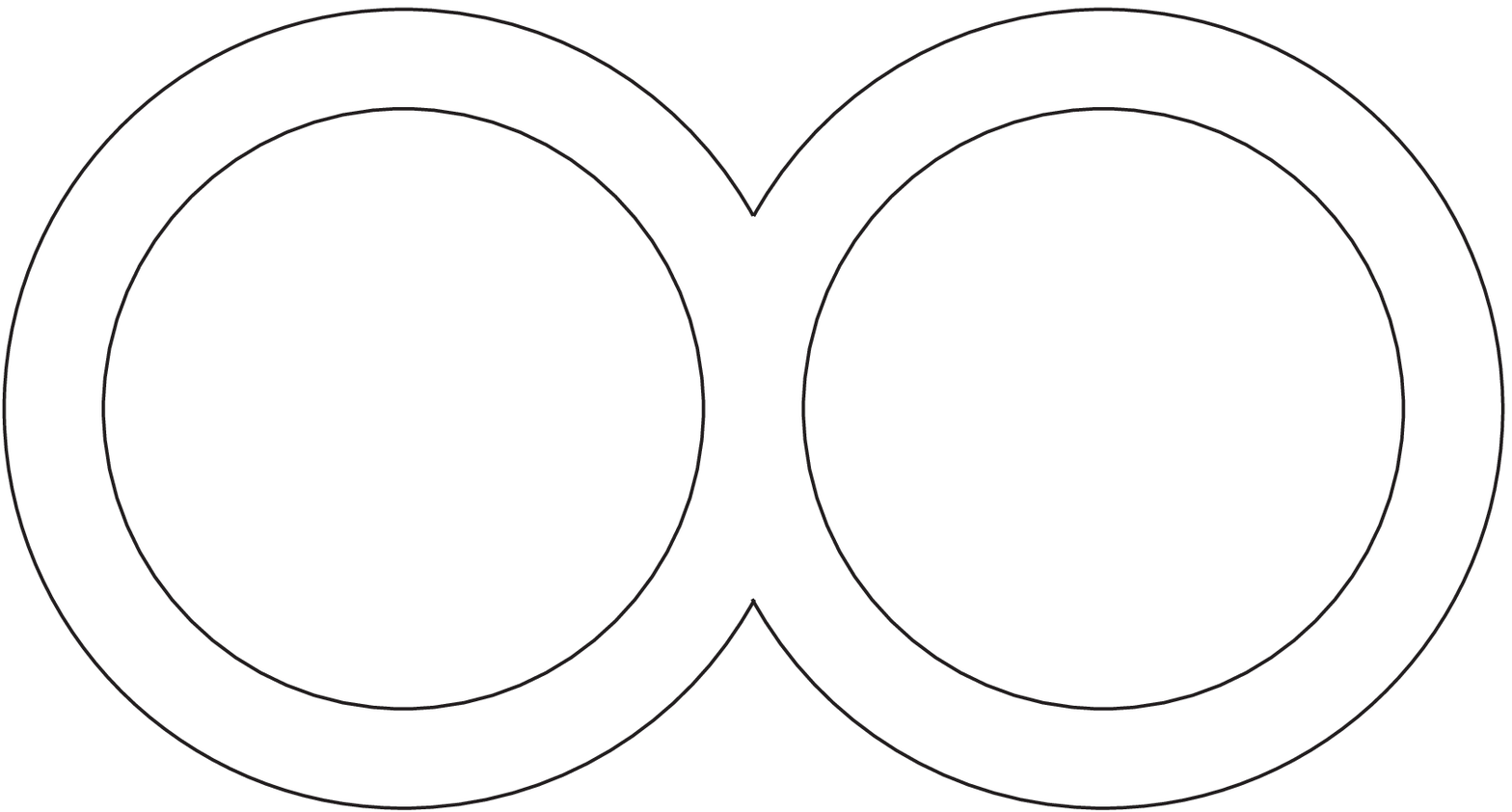, width=.55in}}
+\frac{(-1)^3}{4}\;\raisebox{-.2cm}{\epsfig{file=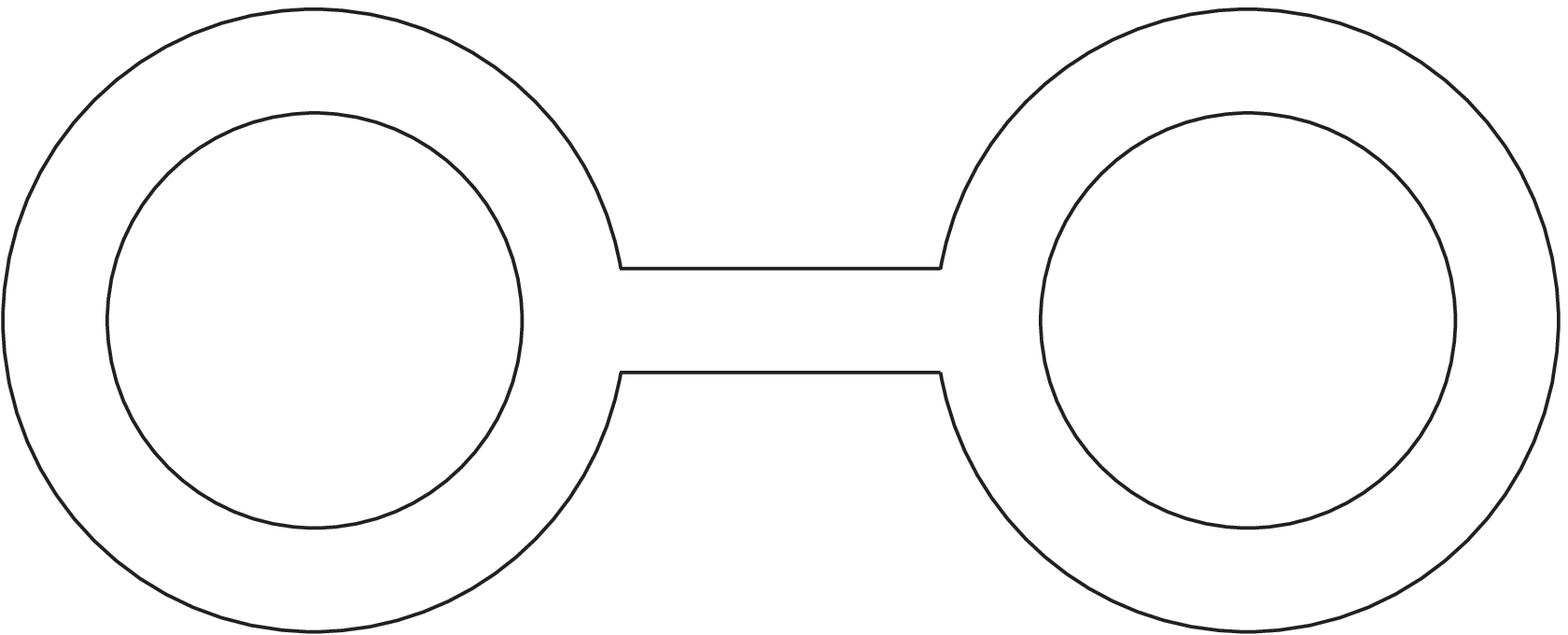, width=.6in}}
+\frac{(-1)^3}{12}\;\raisebox{-.4cm}{\epsfig{file=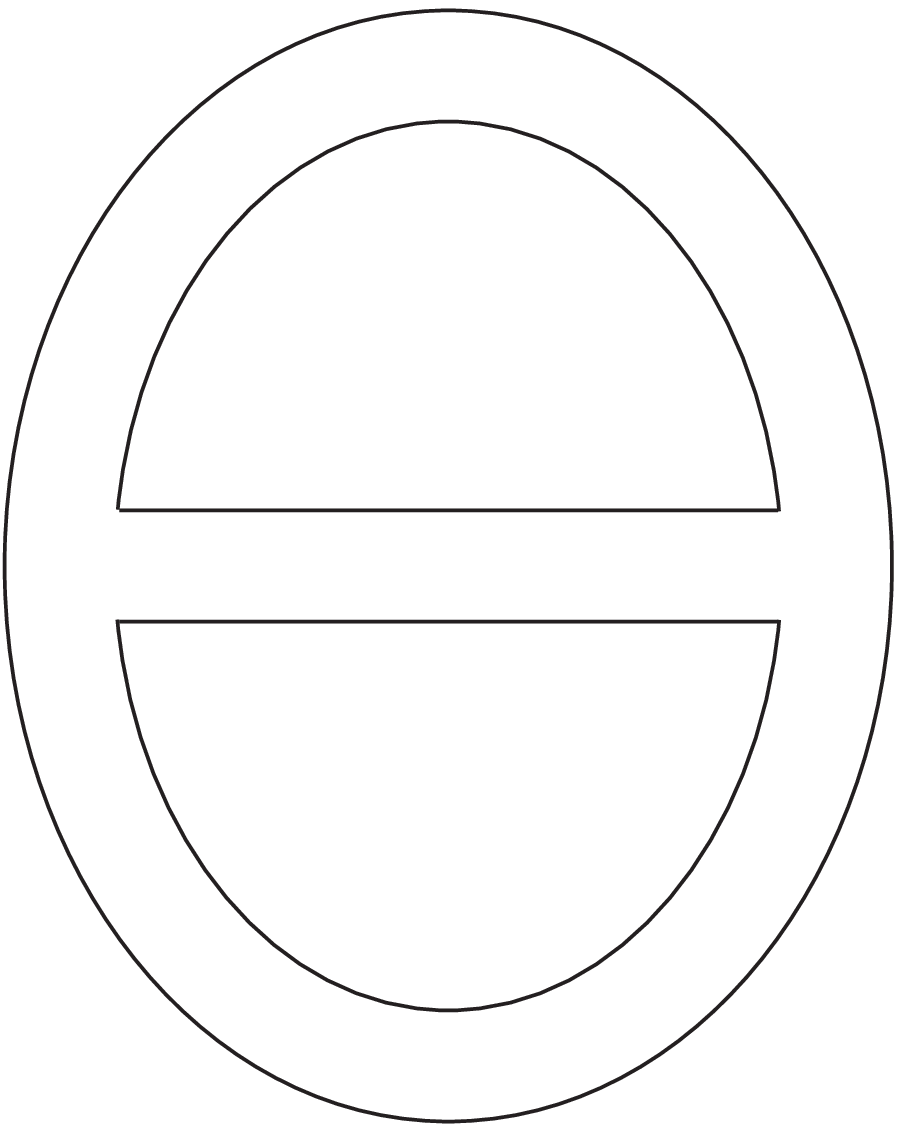, width=.33in}}
\right\}
\\
&+(-2N)^2z\;
\left\{
\frac{(-1)^2}{2}\;\;\raisebox{-.2cm}{\epsfig{file=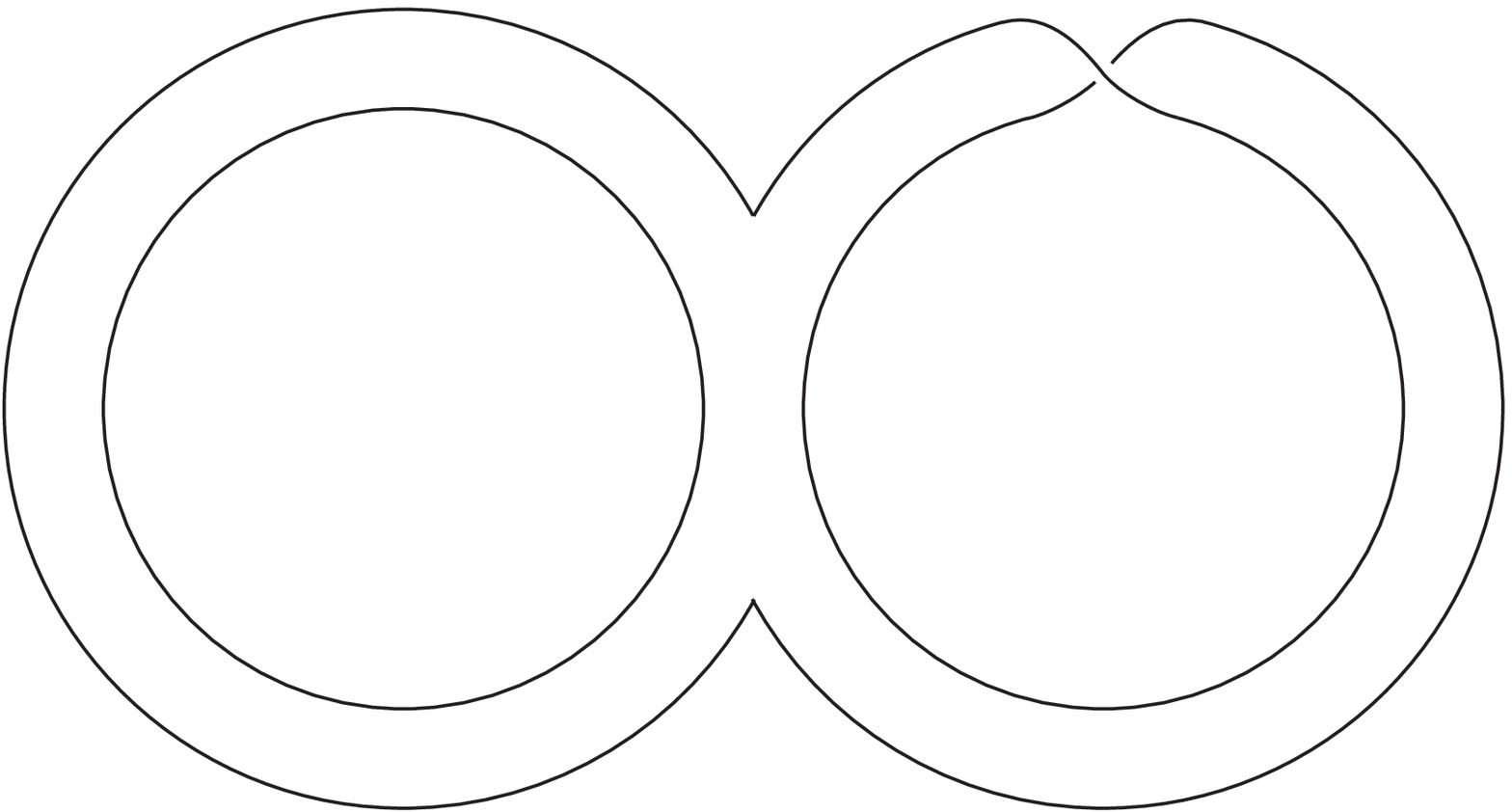, width=.55in}}
+\frac{(-1)^3}{2}\;\raisebox{-.2cm}{\epsfig{file=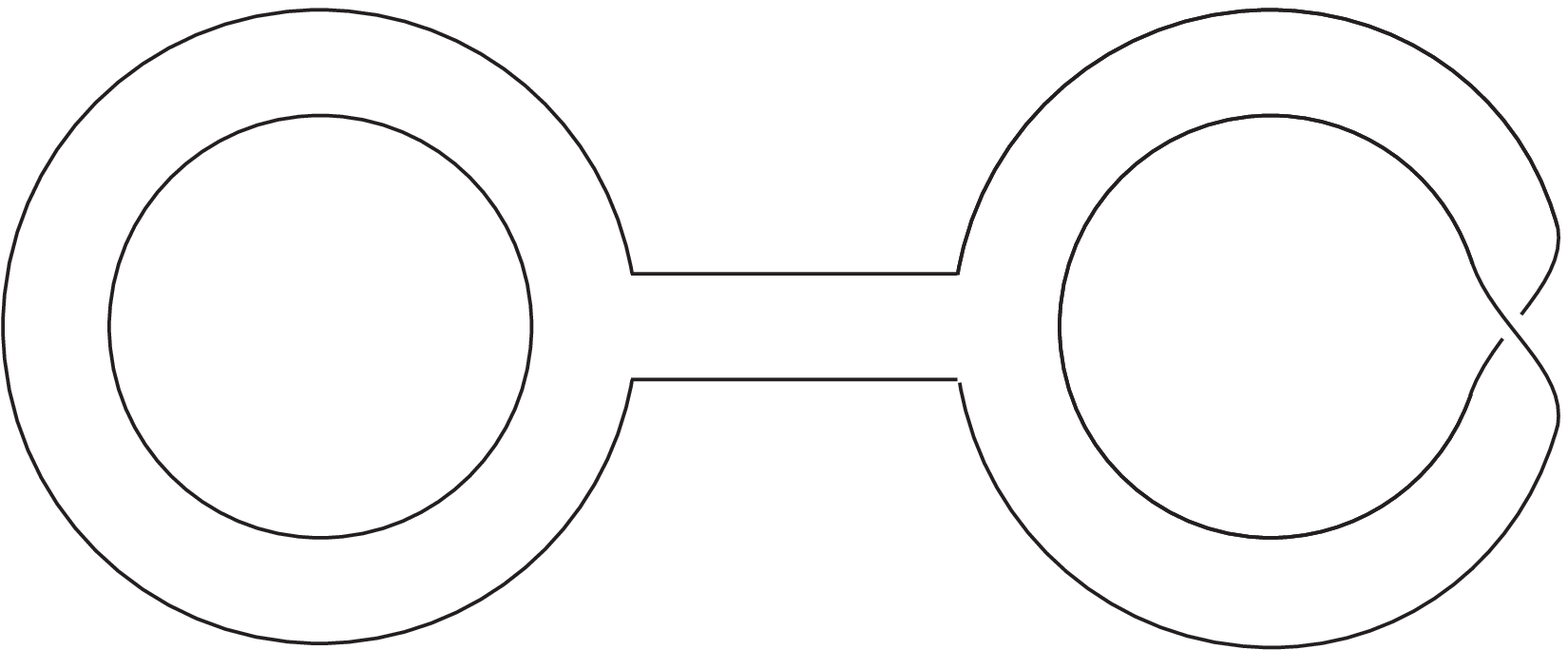, width=.6in}}
+\frac{(-1)^3}{4}\;\raisebox{-.4cm}{\epsfig{file=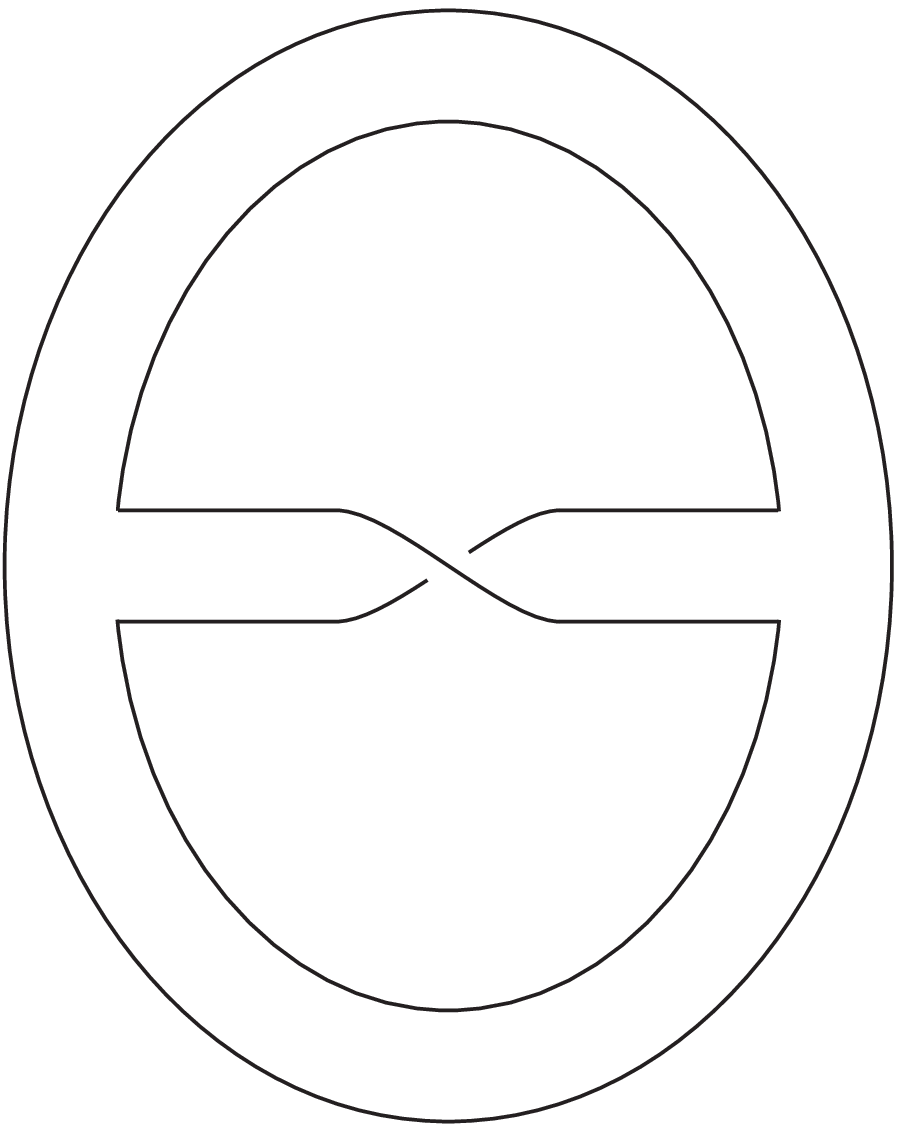, width=.33in}}
\right.
\\
&\qquad\qquad\quad\left.
+\, \frac{(-1)^2}{8}\;\;\,
\raisebox{-.35cm}{\epsfig{file=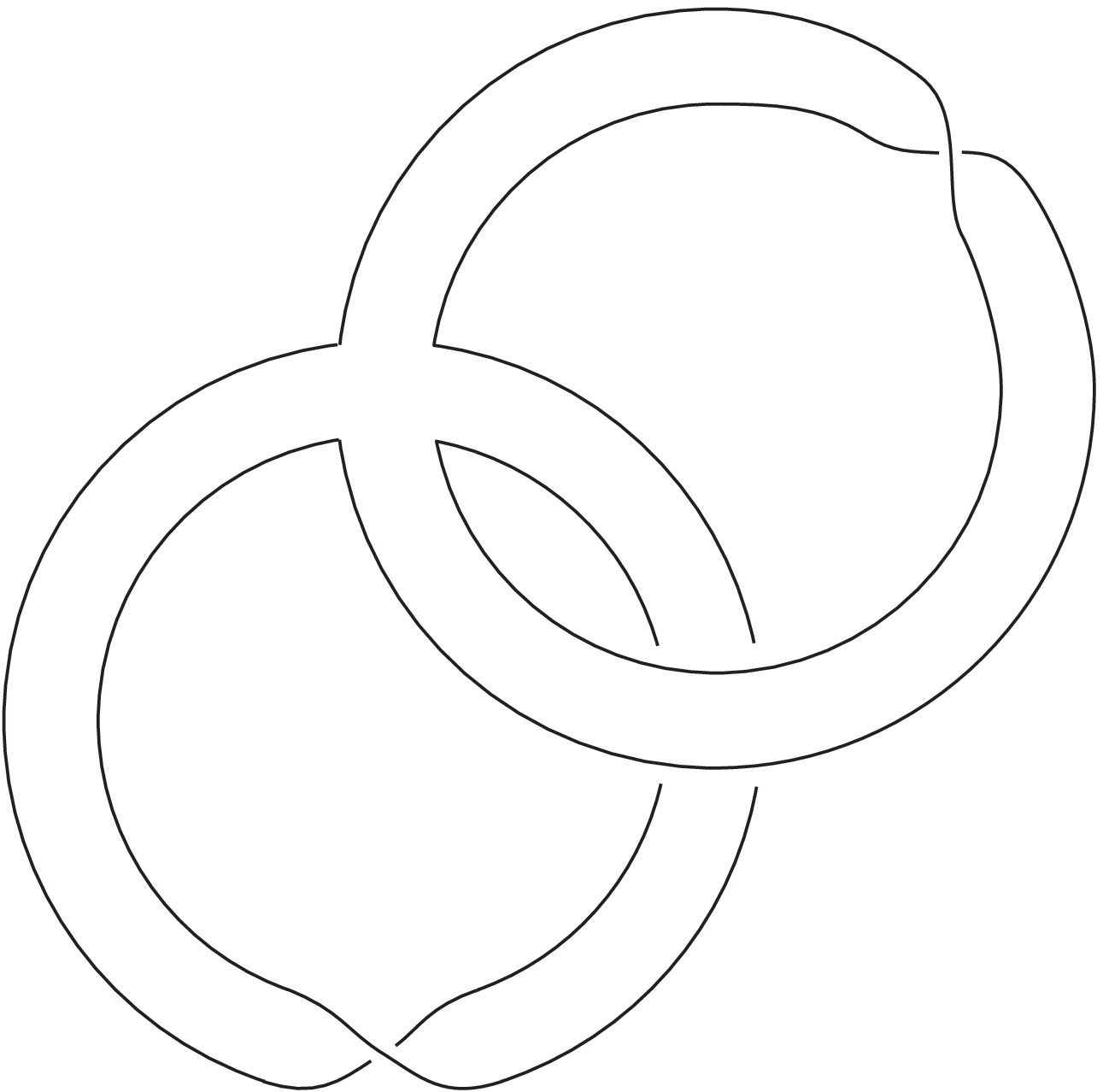, width=.46in}}\,
\right\}
\\
&+(-2N)\, z\;\,
\left\{
\frac{(-1)^2}{4}\;\;\raisebox{-.2cm}{\epsfig{file=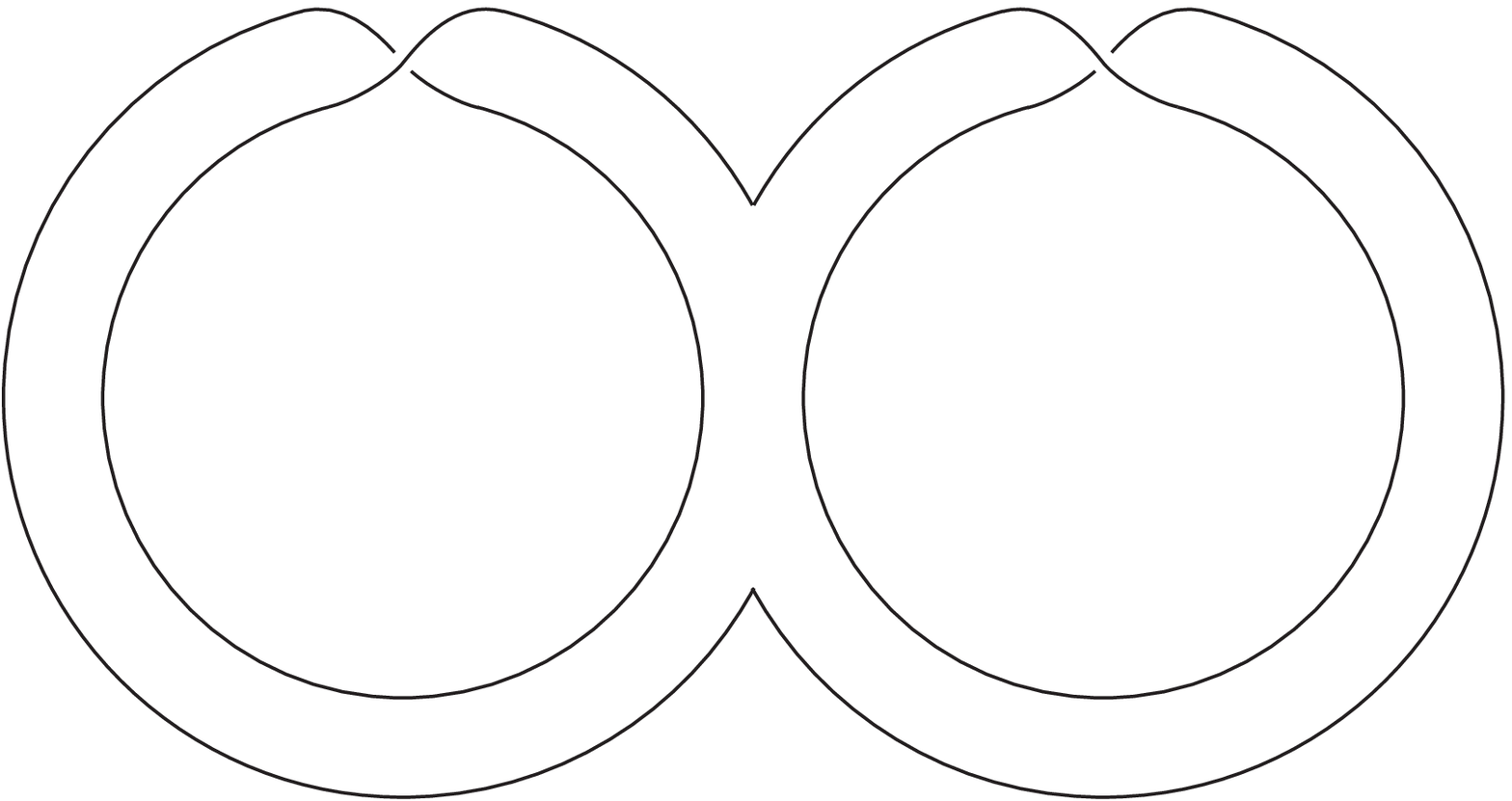, width=.55in}}
+\frac{(-1)^3}{4}\;\raisebox{-.2cm}{\epsfig{file=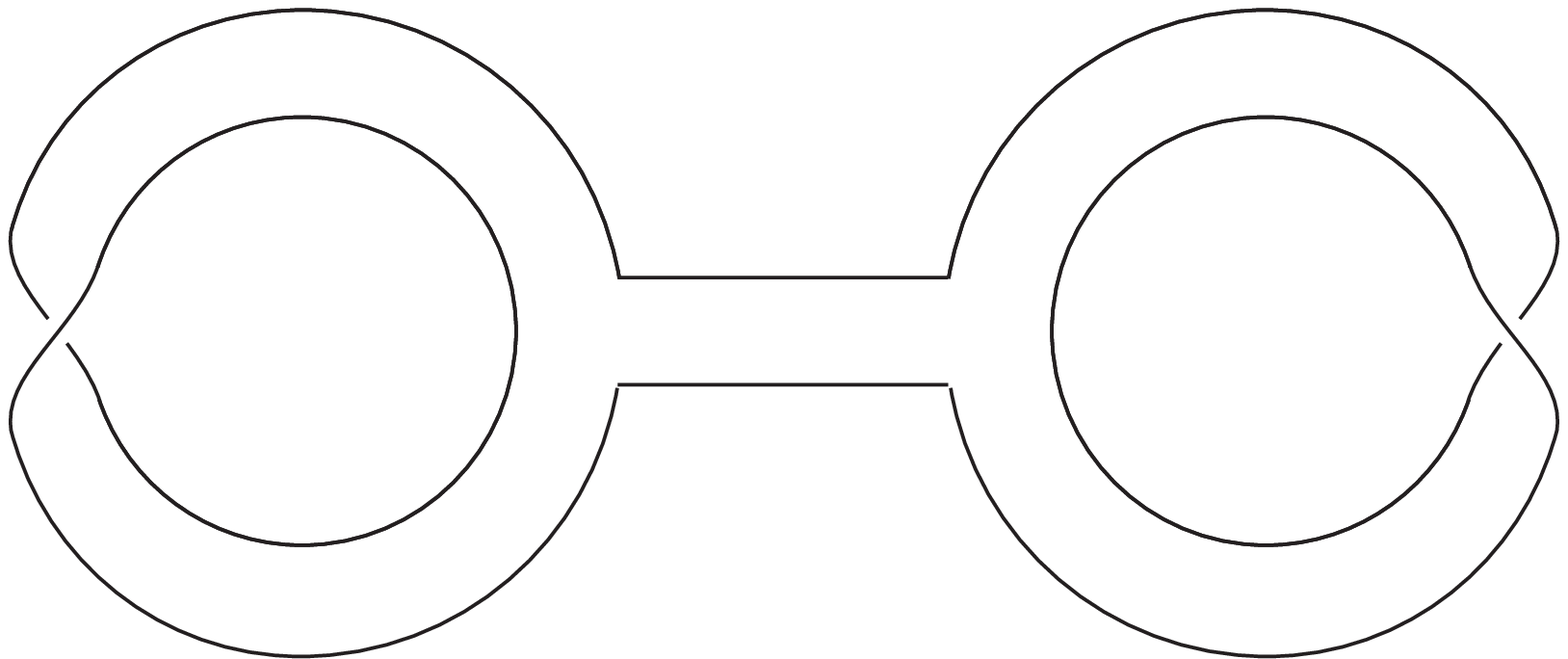, width=.6in}}
+\frac{(-1)^3}{4}\;
\raisebox{-.4cm}{\epsfig{file=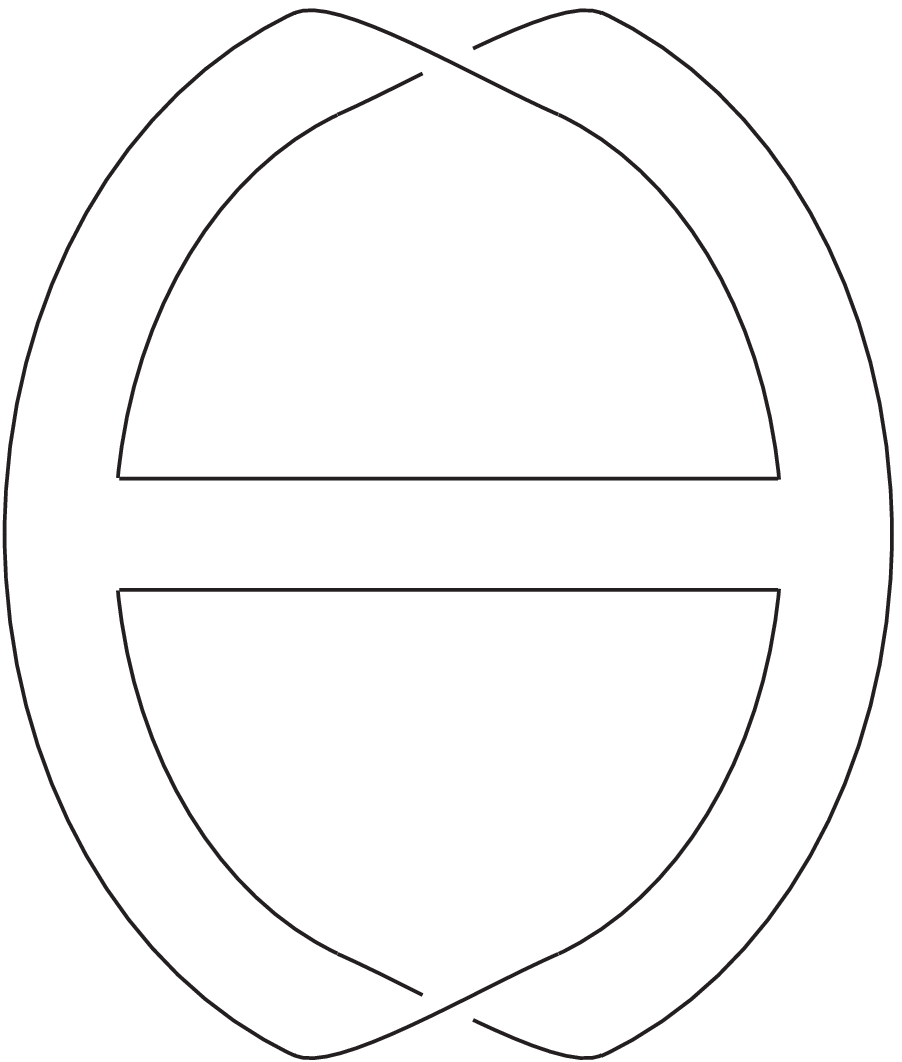, width=.33in}}
\right.
\\
&\qquad\qquad\quad\
+\,\frac{(-1)^2}{4}\;\;\,\raisebox{-.35cm}{\epsfig{file=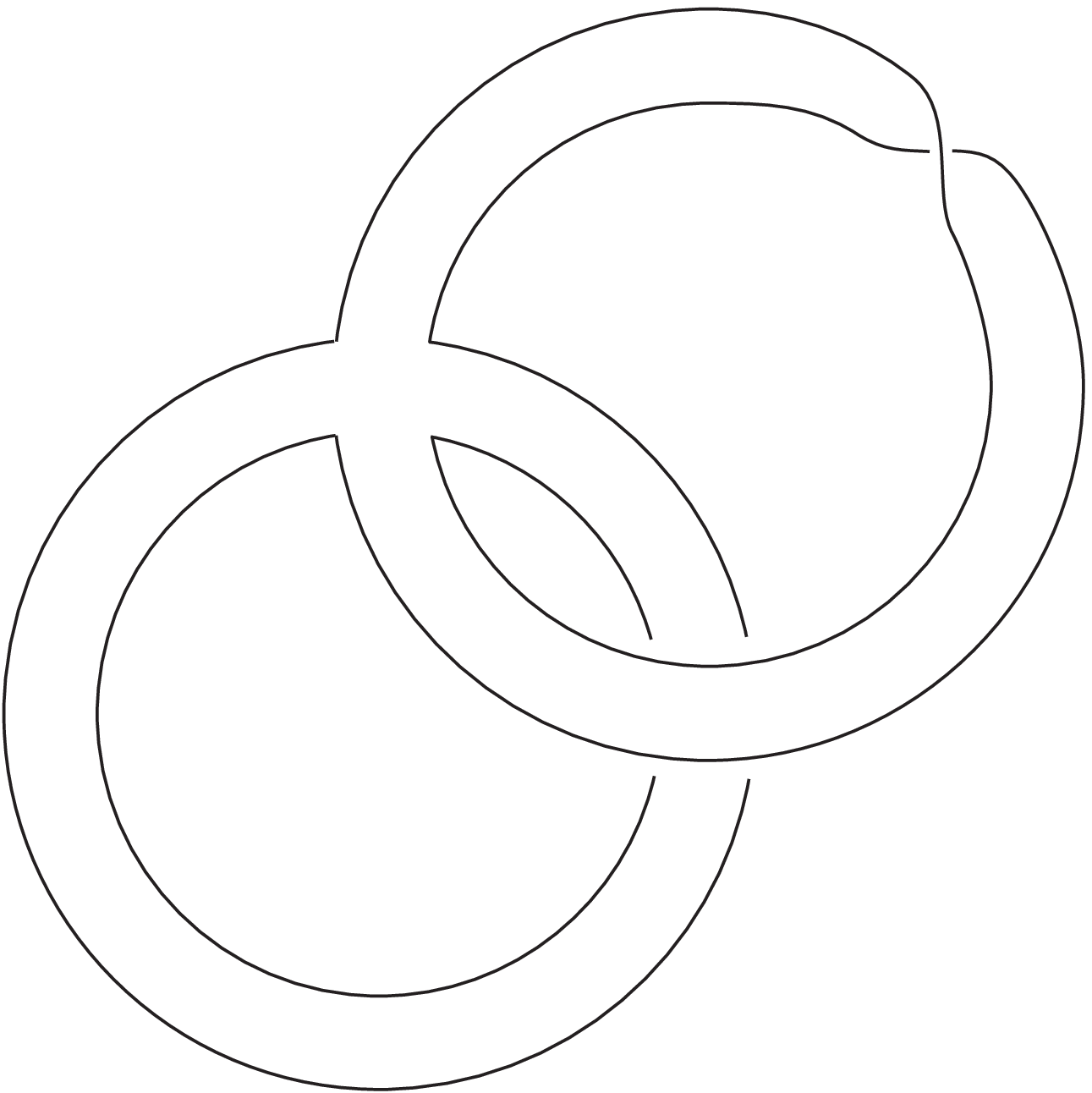, width=.46in}}
\qquad\qquad\qquad\quad\;\;\,
+\frac{(-1)^3}{12}\;
\raisebox{-.35cm}{\epsfig{file=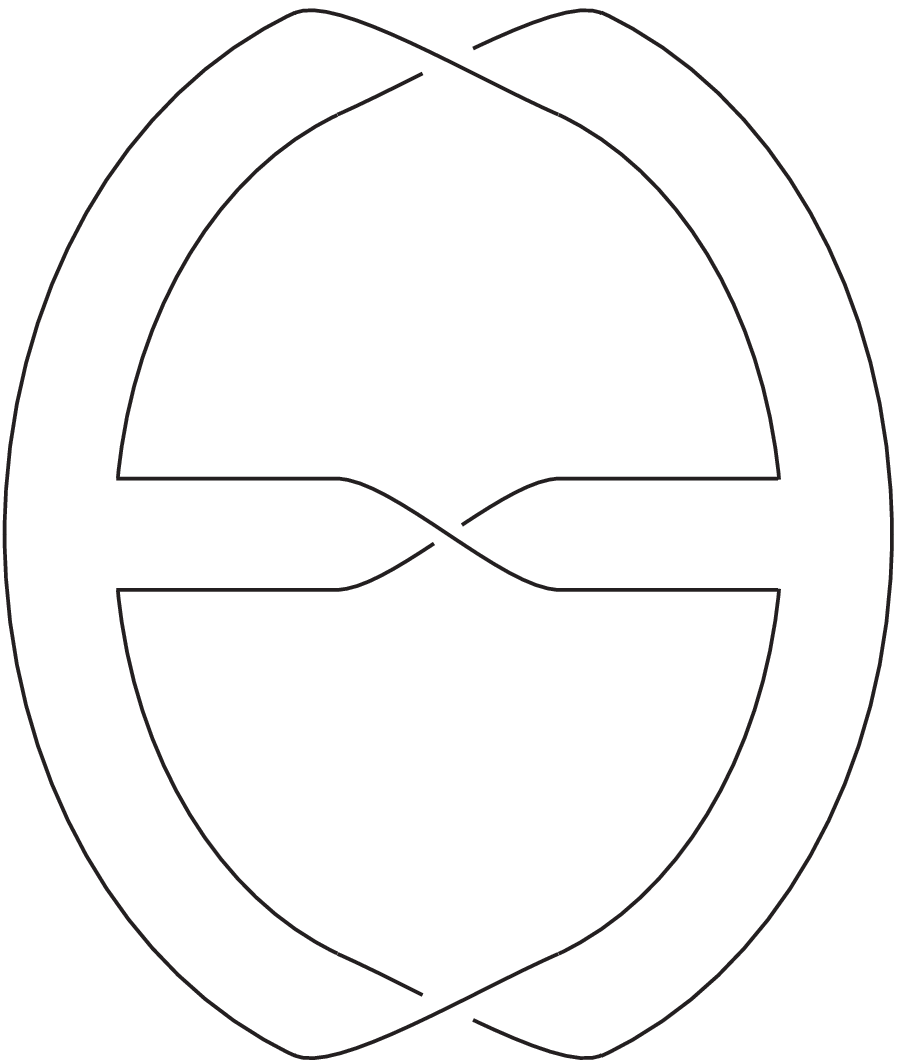, width=.33in}}
\\
&\qquad\qquad\quad\left.\,
+\,\frac{(-1)^2}{8}\;\;\,\raisebox{-.35cm}{\epsfig{file=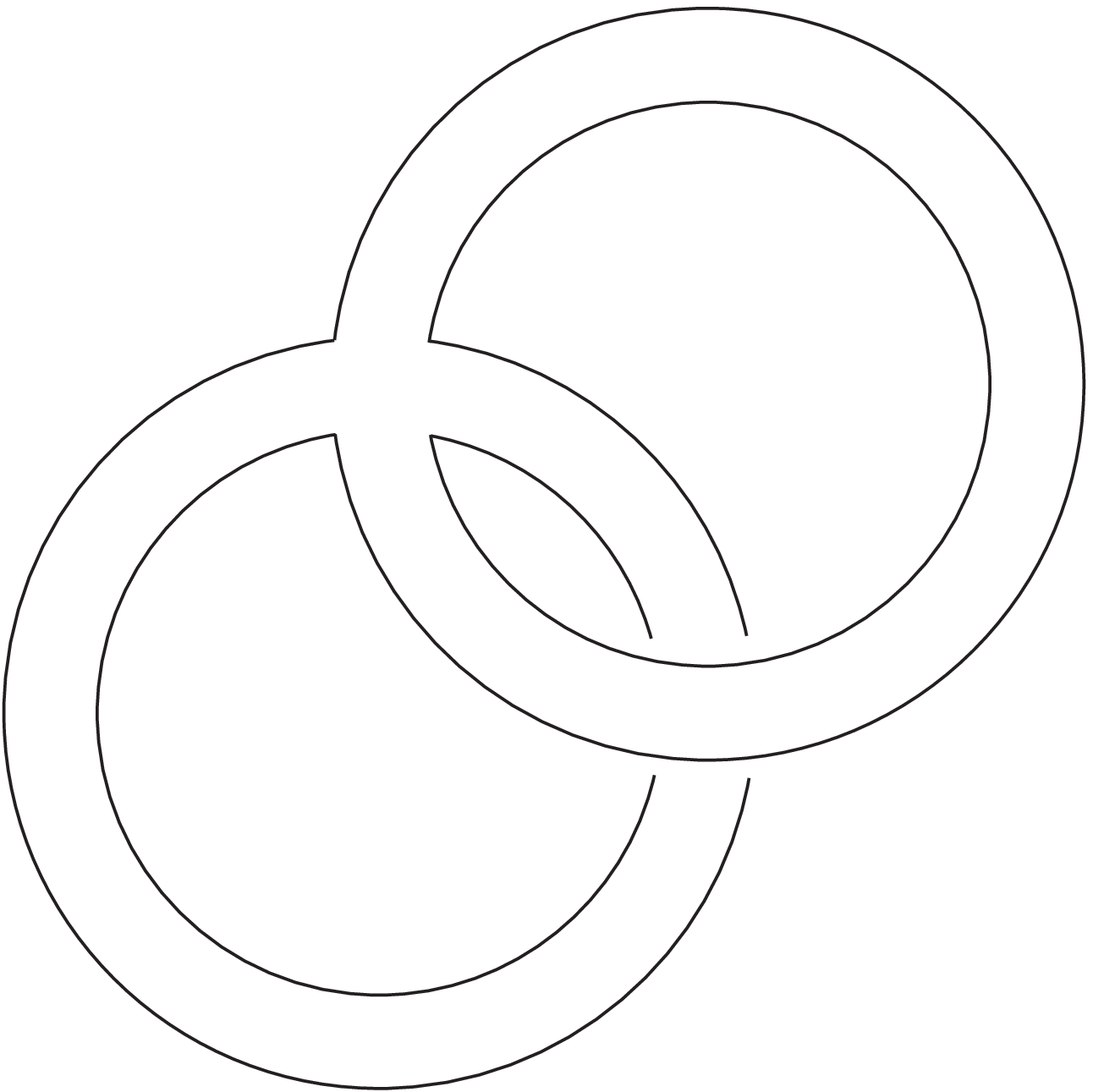, width=.46in}}
\;\right\}\;+\;\mathcal{O}(z^2)
\\
&=\Big(-\frac{1}{12}N-\frac12N^2+\frac23N^3\Big)\ z
+{\mathcal O}(z^2)\, .
\end{split}
\end{equation*}
Needless to say agreement is perfect. In fact, as we shall show 
in Section~\ref{Penner_Model}, 
agreement to all orders amounts to known results for the orbifold
Euler characteristic of the moduli space of real algebraic curves.

\section{Duality for Matrix Integrals}

\label{Duality}

An additional change of variables $X\longrightarrow N^{1/2}X$
in~\eqn{nearly}, absorption of all but a single $N$ in the couplings $t$
as well as the substitution
\begin{equation}
\b=2\a\, ,
\end{equation}
yields
\begin{multline}
\log\left(
\frac{\textstyle\int [dX]_{_{\!\ss(2\a)}}\!\,\exp\Big(\!-\frac{N\a}{2}\ \tr X^2
+\sum_{j=1}^\infty \frac{N\a t_j}{j}\ \tr X^j \Big)}{
\textstyle\int [dX]_{_{\!\ss(2\a)}}\!\,\exp\Big(\!-\frac{N\a}{2}\ \tr 
X^2\Big)}\right)
\\
=
\sum_{\Gamma\in\mathfrak G}
2\ (\a^{1/2}N)^{\chi(S_\Gamma)}\;
\frac{\textstyle 
(3-\a^{-1}-\a)^{^{\ss 1 - \12\Sigma_\Gamma-
\12\chi(S_\Gamma)}} 
(\a^{-1/2}-\a^{1/2})^{^{\ss \Sigma_\Gamma}}
}{|{\rm Aut}(\Gamma)|} \prod_{j}
t_j^{v^{(j)}_\Gamma}\! .
\label{eq: invariant formula}
\end{multline}
This formula is invariant under
\begin{equation}
\a\longrightarrow\a^{-1}\;\mbox{ and }\;N\longrightarrow -\a N\, .
\label{duality}
\end{equation}

\noindent
{\it Remark.}
\begin{enumerate}
\item The duality holds graph by graph\footnote{Physicists
would call this a $T$-duality --  valid order by order
in perturbation theory.}.
\item The $\a=1$ GUE model is self-dual since $\chi(S_\Gamma)$ is
even for orientable graphs.
\item The graphical expansion of the $\a=2$, $N\times N$ 
GSE model is identical to that of the $\a=1/2$ GOE model
if the size of the matrices are doubled and the contribution of every  
M\"obius graph embedded in a non-orientable surface of 
 odd Euler characteristic is multiplied by $-1$. 
\end{enumerate}

One might wonder whether matrix integrals exist whose graphical 
expansion coincides exactly 
with the image of~\eqn{eq: invariant formula} under the
duality~\eqn{duality}. Although we have no definite answer to 
this question at this point, in the next Section we show that the
combination of Poincar\'e duality and the one discovered here underly
the dualities for correlations of characteristic
polynomials~\cite{baker,bh,bh1,mehta}.
	
\section{Characteristic Polynomial Duality}

\label{BZ}

The average of products of characteristic polynomials 
obey dualities between GOE and GSE models~\cite{bh1}: 
\begin{multline}
\frac{\textstyle\int [dS]_{_{\!\ss(1)}}^{N\times N}\,
\exp\Big(\!-\frac{N}{2}\ \tr S^2\Big)\,
\prod_{\ell=1}^k\det_{N\times N}(\lambda_\ell-S)}{
\textstyle\int [dS]_{_{\!\ss(1)}}^{N\times N}\,\exp\Big(\!-\frac{N}{2}\ \tr 
S^2\Big)}
\\
=
\;
\frac{\textstyle\int [dX]_{_{\!\ss(4)}}^{k\times k}\,\exp(-N\, 
 \tr X^2)\,
\mathbb{H}\!\det_{k\times k} ^N\!(\Lambda-\sqrt{-1}\ X)}{
\textstyle\int [dX]_{_{\!\ss(4)}}^{k\times k}\,\exp(-N
\ \tr 
X^2)}\, ,
\label{BHQ}
\end{multline}
as well as a 
self duality for the GUE case\footnote{One might question the reality of
the integrals on the right hand sides of~\eqn{BHC} and ~\eqn{BHQ}
since an explicit $\sqrt{-1}$ 
appears in the determinants. It is clear, however,
from both the graphical expansions below and the original derivation
in~\cite{bh,bh1} that both integrals are real.
}~\cite{bh}:
\begin{multline}
\frac{\textstyle\int [dX]_{_{\!\ss(2)}}^{N\times N}\,
\exp\Big(\!-\frac{N}{2}\ \tr X^2\Big)\,
\prod_{\ell=1}^k\det_{N\times N}(\lambda_\ell-X)}{
\textstyle\int [dX]_{_{\!\ss(2)}}^{N\times N}\,\exp\Big(\!-\frac{N}{2}\ \tr 
X^2\Big)}
\\
=
\;
\frac{\textstyle\int [dY]_{_{\!\ss(2)}}^{k\times k}\,\exp\Big(\!-\frac{N}{2}\ \tr Y^2\Big)\,
\det_{k\times k} ^N(\Lambda-\sqrt{-1}\ Y)}{
\textstyle\int [dY]_{_{\!\ss(2)}}^{k\times k}\,\exp\Big(\!-\frac{N}{2}\ \tr 
Y^2\Big)}\, .
\label{BHC}
\end{multline}
The duality relates expectations of $k$-fold products of distinct
characteristic polynomials of $N\times N$ matrices
to averages over the $N$th power of determinants of certain 
$k\times k$ matrices.
Here, $\Lambda={\rm diag}(\lambda_1,\ldots,\lambda_k)$
is a diagonal $k\times k$ matrix of real entries.
The quaternionic determinant $\mathbb{H}\det$ in~\eqn{BHQ} is defined
by 
\begin{equation}
\mathbb{H}{\det} _{k\times k} M=
{\det} _{2k\times 2k} ^{1/2}\  C(M)\, ,
\end{equation}
where the $2k\times2k$ matrix $C(M)$ is obtained from the $k\times k$
quaternion valued matrix $M$ by replacing the quaternionic units
by their representation in terms of Pauli matrices
$1\rightarrow I_{2\times 2}$, $e_i\rightarrow i\sigma_i$
($i=1,2,3$).

To begin with, we demonstrate that the $N$-$k$ duality for GUE models
follows from the usual ribbon graph expansion along with 
Poincar\'e duality of graphs on a compact oriented surface.
The first step is to represent the determinants as vertices of the
graphical expansion. Let us assume that 
the parameter $\lambda_\ell$ satisfies $\lambda_\ell > \lambda>0$
for every $\ell$ and some positive $\lambda$, 
and let $\Omega_\lambda$ 
denote the
set of all $N\times N$ hermitian matrices whose eigenvalues 
are contained in the bounded interval $[-\lambda, \lambda]$. Then 
for every $X\in \Omega_\lambda$, we have a convergent 
power series expansion in $\lambda_\ell ^{-1}$:
$$
\det\left(I-\frac{X}{\lambda_\ell}\right)
= \exp\left(\tr \log \left(I-\frac{X}{\lambda_\ell}\right)\right)
=\exp\left(-\sum_{j=1} ^\infty \,\frac{1}{j}\, \lambda_\ell ^{-j}
\tr X^j\right).
$$
Therefore, 
\begin{multline}
\label{eq: det}
\int [dX]_{_{\!\ss(2)}}^{N\times N} e^{-\frac{N}{2}\tr X^2}\prod_{\ell=1}^k
\det{}_{\!N\times N}
\left(I-\frac{X}{\lambda_\ell}\right)
\\
= \int_{\Omega_\lambda} [dX]_{_{\!\ss(2)}}^{N\times N}\! 
e^{-\frac{N}{2}\tr X^2}
\exp\left(-\sum_{j=1}^{\infty}\;\frac{1}{j}\;\tr_{\!\!k\times
k}\Lambda^{-j}
\tr_{\!\!N\times N} X^j\right) \\
+ \int_{\Omega_\lambda ^c} 
[dX]_{_{\!\ss(2)}}^{N\times N}\! e^{-\frac{N}{2}\tr X^2}\prod_{\ell=1}^k
\det{}_{\!N\times N}
\left(I-\frac{X}{\lambda_\ell}\right)\;,
\end{multline}
where $\Omega_\lambda ^c$ is the complement of $\Omega_\lambda$
in the space of all $N\times N$ hermitian matrices.
Since $\Omega_\lambda$ is a compact space, the first integral
on the right hand side of~\eqn{eq: det} 
is a convergent power series in $\lambda_1 ^{-1}$, $\lambda_2 ^{-1}$,
$\ldots$, $\lambda_k ^{-1}$. Set $t_j = -\tr \Lambda^{-j}$. Then
${\rm Re}(t_j)<0$, and as $\lambda \rightarrow +\infty$, $t_j\rightarrow 0$.
Thus the ribbon graph  expansion  provides
each coefficient of the
 power series expansion of this integral in $t_j$
as $\lambda \rightarrow +\infty$.  The second integral
on the right hand side of~\eqn{eq: det}
is a polynomial in $\lambda_\ell ^{-1}$
whose coefficients converge to 0 as $\lambda$ goes to infinity
since $\Omega_\lambda ^c$ approaches the empty set and
the integrand is bounded.
 Therefore, we obtain an asymptotic expansion formula
\begin{multline}
\label{eq: NbyN}
\log\left(
\frac{\textstyle\int [dX]_{_{\!\ss(2)}}^{N\times N}\,
\exp\Big(\!-\frac{N}{2}\ \tr X^2\Big)\,
\prod_{\ell=1}^k\det_{N\times N}
\left(I-\frac{X}{\lambda_\ell}\right)}{
\textstyle\int [dX]_{_{\!\ss(2)}}^{N\times N}\,\exp\Big(\!-\frac{N}{2}\ \tr 
X^2\Big)}\right)\\
=
\sum_{\Gamma\in \mathfrak{R}}
\frac{1}{|\Aut_\mathfrak{R}(\Gamma)|}
(-1)^{v_\Gamma} N^{f_\Gamma - e_\Gamma} 
\prod_j (\tr \Lambda^{-j})^{v_\Gamma ^{(j)}}\\
\in (\mathbb{Q}[N])[[\lambda_1 ^{-1}, \ldots,\lambda_k ^{-1}]]\, .
\end{multline}

The computation of
\begin{equation}
\label{eq: kbyk}
\log
\left(
\frac{\textstyle\int [dY]_{_{\!\ss(2)}}^{k\times k}\,\exp\Big(\!-\frac{N}{2}\ \tr Y^2\Big)\,
\det^N\!(I-\sqrt{-1}\ Y\Lambda^{-1})}{
\textstyle\int [dY]_{_{\!\ss(2)}}^{k\times k}\,\exp\Big(\!-\frac{N}{2}\ \tr 
Y^2\Big)}
\right)
\end{equation}
can be performed similarly: First we decompose the space of
all $k\times k$ hermitian matrices into two pieces, one 
consisting of matrices with
eigenvalues in $[-\lambda,\lambda]$, and the other  its complement.
If $\lambda_\ell > \lambda$ for every $\ell$, then 
$\det^N\!(I-\sqrt{-1}\ Y\Lambda^{-1})$ can be expanded
as before. Asymptotically as an element of
$(\mathbb{Q}[N])[[\lambda_1 ^{-1}, \ldots,\lambda_k ^{-1}]]$, 
we have 
\begin{multline}
\label{eq: kbyk2}
\int [dY]_{_{\!\ss(2)}}^{k\times k}\,\exp\Big(\!-\frac{N}{2}\ \tr Y^2\Big)\,
\det (I-\sqrt{-1}\ Y\Lambda^{-1})^N\\
=
\int [dY]_{_{\!\ss(2)}}^{k\times k}\,\exp\Big(\!-\frac{N}{2}\ \tr Y^2\Big)\,
\exp\left(
-N\sum_{j}\frac{(\sqrt{-1})^j}{j}\tr(Y\Lambda^{-1})^j\right).
\end{multline}
The appearance of the term $\tr (Y\Lambda^{-1})^m$ instead of $\tr Y^m$ 
occurring in 
\eqn{eq: kbyk2} replaces 
products of traces over identity matrices 
\begin{equation}
N^{f_\Gamma}=\prod_j(\tr I^{j})^{f_\Gamma^{(j)}}
\end{equation}
incurred in~\eqn{eq: main formula} as one travels around each face of 
the graph $\Gamma$,
by 
\begin{equation}
\prod_j(\tr \Lambda^{-j})^{f_\Gamma^{(j)}}\, .
\end{equation} 
(Recall that $f_\Gamma ^{(j)}$ denotes the number of
$j$-gons in the cell-decomposition of $S_\Gamma$ defined by
the graph $\Gamma$.)
 Therefore, 
\begin{equation}
\label{eq: kbyk final}
\begin{split}
&\log
\left(
\frac{\textstyle\int [dY]_{_{\!\ss(2)}}^{k\times k}\,\exp\Big(\!-\frac{N}{2}\ \tr Y^2\Big)\,
\det^N\!(I-\sqrt{-1}\ Y\Lambda^{-1})}{
\textstyle\int [dY]_{_{\!\ss(2)}}^{k\times k}\,\exp\Big(\!-\frac{N}{2}\ \tr 
Y^2\Big)}
\right)\\
&=
\log\left(
\frac{
\int [dY]_{_{\!\ss(2)}}^{k\times k}\,\exp\Big(\!-\frac{N}{2}\ \tr Y^2\Big)\,
\exp\left(
-N\sum_{j}\frac{(\sqrt{-1})^j}{j}\tr(Y\Lambda^{-1})^j\right)}
{\int [dY]_{_{\!\ss(2)}}^{k\times k}\,\exp\Big(\!-\frac{N}{2}\ \tr Y^2\Big)}\right)\\
&=
\sum_{\Gamma\in \mathfrak{R}}
\frac{1}{|\Aut_\mathfrak{R}(\Gamma)|}
(-1)^{v_\Gamma} N^{v_\Gamma - e_\Gamma} 
(\sqrt{-1})^{2e_\Gamma}
\prod_j (\tr \Lambda^{-j})^{f_\Gamma ^{(j)}}\\
&=
\sum_{\Gamma\in \mathfrak{R}}
\frac{1}{|\Aut_\mathfrak{R}(\Gamma)|}
(-1)^{f_\Gamma} N^{v_\Gamma - e_\Gamma} 
\prod_j (\tr \Lambda^{-j})^{f_\Gamma ^{(j)}},
\end{split}
\end{equation}
where we used
$$
(-1)^{v_\Gamma - e_\Gamma} = (-1)^{\chi(S_\Gamma) - f_\Gamma}
= (-1)^{f_\Gamma}.
$$
Let us denote by $\Gamma^*$ the dual graph 
of a ribbon graph $\Gamma$ drawn on a compact oriented surface
$S_\Gamma$. 
We note that $\Aut_\mathfrak{R} (\Gamma)\cong 
\Aut_\mathfrak{R} (\Gamma^*)$ and 
\begin{equation}
\begin{cases}
v_\Gamma ^{(j)} = f_{\Gamma^*} ^{(j)}\, ,\\
\ e_\Gamma \; =\;  e_{\Gamma^*}\, ,\\
f_\Gamma ^{(j)} = v_{\Gamma^*} ^{(j)}\,  .
\end{cases}
\end{equation}
Since one and two valent vertices are included in the set of ribbon
graphs $\mathfrak{R}$, the map
$$
*: \mathfrak{R}\longrightarrow \mathfrak{R}
$$
is a bijection. [Contrast this situation to the Penner model in
Section~\ref{examples}, where the couplings $t_1=t_2=0$ and Poincar\'e
duality does not apply.] Therefore, 
\begin{equation}
\label{eq: dual graph}
\begin{split}
&\sum_{\Gamma\in \mathfrak{R}}
\frac{1}{|\Aut_\mathfrak{R}(\Gamma)|}
(-1)^{v_\Gamma} N^{f_\Gamma - e_\Gamma} 
\prod_j (\tr \Lambda^{-j})^{v_\Gamma ^{(j)}}\\
=
&\sum_{\Gamma\in \mathfrak{R}}
\frac{1}{|\Aut_\mathfrak{R}(\Gamma^*)|}
(-1)^{v_{\Gamma^*}} N^{f_{\Gamma^*} - e_{\Gamma^*}} 
\prod_j (\tr \Lambda^{-j})^{v_{\Gamma^*} ^{(j)}}\\
=
&\sum_{\Gamma\in \mathfrak{R}}
\frac{1}{|\Aut_\mathfrak{R}(\Gamma)|}
(-1)^{f_\Gamma} N^{v_\Gamma - e_\Gamma} 
\prod_j (\tr \Lambda^{-j})^{f_\Gamma ^{(j)}}. 
\end{split}
\end{equation}
This implies that the matrix integrals~\eqn{eq: NbyN} and
\eqn{eq: kbyk} have the same asymptotic expansion. 

The $N$-$k$ duality in equation~\eqn{BHC}
is a polynomial identity of
degree $Nk$ in
$(\mathbb{Q}[N])[\lambda_1,\lambda_2,\ldots,\lambda_k]$,
where we define $\deg(\lambda_\ell) = 1$. We must now consider also
disconnected graphs, since there is no logarithm.
The coefficient
of the degree $Nk-d$ term of~\eqn{BHC} is therefore determined by
a partition $d=2(e_1+e_2+\cdots+e_m)$ corresponding to the product of
$m$ connected graphs consisting of
$e_i$ edges. The contributions of connected graphs are
computed in~\eqn{eq: NbyN} and
\eqn{eq: kbyk}. We note that the duality~\eqn{eq: dual graph}
holds for every surface even when the number of edges is fixed. Therefore,
the asymptotic equality we have derived implies the polynomial
identity~\eqn{BHC}. In other words, the $N$-$k$ duality 
of~\cite{bh} is a simple consequence of 
the Poincar\'e duality of graphs on a compact oriented
surface.

Our derivation of the characteristic polynomial
duality between the GOE and GSE 
models goes quite similarly. 
Here again we see that the duality is a consequence of our
graphical expansion formula~\eqn{eq: main formula} and
Poincar\'e duality:

Using the same trick for characteristic polynomials
as in the GUE case, from our
expansion formula~\eqn{eq: main formula} we obtain an asymptotic
expansion formula for the GOE side of the duality
\begin{equation}
\label{eq: bh-goe}
\begin{split}
&\log\left(
\frac{\textstyle\int [dS]_{_{\!\ss(1)}}^{N\times N}\,\exp
\Big(\!-\frac{2N}{4}\ \tr S^2\Big)\,
\prod_{\ell=1}^k\det\left(I-\frac{S}{\lambda_\ell}\right)}{
\textstyle\int [dS]_{_{\!\ss(1)}}^{N\times N}\,\exp\Big(\!-\frac{N}{4}\ \tr 
S^2\Big)}
\right)\\
&=
\log\left(
\frac{\int [dS]_{_{\!\ss(1)}}^{N\times N}\,\exp
\Big(\!-\frac{2N}{4}\ \tr S^2\Big)\,
\exp\left(
- \sum_{j=1} ^\infty
\frac{2\, \tr \Lambda^{-j}}{2j}\tr S^j\right)}{
\int [dS]_{_{\!\ss(1)}}^{N\times N} \,\exp\Big(\!-\frac{2N}{4}\ \tr 
S^2\Big)}
\right)\\
&=
\sum_{\Gamma\in\mathfrak{G}}
\frac{1}{|\Aut(\Gamma)|}(-1)^{v_\Gamma} 2^{v_\Gamma - e_\Gamma}
N^{f_\Gamma - e_\Gamma}\prod_j
 (\tr \Lambda^{-j})^{v_\Gamma ^{(j)}}.
\end{split}
\end{equation}
(Note the non-standard normalization of the Gaussian exponent
yields the factor $2^{-e_\Gamma}$.)
Its GSE counterpart requires some care: 
The characteristic polynomial
of a $k\times k$ quaternionic matrix $X$ is defined by
$$
\mathbb{H}\det (\Lambda-X) = 
{\det} ^{1/2}(\Lambda I_{2k\times 2k}-C(X))
$$
and
$$
\tr X^j = \frac{1}{2}\tr C(X)^j.
$$
Thus if all eigenvalues of $X$ are in $[-\lambda, \lambda]$
and  $\lambda_\ell>\lambda>0$, then
$$
\mathbb{H}{\det} (I-X\Lambda^{-1})
= \exp\left(-\frac{1}{2}\sum_{j=1} ^\infty \frac{2}{j}\, 
 \tr (X\Lambda^{-1})^j
\right).
$$
Therefore, we have an asymptotic expansion 
\begin{equation}
\label{eq: bh-gse}
\begin{split}
&\log\left(
\frac{\int [dX]_{_{\!\ss(4)}}^{k\times k}\,\exp(\!-N
\ \tr X^2)\,
\mathbb{H}{\det} _{k\times k} ^N\!(I-\sqrt{-1}\ X\Lambda^{-1})}{
\int [dX]_{_{\!\ss(4)}}^{k\times k}\,\exp(\!-N
\ \tr 
X^2)}\!\right)\\
&=
\log\!\left(\!
\frac{\int [dX]_{_{\!\ss(4)}}^{k\times k}\! \exp\left(-\frac{Nk}{k}
\ \tr X^2\!\right)
\exp\left(
-N\sum_{j=1} ^\infty \frac{2 (\sqrt{-1})^j }{2j}
\tr (X\Lambda^{-1})^{-j}\right)
}{
\int [dX]_{_{\!\ss(4)}}^{k\times k}\! \,\exp\left(-\frac{Nk}{k}
\ \tr X^2\right)}\!\right)\\
&=
\sum_{\Gamma\in \mathfrak{G}}
\frac{1}{|\Aut(\Gamma)|} 
(-1)^{\Sigma_\Gamma + e_\Gamma+v_\Gamma} 
2^{f_\Gamma-e_\Gamma}N^{v_\Gamma-e_\Gamma}
\prod_j (\tr \Lambda^{-j})^{f_\Gamma ^{(j)}}\\
&=
\sum_{\Gamma\in \mathfrak{G}}
\frac{1}{|\Aut(\Gamma)|} 
(-1)^{f_\Gamma} 
2^{f_\Gamma-e_\Gamma}N^{v_\Gamma-e_\Gamma}
\prod_j (\tr \Lambda^{-j})^{f_\Gamma ^{(j)}},
\end{split}
\end{equation}
where we have used the fact that 
$$
(-1)^{\Sigma_\Gamma + e_\Gamma+v_\Gamma} =
(-1)^{f_\Gamma}
$$
that follows from~\eqn{eq: useful}. (In addition the non-standard
Gaussian exponent normalization now accounts for the absence of
explicit factors $k$ in the graphical expansion.)
We now see that~\eqn{eq: bh-goe} and~\eqn{eq: bh-gse} are
equal 
again through the dual construction of a M\"obius graph. 

The polynomial identity~\eqn{BHQ} follows from the equality
of the asymptotic expansions. This time each term of~\eqn{BHQ}
may have contributions from both orientable and non-orientable
graphs, but since the dual graph construction works for each
surface, the equality holds.

\section{The Penner Model}

\label{Penner_Model}

The Penner model for the hermitian matrix integral provides
an effective tool to compute the orbifold Euler characteristic
of the moduli space of smooth algebraic curves defined
over $\mathbb{C}$ with an arbitrary number of marked 
points~\cite{Penner},~\cite{Harer-Zagier}. It was discovered 
in~\cite{Goulden-Harer-Jackson} that the Penner model of
the real symmetric (or GOE) matrix integral yields the orbifold Euler
characteristic of the moduli spaces of
real algebraic curves. Therefore, it is  
natural to ask what topological information
is contained in the corresponding Penner type model
for the quaternionic self-adjoint matrix integral?

In this Section, we show via our duality~\eqn{duality}, 
that the symplectic Penner model is identical to
the GOE case, except for the matrix size and an overall
sign for contributions of non-orientable surfaces. The answer
to the above question is in the negative, the symplectic Penner
model offers no new topological insight. As we shall see, 
of the three main classes of M\"obius graphs --oriented,
non-orientable 
odd $\chi(S_\Gamma)$ and non-orientable even $\chi(S_\Gamma)$ --
only the first two survive the Penner substitution for the couplings,
or in other words, the orbifold Euler characteristic vanishes when
$\chi(S_\Gamma)$ is even. Therefore the third symplectic Penner type
model is not an independent topological quantity.

We also show that the generalized Penner model expressed in terms of
Vandermonde determinants to powers in $2(\mathbb{N}\cup1/\mathbb{N})$
exhibits an extended duality
that agrees with~\eqn{duality} when the power of the Vandermonde
is restricted to $1$, $2$ or $4$. 
Many explicit formul\ae~ and
derivations are reserved for Appendix~\ref{don't-think--type}.

The symplectic Penner model introduced in Section~\ref{examples} reads
\begin{multline}
\lim_{m\rightarrow\infty}
\log\left(
\frac{\textstyle\int [dX]_{_{\!\ss(4)}}\!\,\exp\Big(\!-\sum_{j=2}^{2m}
\frac{z^{j/2-1}}{j}\ \tr X^j \Big)}{
\textstyle\int [dX]_{_{\!\ss(4)}}\!\,\exp\Big(\!-\frac{1}{2}\ \tr
X^2\Big)}\right)
\\
=
\sum_{\Gamma\in\mathfrak G}
\frac{
(-1)^{e_\Gamma}
}{|{\rm Aut}(\Gamma)|}(2N)^{^{\ss f_\Gamma}}
(-1)^{\chi(S_\Gamma)}(-z)^{e_\Gamma-v_\Gamma}\, ,
\label{GSE_penner}
\end{multline}
to be viewed as an element of the formal
power series ring
$(\mathbb{Q}[N])[[z]]$.
This integral is indeed explicitly computable.
Symplectic invariance of the measure and integrand
allows us to diagonalize the matrix variable 
$X\longrightarrow\text{diag}(k_1,k_2,\ldots,k_N)$ so that:
\begin{equation}
\begin{split}
\lim_{m\rightarrow\infty}&\log\left(
\frac{\textstyle\int [dX]_{_{\!\ss(4)}}\!\,\exp\Big(\!
-\sum_{j=2}^{2m} \frac{z^{j/2-1}}{j}\ \tr X^j \Big)}{
\textstyle\int [dX]_{_{\!\ss(4)}}\!\,\exp\Big(\!-\frac{1}{2}\ \tr
X^2\Big)}\right)
\\
&
=\lim_{m\rightarrow\infty}
\log\left(\frac{
\int_{{\mathbb R}^N}\Delta^4(k)\prod_{i=1}^N
\exp\Big(-\sum_{j=2}^{2m} \frac{z^{j/2-1}}{j}\ k_i^j\Big) dk_i}
{\int_{{\mathbb R}^N}\Delta^4(k)\prod_{i=1}^N
\exp\Big(- \frac{k_i ^2}{2}\Big) dk_i}
\right)\, ,
\label{ernie}
\end{split}
\end{equation}
where 
$$
\Delta(k) = \prod_{i<j} (k_i-k_j)
$$
is the Vandermonde determinant.
Using the asymptotic expansion technique 
established in~\cite{Mulase95}, the Selberg integral formula
and the Stirling formula
an explicit asymptotic expansion, even valid for every
$\alpha\in {\mathbb N}$, can be computed for the integral
\begin{equation}
\begin{split}
K(z,N,\alpha)
=\lim_{m\rightarrow\infty}
\log\left(\frac{
\int_{{\mathbb R}^N}\Delta^{2\alpha}(k)\prod_{i=1}^N
\exp\Big(-\sum_{j=2}^{2m} \frac{z^{j/2-1}}{j}\ k_i^j\Big) dk_i}
{\int_{{\mathbb R}^N}\Delta^{2\alpha}(k)\prod_{i=1}^N
\exp\Big(- \frac{k_i ^2}{2}\Big) dk_i}
\right)\, .
\label{eq:K}
\end{split}
\end{equation}
(In what follows $K(z,N,\a)$ should be regarded as 
as the asymptotic expansion, not the underlying integral; an explicit
expression is given in Appendix~\ref{don't-think--type}.)
Specializing to the $\alpha=1$ hermitian case yields a very compact 
result corresponding to
the original formula of Penner~\cite{Penner}:
\begin{equation}
\label{eq:K1}
\begin{split}
K(z,N,1)
&=\sum_{\substack{g\ge 0, n>0\\2-2g-n<0}}
\frac{(2g+n-3)!(2g-1)}{(2g)!n!}b_{2g} N^n (-z)^{2g+n-2}\\
&=\sum_{\Gamma\in\mathfrak{R}}
\frac{(-1)^{e_\Gamma}}{|\Aut_\mathfrak{R}(\Gamma)|}
N^{f_\Gamma}(-z)^{e_\Gamma-v_\Gamma}\, .
\end{split}
\end{equation}
Identifying $n$ with $f_\Gamma$ and $g$ as the genus,
yields the well known
generating function of the Euler characteristic $\chi({\mathfrak
M}_{g,n})$ of the moduli of complex algebraic curves of genus $g$ and $n$ 
marked points.

For the $\alpha=2$ symplectic case a similar simplification occurs
and~\eqn{eq:K} can be written in terms of the $\a=1$ hermitian result
plus additional terms corresponding to non-orientable surfaces
of even genus $g=2q$ with $m+1-2q$ marked points
\begin{equation}
\label{eq:K2genus}
\begin{split}
K(z,N,2)
&=\;\frac{1}{2}K(z,2N,1)\\
&-\;\frac{1}{2}\;\sum_{\substack{q\ge 0,n>0\\ 1-2q-n<0}}
 \frac{(2q+n-2)!(2^{2q-1}-1)}{(2q)!\;n!}b_{2q}
(2N)^{n} (-z)^{2q+n-1} .
\end{split}
\end{equation}
Comparing~\eqn{eq:K2genus} with~\eqn{GSE_penner}, we obtain
\begin{equation}
\label{eq:real}
\sum_{\Gamma:\; f_\Gamma = n,\; g(S_\Gamma)=2q}
\frac{(-1)^{e_\Gamma}}{|\Aut(\Gamma)|}
= \frac{1}{2}\;\frac{(2q+n-2)!(2^{2q-1}-1)}{(2q)!\;n!}\;b_{2q},
\end{equation}
where the summation is over all connected non-orientable
M\"obius graphs with $n>0$ faces that are drawn on a non-orientable
surface of genus $2q$ satisfying a hyperbolicity condition
$1-2q-n<0$. The formula~\eqn{eq:real} is in exact agreement
with the formula for the orbifold Euler characteristic of the
moduli space of smooth {\it real} algebraic curves of genus $2q$
with $n$ marked points that 
can be found in~\cite{Goulden-Harer-Jackson}
and~\cite{Ooguri-Vafa}.

To study the GOE-GSE duality for the Penner model, 
we need the expansion of the 
analog of formula~\eqn{eq:K} valid for the single power of 
the Vandermonde determinant relevant to the GOE model. 
Indeed an
integral formula for
\begin{equation}
\label{eq:J}
\begin{split}
&J(z,N,\gamma)
=\!\lim_{m\rightarrow\infty}\!
\log\!\left(\frac{
\int_{{\mathbb R}^N}|\Delta(k)|^{2/\gamma}\prod_{i=1}^N
\exp\Big(-\sum_{j=2}^{2m} \frac{z^{j/2-1}}{j}\ k_i^j\Big) dk_i}
{\int_{{\mathbb R}^N}|\Delta(k)|^{2/\gamma}\prod_{i=1}^N
\exp\Big(- \frac{k_i ^2}{2}\Big) dk_i}
\right)\, ,
\end{split}
\end{equation}
valid for every 
positive integer $\gamma\in{\mathbb N}$
({\it i.e.} for all powers $2/\mathbb{N}$ of the Vandermonde) 
was derived in~\cite{Goulden-Harer-Jackson} in order to
compute~\eqn{eq:real}.
(See Appendix~\ref{don't-think--type}.) 

The matrix integrals $K(z,N,\a)$ and $J(z,N,\gamma)$ are
closely related: Obviously
\begin{equation}
J(z,N,1)=K(z,N,1)\, .
\end{equation}
The identity
\begin{equation}
\begin{split}
J(2z,2N,2)&=\frac{1}{2}J(z,2N,1)-\Big(K(z,N,2)-\frac{1}{2}K(z,2N,1)\Big)
\\
&=\;\frac{1}{2}J(z,2N,1)\\
&+\;\frac{1}{2}\!\sum_{\substack{q\ge 0,n>0\\ 1-2q-n<0}}
 \frac{(2q+n-2)!(2^{2q-1}-1)}{(2q)!\;n!}b_{2q}
(2N)^{n} (-z)^{2q+n-1} \, ,
\label{eq:J2genus}
\end{split}
\end{equation}
expresses the $\gamma=2$ GOE case in terms of the $\gamma=1$ oriented hermitian
result and a sum over non-orientable contributions with odd Euler
characteristic. 
Notice, as claimed above, in the graphical expansions of 
the orthogonal Penner model $J(2z,2N,2)$
and  the symplectic Penner model $K(z,N,2)$, non-orientable
surfaces of odd genera do not contribute. 
This corresponds to the fact that the orbifold Euler characteristic
of the moduli space of smooth real algebraic curves of
odd genus $2q+1$ with $n$ marked points is 0 for any
value of $q\ge 0$ and $n>0$. 

More importantly, observe that the
GSE and GOE formul\ae~\eqn{eq:K2genus} and~\eqn{eq:J2genus}
almost coincide except that the GOE formula is for matrix size $2N$ 
and the non-orientable odd Euler characteristic terms differ by
an overall sign. (The appearance of $2z$ rather than $z$ will be cured
by the appropriate normalization given below and in the master
formula~\eqn{eq: invariant formula}.)  This is precisely our
duality~\eqn{duality}.

Finally, we show that the duality between 
GOE and GSE extends to
arbitrary positive integers for the two types of 
Penner integrals introduced in this Section. 
Let $r\in\mathbb{N}\cup1/\mathbb{N}$ 
and set
\begin{multline}
\label{eq:general Penner}
I(z,N,r)\\
=\lim_{m\rightarrow\infty}
\log\left(\frac{
\int_{{\mathbb R}^N}|\Delta(k)|^{2r}\prod_{i=1}^N
\exp\Big(-\sum_{j=2}^{2m} 
\frac{k_i^j}{j}\left(\frac{z}{rN}\right)^{j/2-1}\Big) dk_i}
{\int_{{\mathbb R}^N}|\Delta(k)|^{2r}\prod_{i=1}^N
\exp\Big(- \frac{k_i ^2}{2}\Big) dk_i}
\right)\\
\in (\mathbb{Q}[N,N^{-1},r,r^{-1}])[[z]].
\end{multline}
Then we have
\begin{equation}
I(z,N,r)=\left\{
\begin{array}{cll}
K\left(\frac{z}{\alpha N},N,\alpha\right)&,&r=\alpha\in \mathbb{N}\, ,\\
\\
J\left(\frac{\gamma
z}{N},N,\gamma\right)&,&r=\gamma^{-1}\in1/\mathbb{N}\, .
\end{array}
\right.
\end{equation}
{F}rom inspection of the explicit
asymptotic expansion formul\ae~of~\eqn{eq:K_app} and~\eqn{eq:J_app}
presented in Appendix~\ref{don't-think--type},
we obtain an extended duality
\begin{equation}
\label{eq:Penner dual}
I(z,N,r)=
I(z,-r N,r^{-1})
\end{equation}
for an arbitrary positive integer $r$. This is 
in agreement with our duality~\eqn{duality} for $r=1,2$.

\section{The Central Limit Theorem}

\label{central}

To prove a central limit theorem for large matrix size $N$,
we need to show that the leading dependence is Gaussian in the
coupling constants 
$t_j$. More precisely, define the
Gaussian expectation value of $f(X)$ for GOE ($\alpha = 1/2$),
GUE ($\alpha=1$), and GSE ($\alpha=2$) as
\begin{equation}
\langle f(X) \rangle_{(N,\alpha)}=
\frac{\textstyle\int [dX]_{_{\!\ss(2\a)}}\!\,\exp\Big(\!-\frac{N\a}{2}\ \tr
X^2\Big)\; f(X)}
{\textstyle\int [dX]_{_{\!\ss(2\a)}}\!\,\exp\Big(\!-\frac{N\a}{2}\ \tr 
X^2\Big)}\, ,
\end{equation}
and consider
\begin{equation}
V(t, N,\alpha)=
\frac{
\left\langle\exp\Big(\sum_{j} \frac{\a t_j}{j}\ \tr X^j \Big)
\right\rangle_{(N,\alpha)}}
{\exp\left(\sum_{j} \frac{\a t_j}{j}\ \Big\langle\tr X^j\Big
\rangle_{(N,\alpha)} \right)}\, .
\end{equation}
The expansion formula~\eqn{eq: invariant formula} shows that the
contribution of a connected M\"obius graph $\Gamma\in\mathfrak{G}$
to $
\log \left\langle\exp\Big(\sum_{j} \frac{\a t_j}{j}\ \tr X^j \Big)
\right\rangle_{(N,\alpha)}$
is
\begin{equation}
\label{eq: gamma contribution}
2
\a^{\frac{1}{2}\chi(S_\Gamma)}N^{\chi(S_\Gamma)-v_\Gamma}
\frac{\textstyle 
(3-\a^{-1}-\a)^{^{\ss 1 - \12\Sigma_\Gamma-
\12\chi(S_\Gamma)}} 
(\a^{-1/2}-\a^{1/2})^{^{\ss \Sigma_\Gamma}}
}{|{\rm Aut}(\Gamma)|} \prod_{j}
t_j^{v^{(j)}_\Gamma}.
\end{equation}
Since
$$
\sum_{j} \frac{\a t_j}{j}\ \Big\langle\tr X^j\Big
\rangle_{(N,\alpha)}
$$
is the sum of all contributions from $1$-vertex M\"obius graphs, we see that
$\log V(t,N,\alpha)$ has no terms coming from $1$-vertex graphs. In
particular, it has no terms with a positive power of $N$. Indeed,
the power of $N$ in~\eqn{eq: gamma contribution}
is strictly positive only when $\chi(S_\Gamma)= 2$ and
 $v_\Gamma = 1$, {\it i.e.} $\Gamma$ is an orientable planar graph with
one vertex. Therefore, $\lim_{N\rightarrow \infty} \log V(t,N,\alpha)$
consists of contributions from graphs that have two or more vertices and
$\chi(S_\Gamma)-v_\Gamma=0$.~But this is possible only when
$\chi(S_\Gamma) = v_\Gamma = 2$. In other words, $\Gamma$ is an orientable
planar graph with exactly two vertices
contributing (see~\eqn{eq: gamma contribution})
$$
\frac{2\alpha}{|\Aut(\Gamma)|}t_{j_1}t_{j_2},
$$
where $j_1$ and $j_2$ are  the valences of the two vertices of $\Gamma$.
Altogether, we have established the following

\begin{theorem}
\label{thm: CLT}
The Central Limit Theorem for GOE $(\alpha=1/2)$, GUE
$(\alpha=1)$ and GSE $(\alpha=2)$ ensembles is 
\begin{equation}
\begin{split}
\lim_{N\rightarrow \infty}
\log
&\left(
\frac{
\left\langle\exp\Big(\sum_{j} \frac{\a t_j}{j}\ \tr X^j \Big)
\right\rangle_{(N,\alpha)}}
{\exp\left(\sum_{j} \frac{\a t_j}{j}\ \Big\langle\tr X^j\Big
\rangle_{(N,\alpha)} \right)}
\right)\\
&= 
\sum_{\substack{\Gamma \text{ connected, oriented, planar}\\
2\text{-vertex ribbon graph}}}
\frac{\alpha}{|\Aut_\mathfrak{R}(\Gamma)|}t_{j_1}t_{j_2},
\end{split}
\end{equation}
where $j_1$ and $j_2$ are  the valences of the two vertices of 
the ribbon graph $\Gamma$.
\end{theorem}

We notice that the  formula is the same for all three
ensembles except for the overall factor of $\alpha$.
In particular, only oriented planar ribbon graphs 
contribute in the large $N$ limit. This mechanism
was observed long ago by 't Hooft in the hope that large $N$ quantum 
chromodynamics could be solved exactly~\cite{'tHooft:1974jz}.

\section{Conclusions}

\label{conclusions}

Let us tabulate the patina of results gathered here:
\begin{itemize}
\item
The asymptotic expansion of the three Gaussian random matrix ensembles
is expressed as a sum over M\"obius graphs.
\item
These expansions are related by a duality
\begin{equation}
\a\longrightarrow\a^{-1} \mbox{ and } N\longrightarrow-\a N
\label{dualitee}
\end{equation}  
The $\a=1$ GUE model is self-dual and sums over only ribbon graphs.
The duality between $\a=1/2$ GOE and $\a=2$ GSE models amounts to 
an equality of graphical expansions up to a factor
$$(-1)^{\chi(S_\Gamma)}\, $$
for any graph $\Gamma$. 
\item
When specialized to Penner model couplings, the Selberg integral
representation yields an asymptotic expansion for all $\a\in
\mathbb{N}\cup1/\mathbb{N}$ and the duality~\eqn{dualitee} holds
for this extended set of $\a$'s.
\end{itemize}

\vspace{.2cm}
\noindent
Therefore, the first and probably most interesting question one might pose, 
is whether our graphical expansion
formul\ae~can also generalized to the extended set of 
$\a\in\mathbb{N}\cup1/\mathbb{N}$. {\it I.e.};

\begin{picture}(300,200)(-20,0)
\thicklines
\put(150,150){\oval(110,50)}
\put(108,148){\begin{tabular}{c}Matrix Integrals\\$\a=1/2,1,2$\end{tabular}}
\put(50,50){\oval(110,50)}
\put(10,48){\begin{tabular}{c}Penner Model\\
$\a\in \mathbb{N}\cup1/\mathbb{N}$\end{tabular}}
\put(250,50){\oval(110,50)}
\put(195,48){\begin{tabular}{c}Graphical Expansion\\
$\a=1/2,1,2,??$
\end{tabular}}
\put(100,120){\vector(-1,-1){35}}
\put(200,120){\vector(1,-1){35}}
\end{picture}

\vspace{.2cm}
\noindent
Let us briefly postpone a discussion of this issue while enumerating
several other questions for which we have no immediate answers:
\begin{enumerate}
\item Do there exist matrix models where the duality holds
exactly, without a factor $(-1)^{\chi(S_\Gamma)}$?
\item What is the significance of the minus sign in the transformation 
$$N\rightarrow -\a N\, ?$$
Is there an interpretation where traces over $N\times N$ matrices
are replaced by a supertrace and in turn bosonic matrix integrals by
fermionic ones?
\item Why the factor $\a$ in the transformation
$$N\rightarrow -\a N\, ?$$
Is there a generalization of the GSE dual to the GOE at odd values
of $N$?
\end{enumerate}

After this disquisitive interlude, we return to the postponed
question. Let us examine the generality of Lemma~\ref{puncture}
which claimed that a topological invariant of a punctured surface 
$S_\Gamma$ was obtained by counting signed configurations of
units $\{1,e_1\ldots,e_{2\a-1}\}$ on the associated graph $\Gamma$.
Its proof relied on the following: (i) The units all square to $\pm
1$ and $\{\pm 1,\pm e_i\}$ is a group. 
(ii) At any vertex, their product was $\pm 1$ and therefore real.
(iii) Units whose square was ($-$)1 were joined by Wick contractions
of (anti)symmetric matrices. 

Therefore generalized matrix models whose graphical expansion is
completely determined via our methods can be written down
based on a larger group of ``imaginary'' units $\{\pm 1,\pm f_a;\pm
e_i:f_a^2=1,e_i^2=-1\}$. 
Simple examples
are generated by considering elements drawn from Clifford algebras.
{\it I.e.}, the Pauli matrix representation of the quaternions can
be generalized to larger sets of higher dimensional Dirac matrices.
Although these Clifford type models seem not to generate
theories with the new values of $\alpha$ exhibited in the Penner
model, it would be interesting to investigate whether new matrix
models of this type can indeed be constructed.

\section*{Acknowledgments}

The authors thank G.~Kuperberg, M.~Penkava, A.~Schilling,
A.~Schwarz, A.~Soshnikov, W.~Thurston and J.~Yu for 
both stimulating and useful discussions.

\begin{appendix}

\section{Quaternionic Feynman Calculus}

\label{Qcalculus}

Since the quaternions are the last real associative division algebra,
it is natural to develop a manifestly quaternionic Feynman calculus.
Again, let us consider GOE, GUE and GSE models all at once via the
unified notation
\begin{equation}
X=S+\sum_{i=1}^{\beta-1} e_i A_i
\end{equation}
where the $N\times N$  matrices $X$ are real, complex or quaternionic
self-adjoint
\begin{equation}
X^\dagger\equiv\overline X^\top=X\, ,\qquad
\overline e_i\equiv -e_i\, ,
\end{equation} 
depending on the value of $\beta=$ $1$, $2$ or $4$, respectively.

To begin, we need a shift identity
\begin{equation}
\exp\left(\frac{1}{2}\,\tr\!\Big(Q^\top X+X Q^\top\Big)\right)\,f(Q)=f(X+Q)\, 
,\label{id}
\end{equation}
where the ``quantum variable''
\begin{equation}
Q\equiv\wh S+\sum_i e_i \wh A_i=Q^\dagger
\end{equation}
and the $N\times N$ matrix of derivatives $\d$ is given by
\begin{equation}
\d_{ab}=\left\{\begin{array}{ll}\frac{1}{2}\Big(\frac{\partial}{\partial
\wh S^{ab}}-\sum_i e_i \frac{\d}{\d \wh A^{ab}_i}\Big)\, ,&a\neq b\, ,\\ \\
\frac{\d}{\d \wh S^{aa}}\, ,&a=b\, .
\end{array}
\right. 
\label{d}
\end{equation}
The identity~\eqn{id} holds thanks to the commutation relations
\begin{equation}
\Big[\frac{1}{2}\tr\!\Big(\d^\top X+X \d^\top\Big),Q\Big]=X\, .
\end{equation}
Note also that
\begin{equation}
\Big[\d,\frac{1}{2}\,\tr\!\Big(Q^\top X+X Q^\top\Big)\Big]=Q\, ,
\end{equation}
although $[\d, Q]=I$ only for the real and complex cases $\beta
=1,2$.

We may now rewrite matrix integration as differentiation\footnote{The result 
is equivalent to the one obtained using a Schwinger source term. The
reformulation presented here is often called the background field
formalism, a simple account may be found in the on-line
textbook~\cite{Siegel}.}
\begin{multline}
\int [dX]_{_{\!\ss(4)}}\!\,\exp\Big(\!-\frac{1}{2}\ \tr X^2
+\sum_{j=1}^\infty \frac{ t_j}{j}\ \tr X^j \Big)
\\
\hspace{-2.7cm}
=\left.\int [dX]_{_{\!\ss(4)}}\!\!\!\,\exp\Big(\!-\frac{1}{2}\ \tr X^2
+\sum_{j=1}^\infty \frac{ t_j}{j}\ \tr (X+Q)^j \Big)\right|_{Q=O}
\\
\!\!=\!
\left.\int \![dX]_{_{\!\ss(4)}}\!\!\!\,
\exp\!\Big(\!-\frac12 \tr \Big(X-\d^\top\Big)^2
\!+\frac12\ \tr {\d^\top}^2\Big)
\exp\!\Big(\sum_{j=1}^\infty \frac{t_j}{j}\ \tr Q^j \Big)\!\right|_{Q=O}
\\
=\!\left.
\left(\!\int [dX]_{_{\!\ss(4)}}\!\!\!\,\exp\!\Big(\!-\frac{1}{2}\ \tr 
X^2\Big)\right)\;\exp\!\Big(\frac12\ \tr {\d}^2\Big)
\exp\!\Big(\sum_{j=1}^\infty \frac{t_j}{j}\ \tr Q^j \Big)\right|_{Q=O} .\;
\label{hamlet}
\end{multline}
The first factor on the last line is just an overall normalization 
while the two exponentials can be expanded in terms of Feynman diagrams:
the $n$-th order term in the expansion of each exponential
is interpreted as either $n$ edges or $n$ vertices, respectively.

Let us give some details: 
The operator $\12\ \tr {\d}^2$ acting on a quantum variable, 
yields
\begin{equation}
[\12\ \tr {\d}^2,Q]=\d^\top\, ,
\end{equation}
which is represented graphically as attaching a ribbon edge to a vertex
since the operator $\12\ \tr {\d}^2$ is to be viewed as an edge.
Note that by this rule,  a vertex emitting a $Q_{ab}$ is replaced with one
emitting $\d_{ba}$ which amounts to a twist.
Attaching the other end to an adjacent vertex yields 
\begin{equation}
\Big(\d_{ab}Q_{cd}\Big)=
\12\ \beta\,\delta_{ac}\delta_{bd}+
\12\ (2-\beta)\,\delta_{ad}\delta_{bc}\, .
\label{prop}
\end{equation}
Note that the brackets on the left hand sides above
indicate that we are computing the 
derivative rather allowing it to continue acting to the right as an
operator. In particular, observe that for the GUE case, $\beta=2$
so no twisted ribbon graphs can appear.

\begin{figure}
\begin{picture}(300,200)(40,0)
\thicklines
\put(40,100){\epsfig{file=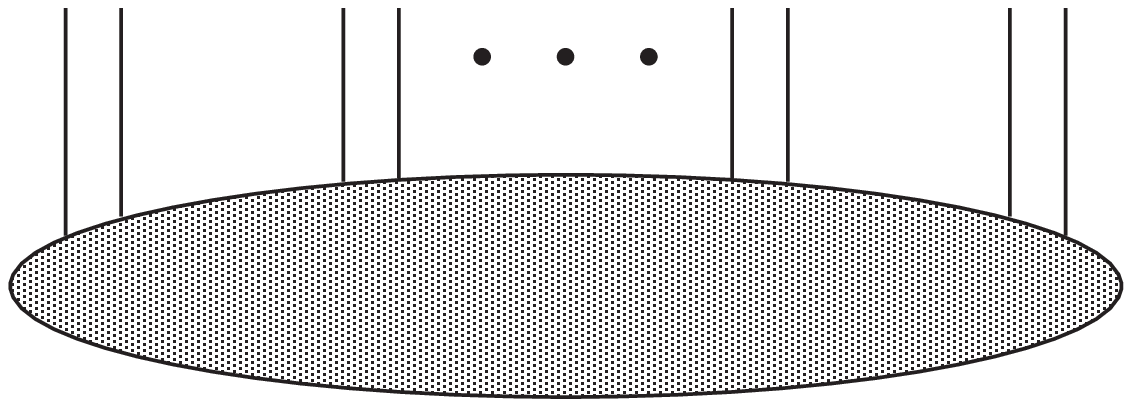,width=1.3in}}
\put(169,100){\epsfig{file=figsnail1.eps,width=1.3in}}
\put(50,10){\epsfig{file=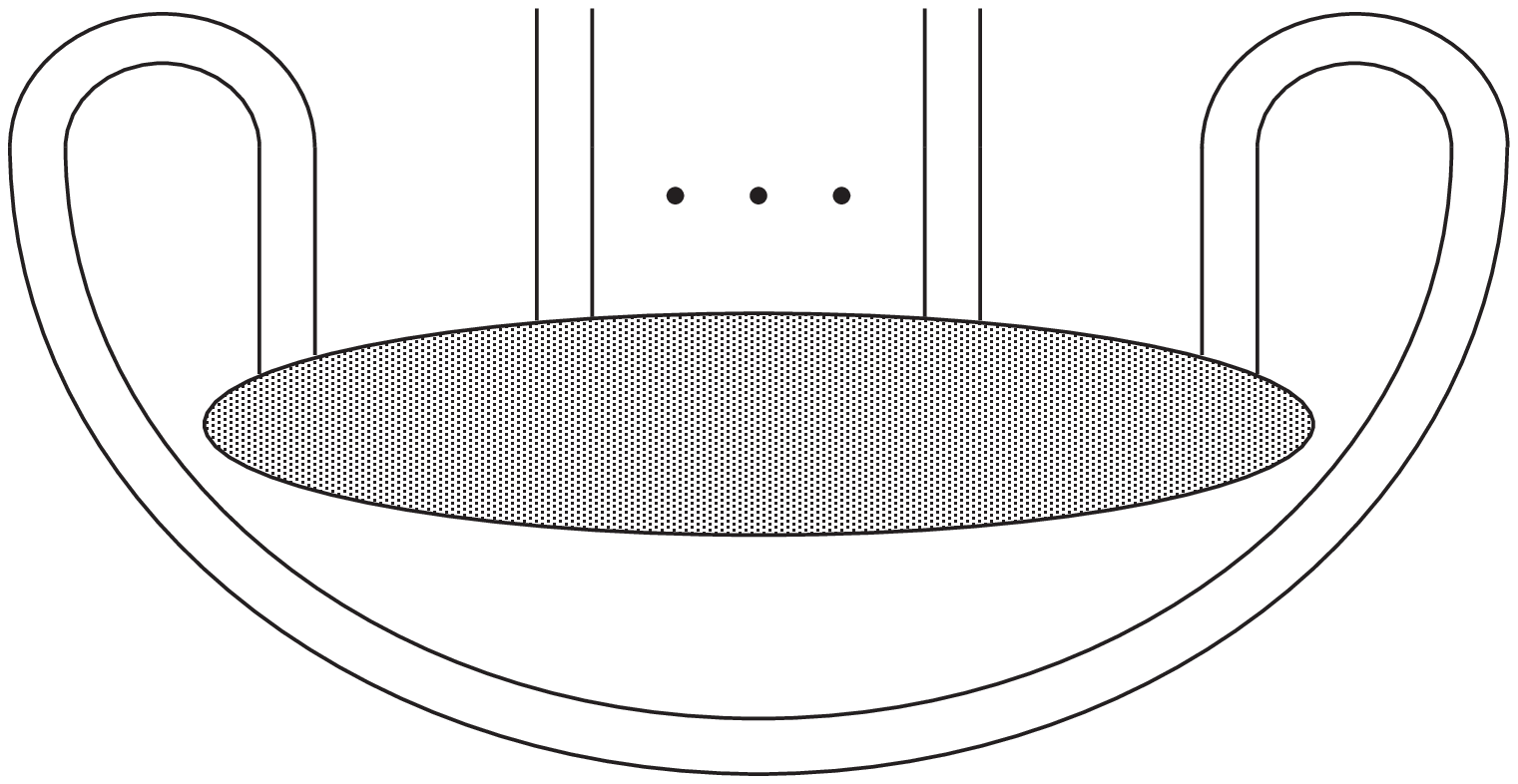,width=1.3in}}
\put(160,10){\epsfig{file=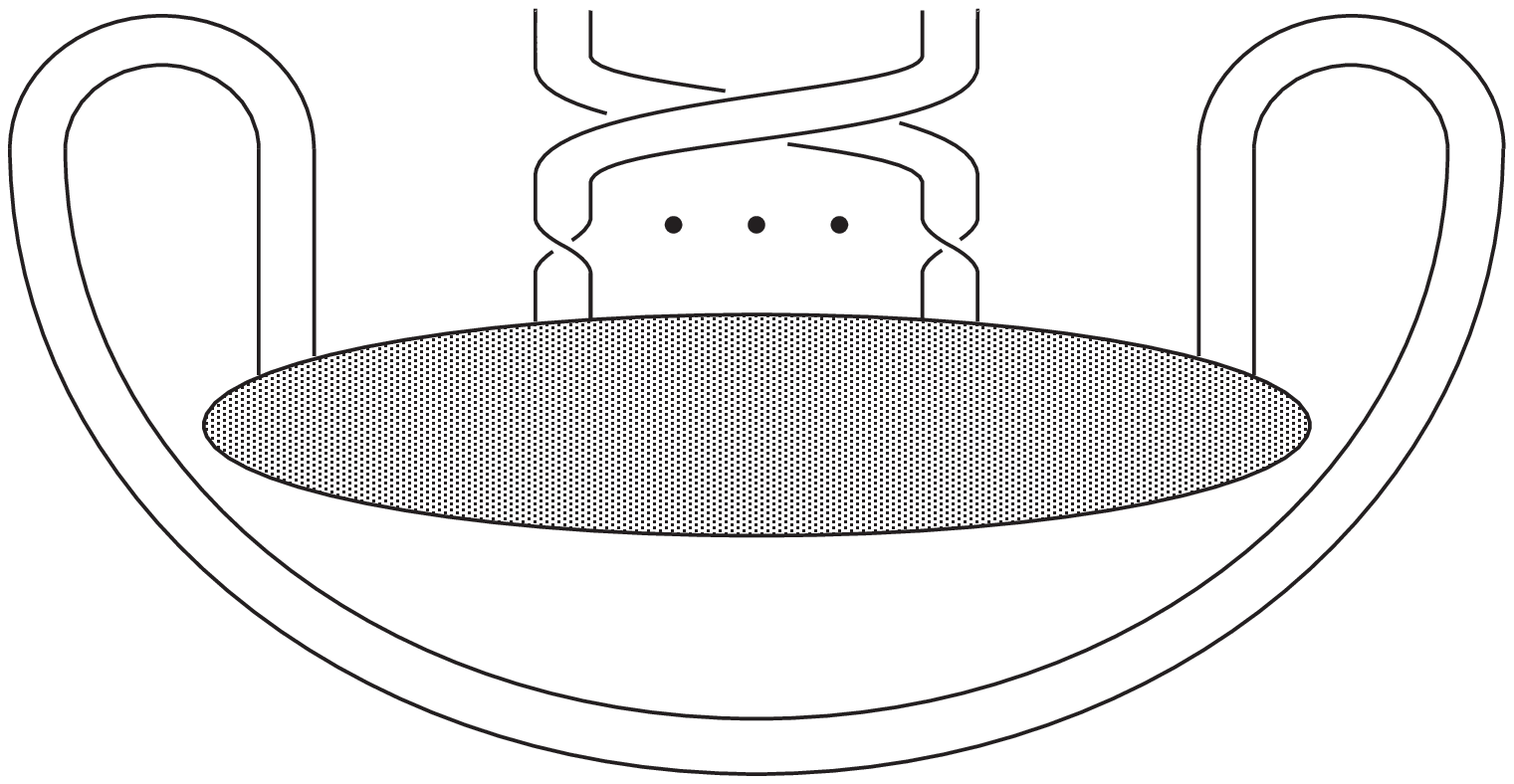,width=1.3in}}
\put(271,10){\epsfig{file=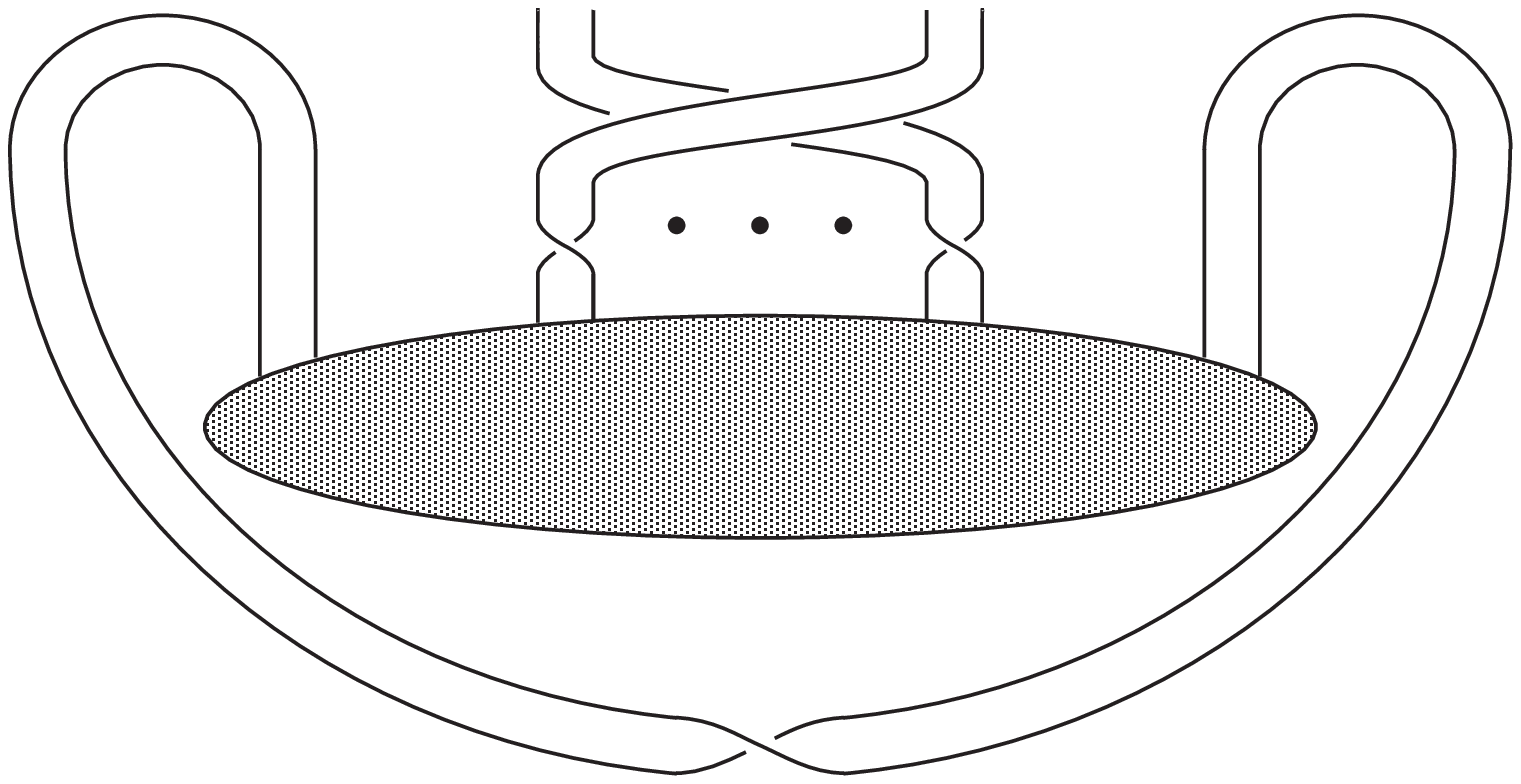,width=1.3in}}
\put(34,145){$\;\;\,\partial^\top\,\Big(\;\quad \mathcal{Q}\;\, \quad\;\;\, \;Q\, \Big)$}
\put(165,145){$\Big(\partial^\top\;\quad\;\;\, \mathcal{Q}\quad\; \Big)\;\;\, 
 Q\,$}
\put(69,133){$\overbrace{\qquad\quad}$}
\put(197,133){$\overbrace{\qquad\quad}$}
\put(145,120){$\Large -$}
\put(30,40){$\Large =$}
\put(147,40){$\Large -$}
\put(258,40){$\Large +$}
\put(92,72){$\mathcal{Q}$}
\put(180,72){$\frac{2-\b}{2}\;\overline{\mathcal{Q}}$}
\put(290,72){$\frac{2-\b}{2}\;\overline{\mathcal{Q}}$}
\put(83,60){$\overbrace{\;\quad\quad}$}
\put(193,60){$\overbrace{\;\quad\quad}$}
\put(303,60){$\overbrace{\;\quad\quad}$}
\end{picture}
\caption{Quaternionic Feynman rule. Observe how connecting vertices
with edges changes vertices yet to be connected.
}
\label{elephants}
\end{figure}

For the GOE and GUE models we are done, one simply attaches all possible
edges to vertices using the above rules and finds the usual known results. 
The symplectic case is more
subtle however, thanks to the quaternionic 
non-commutativity of $\d$ and $Q$. In
particular 
\begin{equation}
\Big(\d_{ab} f(Q) \ Q_{cd}\Big)\neq\Big(\d_{ab} f(Q)\Big)\ Q_{cd}+ f(Q)\ 
\Big(\d_{ab}Q_{cd}\Big)\, .
\end{equation}
for some function $f$ of the quaternionic matrix $Q$.
{\it I.e.}, the quaternion valued operator $\d$ does not  satisfy the
Leibniz rule. However, a generalized Leibniz rule does apply: 
First note that
for any 
\begin{equation}
{\mathcal Q}={\mathcal S} + \sum_\a e_\a {\mathcal A}_\a
\end{equation}
where the real matrices ${\mathcal S}$ and ${\mathcal A}_\a$
need not have any definite symmetry properties 
(so that ${\mathcal Q}$ is not necessarily self adjoint)
we have\footnote{When $\beta=1$, the sum on the left hand side is empty
and equal to zero, while the right hand side vanishes for real
${\mathcal Q}$.}
\begin{equation}
\sum_\a e_\a  {\mathcal Q} \ e_\a =
-\ {\mathcal Q}+(2-\beta)\ \overline{\mathcal Q}\; . 
\end{equation}
Therefore we have the generalized Leibniz rule
\begin{equation}
\d_{ab}\Big( {\mathcal Q} \ Q_{cd}\Big)-\Big(\d_{ab} {\mathcal Q}\Big)\ Q_{cd}
=\delta_{ac}\delta_{bd}\ {\mathcal Q}-\12(2-\beta)(\delta_{ac}\delta_{bd}
-\delta_{ad}\delta_{bc})\overline{\mathcal Q}\, .
\end{equation}
Note that for $\b=1,2$ the right hand side is equal to ${\mathcal Q} \Big(
\d_{ab} Q_{cd}\Big)$, expressing the commutativity of 
real and complex numbers. This relation is the central 
graphical rule for our quaternionic Feynman calculus and is depicted
in Figure~\ref{elephants}.
Note that it reverts to the rule~\eqn{prop} when ${\mathcal Q}=1$.

It is now possible to compute any term in the expansion
of~\eqn{hamlet} in terms of graphs. 
However, for quaternionic matrices, 
when connecting vertices with ribbon edges, 
intermediate vertices and unconnected edges may be twisted and/or flipped
according
to~\eqn{elephants}. It is easy, but  tedious to verify that
the results for simple graphs coincide with our general 
formula~\eqn{eq: invariant formula}.

\section{Generalized Penner Model}

\label{don't-think--type}

In this Appendix, we derive the asymptotic expansion formula
\begin{equation}
\begin{split}
&K(z,N,\alpha)
=\lim_{m\rightarrow\infty}
\log\left(\frac{
\int_{{\mathbb R}^N}\Delta^{2\alpha}(k)\prod_{i=1}^N
\exp\Big(-\sum_{j=2}^{2m} \frac{z^{j/2-1}}{j}\ k_i^j\Big) dk_i}
{\int_{{\mathbb R}^N}\Delta^{2\alpha}(k)\prod_{i=1}^N
\exp\Big(- \frac{k_i ^2}{2}\Big) dk_i}
\right)\\
&= \sum_{m=1} ^\infty
\frac{ b_{2m}}{2m\ (2m-1)}\; N z^{2m-1}+
\sum_{m=1} ^\infty \frac{1}{4m} (-1)^m \alpha^m N^m z^m\\
&+\;\frac{1}{2}\;\sum_{m=1} ^\infty 
\sum_{q=0}^{[\frac{m}{2}]}
\frac{(-1)^m (m-1)!b_{2q}}{(2q)!(m+1-2q)!}
\alpha^m (\alpha^{1-2q} -1)N^{m+1-2q} z^m\\
&-\sum_{m=1} ^\infty 
\sum_{q=0}^{[\frac{m}{2}]}
\sum_{s=0}^{[\frac{m+1}{2}]-q}
 \frac{(-1)^m(m-1)!b_{2q}b_{2s}}{(2q)!(2s)!(m+2-2q-2s)!}
\alpha^{m+1-2q} N^{m+2-2q-2s} z^m\, , 
\label{eq:K_app}
\end{split}
\end{equation}
which is valid for every positive integer $\alpha$.
Here the  $b_n$'s are the Bernoulli numbers defined by
$$
\sum_{n=0}^\infty \; b_n\;\frac{t^n}{n!}=\frac{t}{e^t-1}\, .
$$
The key techniques are the Selberg integration formula,
Stirling's formula for $\Gamma(1/z)$ and the asymptotic
analysis of \cite{Mulase95}. First we note that as an
asymptotic series in $z$ when $z\rightarrow 0$ while keeping
$z>0$, we have
\begin{equation}
\begin{split}
&\lim_{m\rightarrow\infty}
\log\left(
\int_{{\mathbb R}^N}\Delta^{2\alpha}(k)\prod_{i=1} ^N
\exp\Big(-\sum_{j=2} ^{2m} \frac{z^{j/2-1}}{j}\ k_i ^j\Big) dk_i
\right)\\
&
= \log\left(
\left(z^{\frac{1}{2}} e^{\frac{1}{z}} z^{\frac{1}{z}}\right)^n
z^{\frac{\alpha N(N-1)}{2}}
\int_{[0,\infty)^N} \Delta^{2\alpha}(k) \prod_{i=1} ^{N}
e^{k_i} k_i ^{\frac{1}{z}} dk_i\right).
\end{split}
\end{equation}
(For the mechanism changing the integration from $\mathbb{R}^N$
to $[0,\infty)^N$, we refer to~\cite{Mulase95}.) This integral
can be calculated by the Selberg integration formula:
$$
\int_{[0,\infty)^N} \Delta^{2\alpha}(k) \prod_{i=1} ^{N}
e^{k_i} k_i ^{\frac{1}{z}} dk_i =
\prod_{j=0} ^{N-1}
\frac{\Gamma(1+\alpha + j\alpha)\Gamma(1+\frac{1}{z}+j\alpha)}
{\Gamma(1+\alpha)}.
$$
Therefore, 
we have
\begin{multline}
\label{eq: asym1}
\lim_{m\rightarrow\infty}
\log\left(
\int_{{\mathbb R}^N}\Delta^{2\alpha}(k)\prod_{i=1} ^N
\exp\Big(-\sum_{j=2} ^{2m} \frac{z^{j/2-1}}{j}\ k_i ^j\Big) dk_i
\right)\\
=
c+\frac{N}{2}\log z + \frac{N}{z}+\frac{N}{z}\log z
+ \frac{\alpha N(N-1)}{2}\log z \\
\qquad+\log\prod_{j=0} ^{N-1}
\Gamma\left(1+\frac{1}{z}+j \alpha\right),
\end{multline}
where $c$ is the constant term independent of $z$.
Since \eqn{eq:K_app} does not have any constant term
relative to $z$, here and below we ignore all constant 
terms independent of $z$ (but possibly $N$ dependent).
The  product of $\Gamma$-functions can be calculated
by the recursion formula,
noticing that $\alpha$ is an integer:
\begin{equation}
\begin{split}
&\prod_{j=0} ^{N-1}
\Gamma\left(1+\frac{1}{z}+j \alpha\right)\\
&=
\Gamma(1/z)^N \prod_{i=0}^{N-1}\prod_{j=0}^{i\alpha}
\left(\frac{1}{z}+i\alpha -j\right)\\
&=
\Gamma(1/z)^N \left(\frac{1}{z}\right)^N\prod_{i=0} ^{N-1}
\prod_{j=1}^\alpha \left(\frac{1}{z}+i\alpha +j\right)^{N-i-1}\\
&=
\Gamma(1/z)^N \left(\frac{1}{z}\right)^N\prod_{i=1} ^{N-1}
\prod_{j=0}^{\alpha-1} \left(\frac{1+z(1+(i-1)\alpha +j)}{z}\right)^{N-i}.
\end{split}
\end{equation}
We now apply Stirling's formula for $\log\Gamma(1/z)$ to obtain,
up to a constant term:
\begin{equation}
\label{eq:stirling}
\begin{split}
&\log\prod_{j=0} ^{N-1}
\Gamma\left(1+\frac{1}{z}+j \alpha\right)\\
&=
-\frac{N}{z}\log z -\frac{N}{z}+\frac{N}{2}\log z +
\sum_{m=1} ^\infty \frac{b_{2m}}{2m(2m-1)} N z^{2m-1}\\
&-N\log z -\frac{\alpha N(N-1)}{2}\log z\\
&+\sum_{m=1} ^\infty \sum_{i=1} ^{N-1}
\sum_{j=0} ^{\alpha-1} \frac{1}{m}(-1)^{m-1}(N-i)
(1+(i-1)\alpha +j)^m z^m.
\end{split}
\end{equation}
We note that all negative powers of $z$ and $\log z$ related terms
in~\eqn{eq: asym1} cancel out using~\eqn{eq:stirling}. Finally, we obtain
\begin{multline}
\label{eq:powersum}
K(z,N,\alpha)= \sum_{m=1} ^\infty \frac{b_{2m}}{2m(2m-1)} N z^{2m-1}\\
+\sum_{m=1} ^\infty \sum_{i=0} ^{N-1}
\sum_{j=1} ^{\alpha} \frac{1}{m}(-1)^{m-1}(N-1-i)
(i\alpha +j)^m z^m.
\end{multline}
This last sum of powers can be calculated using Bernoulli
polynomials, from which~\eqn{eq:K_app} follows.

Using a formula
for Bernoulli numbers
$$
(1-2n)b_{2n}=\sum_{q=0} ^n \binom{2n}{2q}b_{2q}b_{2n-2q}
=\sum_{q=0} ^n \binom{2n}{2q}b_{2q}b_{2n-2q}2^{2q},
\quad n\ne 1,
$$
and noting that $b_2=1/6$,
we recover  the original formula of Penner for 
$\alpha=1$~\cite{Penner}:
\begin{equation}
\begin{split}
K(z,N,1)&= -\sum_{m=1} ^\infty
\frac{ b_{2m}}{2m}\; N z^{2m-1}\\
&+\sum_{m=1} ^\infty 
\sum_{q=0}^{[\frac{m}{2}]}
\frac{(m-1)!(2q-1)}{(2q)!(m+2-2q)!}b_{2q}
N^{m+2-2q} (-z)^m\\
&=\sum_{\substack{g\ge 0, n>0\\2-2g-n<0}}
\frac{(2g+n-3)!(2g-1)}{(2g)!n!}b_{2g} N^n (-z)^{2g+n-2}\\
&=\sum_{\Gamma\in\mathfrak{R}}
\frac{(-1)^{e_\Gamma}}{|\Aut_\mathfrak{R}(\Gamma)|}
N^{f_\Gamma}(-z)^{e_\Gamma-v_\Gamma}.
\end{split}
\end{equation}
For $\alpha=2$,~\eqn{eq:K_app} simplifies again:
\begin{equation}
\label{eq:K2}
\begin{split}
K(z,N,2)
&= -\sum_{m=1} ^\infty
\frac{ b_{2m}}{2m}\; N z^{2m-1}\\
&+\;\frac{1}{2}\;\sum_{m=1} ^\infty 
\sum_{q=0}^{[\frac{m}{2}]}
\frac{(m-1)!(2q-1)}{(2q)!(m+2-2q)!}b_{2q}
2^{m+2-2q}N^{m+2-2q} (-z)^m\\
&-\;\frac{1}{2}\;\sum_{m=1} ^\infty 
\sum_{q=0}^{[\frac{m}{2}]}
 \frac{(m-1)!(2^{2q-1}-1)}{(2q)!(m+1-2q)!}b_{2q}
2^{m+1-2q} N^{m+1-2q} (-z)^m .
\end{split}
\end{equation}
Note that the first two lines of~\eqn{eq:K2} are identical
to the Penner model $\frac{1}{2}K(z,2N,1)$. 

The following integral formula, again valid for every 
positive integer
$\gamma\in{\mathbb N}$, has been established in 
\cite{Goulden-Harer-Jackson}:
\begin{equation}
\label{eq:J_app}
\begin{split}
&J(z,N,\gamma)
=\lim_{m\rightarrow\infty}
\log\left(\frac{
\int_{{\mathbb R}^N}|\Delta(k)|^{2/\gamma}\prod_{i=1}^N
\exp\Big(-\sum_{j=2}^{2m} \frac{z^{j/2-1}}{j}\ k_i^j\Big) dk_i}
{\int_{{\mathbb R}^N}|\Delta(k)|^{2/\gamma}\prod_{i=1}^N
\exp\Big(- \frac{k_i ^2}{2}\Big) dk_i}
\right)\\
&=\sum_{m=1} ^\infty
\frac{ b_{2m}}{2m\ (2m-1)}\; \frac{N}{\gamma} \left(
\frac{z}{\gamma}\right)^{2m-1}+
\sum_{m=1} ^\infty \frac{1}{4m} (-1)^m N^m
\left(\frac{z}{\gamma}\right)^m \\
&-
\;\frac{1}{2}\;\sum_{m=1} ^\infty 
\sum_{q=0}^{[\frac{m}{2}]}
\frac{(-1)^m (m-1)!b_{2q}}{(2q)!(m+1-2q)!}
 \left(1-\frac{1}{\gamma^{1-2q}}\right)N^{m+1-2q} 
\left(\frac{z}{\gamma}\right)^m\\
&-
\sum_{m=1} ^\infty 
\sum_{q=0}^{[\frac{m}{2}]}
\sum_{s=0}^{[\frac{m+1}{2}]-q}\!\!
 \frac{(-1)^m(m-1)!b_{2q}b_{2s}}{(2q)!(2s)!(m+2-2q-2s)!}\cdot
\frac{1}{\gamma^{1-2s}} N^{m+2-2q-2s} 
\left(\frac{z}{\gamma}\right)^m \! .
\end{split}
\end{equation}
Our duality~\eqn{eq:Penner dual} for the generalized Penner model
follows from comparing \eqn{eq:K_app} with
\eqn{eq:J_app}.

\end{appendix}

\bibliographystyle{amsplain}
\bibliography{Q}

\end{document}